\documentclass[12pt,letterpaper]{article}
\usepackage{epsfig,rotating,setspace,latexsym,amsmath,epsf,amssymb,amsfonts,bm,theorem,cite,caption,subcaption,enumerate,longtable,accents}
\usepackage{algorithm,algorithmic,graphicx,epsf,authblk,epstopdf,url,color,multirow}
\usepackage{mathtools}
\usepackage{comment}
\usepackage{graphicx}
\usepackage{amsmath}

\newtheorem{theorem}{Theorem}

\newtheorem{definition}{Definition}
\newtheorem{remark}{Remark}
\newtheorem{lemma}{Lemma}

\newenvironment{Proof}[1]{\medskip\par\noindent{\bf Proof:\,}\,#1}{{\mbox{\,$\blacksquare$}\par}}

\newcommand{\bF}{{\mathbb{F}}}

\allowdisplaybreaks

\begin{document}

\title{Private Read Update Write (PRUW) in Federated Submodel Learning (FSL): Communication Efficient Schemes With and Without Sparsification\thanks{This work was supported by ARO Grant W911NF2010142, and presented in part at IEEE ICC 2022.}}

\author{Sajani Vithana \qquad Sennur Ulukus\\
	\normalsize Department of Electrical and Computer Engineering\\
	\normalsize University of Maryland, College Park, MD 20742 \\
	\normalsize {\it spallego@umd.edu} \qquad {\it ulukus@umd.edu}}

\date{}

\maketitle

\begin{abstract}
We investigate the problem of private read update write (PRUW) in relation to private federated submodel learning (FSL), where a machine learning model is divided into multiple submodels based on the different types of data used to train the model. In PRUW, each user downloads the required submodel without revealing its index in the reading phase, and uploads the updates of the submodel without revealing the submodel index or the values of the updates in the writing phase. In this work, we first provide a basic communication efficient PRUW scheme, and study further means of reducing the communication cost via sparsification. Gradient sparsification is a widely used concept in learning applications, where only a selected set of parameters is downloaded and updated, which significantly reduces the communication cost. In this paper, we study how the concept of sparsification can be incorporated in private FSL with the goal of reducing the communication cost, while guaranteeing information theoretic privacy of the updated submodel index as well as the values of the updates. To this end, we introduce two schemes: PRUW with top $r$ sparsification and PRUW with random sparsification. The former communicates only the most significant parameters/updates among the servers and the users, while the latter communicates a randomly selected set of parameters/updates. The two proposed schemes introduce novel techniques such as parameter/update (noisy) permutations to handle the additional sources of information leakage in PRUW caused by sparsification. Both schemes result in significantly reduced communication costs compared to that of the basic (non-sparse) PRUW scheme.      
\end{abstract}

\section{Introduction}

Many engineering applications at present are driven by various forms of learning techniques. These learning models require a large amount of data and processing power in order to provide accurate outcomes. The increasing data and processing power requirements as well as the privacy concerns of data providers hinder the usage of central serves that perform both data collection/storage and processing in learning applications. One of the main solutions to both processing power limitations and privacy concerns is federated learning (FL) \cite{FL1,FL2,Advances,magazine}. In FL, a central server stores the learning model and shares it with different users containing data that can be used to train the model. Consequently, the users train the model using their own data and communicate only the updates (gradients) with the central server, which then aggregates these gradients and updates the existing learning model. In this way, the processing power requirements are decentralized and the users' data privacy is \emph{partially} preserved. Privacy in basic FL is only \emph{partial}, as it has been shown that even the gradients shared by the users in FL leak information about the users' private data \cite{comprehensive,MembershipInterference,featureLeakage,SecretSharer,InvertingGradients,DeepLeakage,BeyondClassRepresentatives}. Different methods have been developed to minimize this information leakage in FL such as classical cryptographic protocols as in secure aggregation \cite{PracticalSecureAgg} and differential privacy \cite{DP} via noise addition, data sampling and data shuffling, e.g., \cite{reinforcement,avgDP,cpSGD,PrivacyAmp,cross,language,DPFL,shuffle,PrivacyBlanket,shuffledDPFL,client,recent,aggregation}. In this work, we address the privacy problem as well as the communication cost problem in FL as described next.

The communication cost of FL is considerably high since large machine learning models and the corresponding gradients need to be communicated between the central server and the users in multiple rounds. Furthermore, FL requires each user to download and train the entire machine learning model even in cases where the users engaged in the FL process are as small as mobile phones that do not contain all types of data required to train the entire model. As solutions to these communication inefficiencies, several methods such as gradient sparsification \cite{sparse1,GGS,adaptive,conv,conv2,overtheair,rtopk,timecorr}, gradient quantization \cite{qsl,fedpaq,qsgd,constraints} and federated submodel learning (FSL) \cite{billion,secureFSL,paper1,rw_jafar,ourICC,dropout,pruw,sparse,rd} have been introduced. In gradient sparsification, the users only communicate a selected set of gradients (most significant/randomly chosen) as opposed to sending all gradient updates corresponding to all parameters in the model to the central server. In gradient quantization, the values of the gradients are quantized and represented with fewer bits. In FSL, the machine learning model is divided into multiple submodels based on the different types of data used to train the entire model, and each user only downloads and updates the submodel that can be updated by the user's own local data. This saves communication cost and makes the distributed learning process more efficient.  

FSL comprises two main phases, namely, the \emph{reading phase} in which a user downloads the required submodel and the \emph{writing phase} in which the user uploads the update of the relevant submodel. Although FSL is efficient in terms of communication cost and processing power of local users, it introduces an important issue with respect to user privacy. The submodel that a given user updates may leak information on the type of data the user has. Moreover, as in FL, the values of the updates uploaded by a user may leak information about the local data of the user. Consequently, in order to guarantee the privacy of a user, two quantities need to be kept private from the central model (databases that contain all submodels), namely, 1) the index of the submodel updated by each user, and 2) the value of the update. In general, the problem of reading a required section of a given storage system and writing back to the same section while guaranteeing the privacy of the section read/written as well as the content written is known as private read update write (PRUW). Private FSL is a specific application of PRUW. The \emph{reading phase} of private FSL requires the user to hide the index of the submodel it reads (downloads). From an information theoretic privacy point of view, this is equivalent to the problem of private information retrieval (PIR), see e.g.,\cite{original,PIR,leaky,ChaoTian,semanticPIR,colluding,coded,incorrectconjecture,codedcolluded,sideinfo,singleDB,byzantine,SecureStorage,XSTPIR,CodeColludeByzantinePIR,smallfields,Kumar_PIRarbCoded,YamamotoPIR,VardyConf2015,MultiroundPIR,SPIR,MMPIR,evesdroppers,cache,PSI}. The \emph{writing phase} requires the user to write (upload) the updates back to the relevant submodel without revealing the submodel index or the values of the updates.

Existing works on PRUW (and private FSL) \cite{billion,secureFSL,paper1,rw_jafar,ourICC,dropout,pruw,sparse,rd} provide schemes with different notions of privacy.  References \cite{billion,secureFSL} consider locally differential privacy, in which a predetermined amount of information of the user is leaked to the databases. Reference \cite{aggregation} presents a group-wise aggregation scheme (related to the writing phase in FSL) based on local differential privacy. The schemes in \cite{paper1,rw_jafar} consider information theoretic privacy of the submodel index and the values of the updates. However, they are less efficient in terms of the communication cost compared to the schemes presented in \cite{dropout,ourICC} (and their recent variants in \cite{sparse,rd,pruw}) for the same notion of privacy.

Based on the existing works on PRUW that consider information theoretic privacy, the lowest known communication cost is achieved in a setting where the equal sized submodels are stored in $N$ non-colluding databases, that are accessible by each individual user. In this setting, an asymptotic reading and writing cost (equal reading and writing costs) of twice the size of the submodel can be achieved. The corresponding achievable schemes are described in \cite{ourICC,dropout},\footnote{A basic version of these schemes was first introduced in \cite{rw_jafar}, which we subsequently improved in \cite{ourICC}. The authors of \cite{rw_jafar} also improved the basic scheme in \cite{dropout} (also extended to drop-outs and coded storage). The improvements of the basic scheme from \cite{rw_jafar} to \cite{ourICC} and from \cite{rw_jafar} to \cite{dropout} are independent and concurrent.} which are based on cross subspace allignment (CSA)\cite{CSA}. In this setting, it is assumed that each individual user updates all parameters of the required submodel. However, the communication cost can be further reduced by only downloading and updating a selected number of parameters within the submodel. This is in fact gradient sparsification in FSL. Sparsification in FSL can be performed in two main ways, namely, top $r$ and random sparsification. In top $r$ sparsification, only a given fraction of the most significant parameters are downloaded and updated in the reading and writing phases. In random sparsification, a random set of parameters is read in the reading phase, and the same/different random set of parameters is updated in the writing phase. A given amount of distortion is introduced in both sparsification methods, which in general has little or no impact on the accuracy of the model. In fact, sparsification is a widely used technique in most learning tasks to reduce the communication cost, which even performs better than the non-sparse models in certain cases.

In this work, we present two schemes that perform PRUW with sparsification in relation to private FSL. Each of the two schemes correspond to the two means of sparsification: top $r$ and random, respectively. In private FSL with top $r$ sparsification, each user only updates the most significant $r$ fraction of parameters in the writing phase, and downloads only another fraction $r'$ of parameters in the reading phase.\footnote{The $r'$ fraction of parameters can be the union of the sparse sets of parameters updated by all users in the previous iteration, or they can be chosen in a specific way as in \cite{GGS}.} This ensures that the most significant gradient variations in the training process are communicated while incurring significantly reduced communication costs compared to non-sparse training. With no constraints on the storage costs, the asymptotic reading and writing costs with top $r$ sparsification can be as low as $2r$ times the size of a submodel, where $r$ is typically around $10^{-2}$ and $10^{-3}$. The main challenge in top $r$ sparsification in private FSL is satisfying the privacy constraint on the values of updates. Note that the users are unable to simply send the sparse updates by specifying their positions directly, as it reveals the values of the updates (zero) of the parameters whose positions are not specified, which violates the privacy constraint on the values of updates in private FSL. In other words, PRUW with top $r$ sparsification requires three components to be kept private: 1) updated submodel index, 2) values of sparse updates, 3) positions of sparse updates. In this work, the privacy of the first two components is ensured using similar techniques as in \cite{ourICC,dropout,pruw,sparse,rd} and the third component is kept private using a (noisy) parameter shuffling mechanism, which in turn requires extra storage space in databases.

In this paper, we also propose a scheme for PRUW with random sparsification, which focuses on finding the optimum reading and writing subpacketizations, where only a single bit is read and written per subpacket per database. In the reading phase, the collection of these single bits from all databases is used to decode the sparse parameters of each subpacket. In the writing phase, the sparse updates of each subpacket are combined into a single bit and sent to the databases. Each database privately decomposes these single bits into their respective updates and places them at the relevant positions using some fixed queries. In this scheme, a randomly selected set of parameters in each \emph{writing subpacket} is always updated and the same/different set of parameters of each \emph{reading subpacket} is always downloaded, which results in a given amount of distortion. The analysis of the costs in this scheme is formulated in terms of a rate-distortion trade-off, where we use the proposed scheme to achieve the minimum reading and writing costs for given amounts of allowed distortions in the reading and writing phases. This scheme achieves slightly lower reading and writing costs compared to the top $r$ sparsification scheme for similar sparsification rates, while not requiring any additional storage in databases. However, the sparsification is random in this case, which does not promote the most significant updates/parameters in the sparsification process. This may have an adverse impact on the convergence time of the training process and the accuracy of the model.

The main contributions of this work are as follows: 1) basic PRUW scheme with non-colluding databases that is over-designed with the optimum number of random noise terms in storage to minimize the communication cost; 2) introduction of the concept and system models for sparsification in PRUW; 3) scheme for PRUW with top $r$ sparsification that satisfies information theoretic privacy of submodel index and values of parameter updates (including the positions of the sparse updates); 4) scheme for PRUW with random sparsification; and 5) characterization of the rate-distortion trade-off in PRUW.

\section{Basic PRUW}\label{scheme}

In this section, we formally describe the PRUW problem setting and explain the PRUW scheme presented in \cite{ourICC} and \cite{dropout} in detail, for the special case of non-colluding databases with uncoded data storage. This is the basic scheme which the schemes proposed in Section~\ref{topr} and Section~\ref{rndm} will be built on.

\subsection{PRUW Problem Setting}\label{basicproblem}

We consider $N$ non-colluding databases storing $M$ independent submodels. Initially, each submodel consists of random symbols picked from a finite field $\bF_q$, such that, 
\begin{align}
    H(W_k^{[0]})&=L, \quad k\in\{1,\dotsc,M\},\\
    H(W_1^{[0]},\dotsc,W_M^{[0]})=\sum_{k=1}^M H(W_k^{[0]})&=LM,
\end{align}
where $W_{k}^{[0]}$ is the initial version of the $k$th submodel and $L$ is the length of a submodel. At any given time instance, a single user reads, updates and writes a single submodel of interest, while keeping the submodel index and the value of the update private. The submodels are generated in such a way that any given user is equally probable to update any given submodel at a given time instance. The process of updating consists of two phases, namely, the reading phase where the user downloads the required submodel and the writing phase where the user uploads the incremental update back to the databases. 

In the reading phase, the user sends queries to the databases to download the required submodel. The user (at time $t$) has no prior information on the submodels contained in the databases. Therefore, the queries sent by the user at time $t$ to the databases in the reading phase are independent of the existing submodels,  
\begin{align}
    I(Q_1^{[t]},\dotsc,Q_N^{[t]};W_1^{[t-1]},\dotsc,W_M^{[t-1]})=0, \quad  t\in\mathbb{Z}^+, 
\end{align}
where $Q_n^{[t]}$, $n\in\{1,\dotsc,N\}$ are the queries sent by the user to the databases and $W_k^{[t-1]}$, $k\in\{1,\dotsc,M\}$ are the existing versions of the submodels (before updating) at time $t$. After receiving the queries, each database generates an answer and sends it back to the user. This answer is a function of its existing storage and the query received, 
\begin{align}
    H(A_n^{[t]}|Q_n^{[t]},S_n^{[t-1]})=0,\quad n\in\{1,\dotsc,N\},
\end{align}
where $A_n^{[t]}$ is the answer sent by database $n$ at time $t$ and $S_n^{[t-1]}$ is the existing storage (before updating) of database $n$ at time $t$.

In the writing phase, the user sends the incremental updates of the updated submodel to each database. Any PRUW scheme contains a specific mechanism that privately places these updates at correct positions in each database, since the submodel index and the value of the update are kept private from the databases. 

\begin{figure}[t]
    \centering
    \begin{subfigure}[b]{0.42\textwidth}
        \centering
        \includegraphics[width=\textwidth]{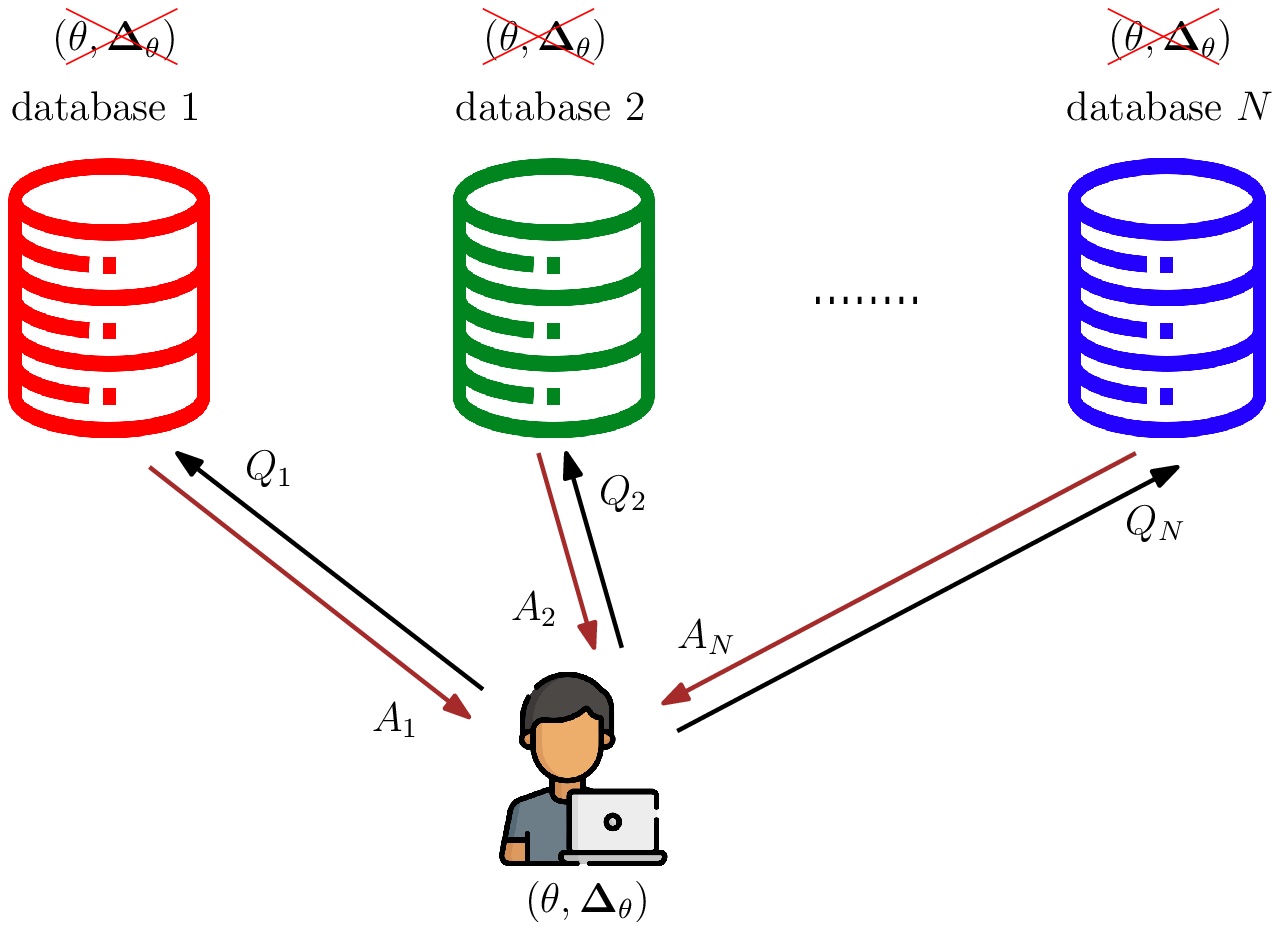}
        \caption{Reading phase.}
        \label{fig:read}
    \end{subfigure}
    \hfill
    \begin{subfigure}[b]{0.42\textwidth}
        \centering
        \includegraphics[width=\textwidth]{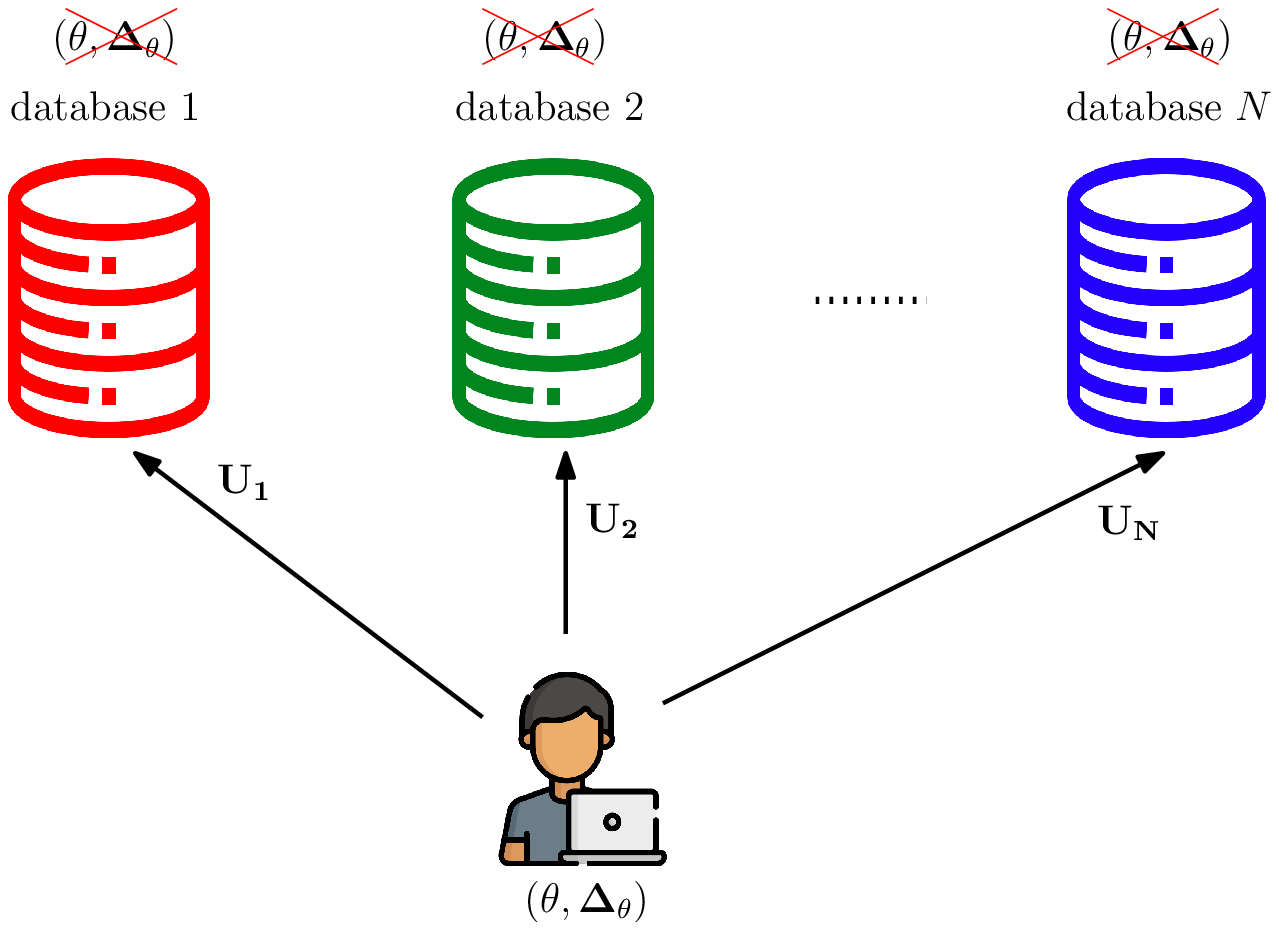}
        \caption{Writing phase.}
        \label{fig:write}
    \end{subfigure}
    \caption{A user reads a submodel, updates it, and writes it back to the databases.}
    \label{fig:model}
\end{figure}

Any information that is communicated in both phases takes place only between a single user and the system of databases. Users that update the model at different time instances do not communicate with each other. The problem is designed to study the PRUW procedure involving a single user at a given time instance. The same process is independently carried out at each time instance with different users. The system model is illustrated in Figure~\ref{fig:model}. 

Next, we formally define the privacy, security and correctness conditions under which a PRUW setting operates. 

\emph{Privacy of the submodel index:} No information on the index of the submodel updated by any given user is allowed to leak to any of the databases, even with the availability of all past data of the database such as past storages, queries and updates. That is, for each database $n$, $n\in\{1,\dotsc,N\}$,
\begin{align}\label{prvcy_id}
    I(\theta;Q_n^{[t]},U_n^{[t]}|Q_n^{[1:t-1]},S_n^{[0:t-1]},U_n^{[1:t-1]})=0, \quad  t\in \mathbb{N},
\end{align}
where $\theta$ is the index of the submodel updated by the user at time $t$, $Q_n$, $S_n$ and $U_n$ are the query, storage and update communicated between the user and database $n$ at the corresponding time instances indicated in square brackets.

\emph{Privacy of the values of updates:} No information on the values of the updates $\Delta_\theta^{[t]}$ is allowed to leak to any of the databases, given all past data of the database. That is, for each database $n$, $n\in\{1,\dotsc,N\}$,
\begin{align}\label{prvcy_u}
    I(\Delta_\theta^{[t]};Q_n^{[t]},U_n^{[t]}|Q_n^{[1:t-1]},S_n^{[0:t-1]},U_n^{[1:t-1]})=0, \quad  t\in \mathbb{N}.
\end{align}
We assume that the distribution of each update $\Delta_{\theta}^{[t]}$ is uniform.

\emph{Security of the stored data:} No information on the submodels is allowed to leak to any of the databases. That is, for each database $n$, $n\in\{1,\dotsc,N\}$,
\begin{align}\label{security}
    I(W_{1:M}^{[t]};S_n^{[t]})=0, \quad  t\in \mathbb{Z}^+_0.
\end{align}

\emph{Correctness in the reading phase:} In the reading phase, the user must be able to correctly decode the required submodel using the queries sent and the answers received from all databases. That is,
\begin{align}
    H(W_\theta^{[t-1]}|Q_{1:N}^{[t]},A_{1:N}^{[t]},\theta)=0,\quad t\in\mathbb{N},
\end{align}
where $W_\theta^{[t-1]}$ is the submodel (before updating) required by the user at time $t$. 

\emph{Correctness in the writing phase:} At time $t$, $t\in\mathbb{N}$, all submodels stored in each database must be correctly updated as,
\begin{align}
    W_{m}^{[t]}=\begin{cases}
        W_{m}^{[t-1]}+\Delta_{m}^{[t]}, & \text{if $m=\theta$,}\\
        W_m^{[t-1]}, & \text{if $m\neq\theta$.}
    \end{cases}
\end{align}

A PRUW scheme is a scheme that satisfies the above privacy, security and correctness requirements. The reading and writing costs are defined as $C_R=\frac{\mathcal{D}}{L}$ and $C_W=\frac{\mathcal{U}}{L}$, respectively, where $\mathcal{D}$ is the total number of bits downloaded from all databases when retrieving the required submodel, $\mathcal{U}$ is the total number of bits sent to all databases in the writing phase and $L$ is the size of each submodel. The total cost is defined as $C_T=C_R+C_W$.

\subsection{Main Result}\label{main_p1}

\begin{theorem}\label{thm0}
Following reading and writing costs are achievable in a PRUW system in FSL with $N\geq4$ non-colluding databases containing $M$ submodels.
\begin{align}
    C_R&=\begin{cases}
    \frac{2}{1-\frac{2}{N}}, & \text{if $N$ is even}\\
    \frac{2}{1-\frac{3}{N}}, & \text{if $N$ is odd}
    \end{cases}\label{read_p1}\\
    C_W&=\begin{cases}
    \frac{2}{1-\frac{2}{N}}, & \text{if $N$ is even}\\
    \frac{2-\frac{2}{N}}{1-\frac{3}{N}}, & \text{if $N$ is odd}
    \end{cases}\label{write_p1}.
\end{align}
\end{theorem}

\begin{remark}
PRUW in FSL can be carried out by downloading/uploading approximately twice as many bits as the size of a submodel.
\end{remark}

\subsection{Basic PRUW Scheme}\label{basicscheme}

This scheme can be applied to any PRUW system with $N\geq4$ non-colluding databases. In this scheme, the privacy-security requirement is satisfied by adding random noise terms within the field $\bF_q$ to the queries, updates and storage. This is because the noise added queries, updates and storage are uniformly distributed and independent of their original versions. This is known as Shannon's one-time-pad and also as crypto lemma \cite{OTP,wireless,DOF}. Furthermore, if $k\in \bF_q$ is a constant and $Z\in\bF_q$ is random noise, $k Z$ is also random noise (uniformly distributed in $\bF_q$) if $k$ and $q$ are coprime. 

Based on the crypto lemma, any given random variable $A$ that takes values in $\bF_q$ with an arbitrary distribution is independent of the uniformly distributed random variable $A+Z_1$, where $Z_1$ is random noise. Applying the crypto lemma again on $A+Z_1$ with another random noise symbol $Z_2$ results in $(A+Z_1)+Z_2$ being uniformly distributed. Moreover, since $(A+Z_1)+Z_2=A+(Z_1+Z_2)$ and $Z_1+Z_2$ is uniformly distributed (again from crypto lemma), $A+Z_1+Z_2$ is independent of $A$. Therefore, by induction, for any $r\in\mathbb{N}$, $A+\sum_{i=1}^r Z_i$ is uniformly distributed and independent of $A$, where $Z_i$s are random noise symbols.

With the above argument, the privacy and security requirements are satisfied by adding $T_1\geq 1$, $T_2\geq1$ and $T_3\geq1$ random noise terms to the submodel parameters in storage, queries and updates, respectively. The scheme provides the optimum values of $T_1$, $T_2$ and $T_3$ that minimize the total cost. In other words, this scheme is over-designed with extra noise terms to make the PRUW process more cost efficient.

We now present the basic scheme with arbitrary values of $T_1$, $T_2$, $T_3$ satisfying all $T_i\geq1$. The optimum values of $T_1$, $T_2$, $T_3$ that minimize the total cost, i.e., $T_1^*$, $T_2^*$, $T_3^*$, are derived later in this section. Let $\ell$ be the subpacketization of the scheme, i.e., the scheme is defined on a set of $\ell$ bits of each submodel, which is called a \emph{subpacket}, and is applied repeatedly in the same way on all subpackets in the model. We choose $\ell=N-T_1-T_2$. An additional constraint given by $\frac{N+T_3-1}{2}\leq T_1\leq N-T_2-1$ must be satisfied by $T_1$, $T_2$, $T_3$ for a given $N$.\footnote{These conditions will be evident as the description of the scheme progresses.} 

\subsubsection{General Scheme}\label{generalscheme1}

\textbf{Storage and initialization:} The storage of a single subpacket of all submodels in database $n$ is given by, 
\begin{align}
    S_n=\begin{bmatrix}\label{storage3}
    \begin{bmatrix}
        W_{1,1}+ (f_1-\alpha_n)(Z_{1,0}^{[1]}+\alpha_nZ_{1,1}^{[1]}+\dotsc+\alpha_n^{T_1-1}Z_{1,T_1-1}^{[1]})\\
        W_{2,1}+ (f_1-\alpha_n)(Z_{2,0}^{[1]}+\alpha_nZ_{2,1}^{[1]}+\dotsc+\alpha_n^{T_1-1}Z_{2,T_1-1}^{[1]})\\
        \vdots\\
        W_{M,1}+ (f_1-\alpha_n)(Z_{M,0}^{[1]}+\alpha_nZ_{M,1}^{[1]}+\dotsc+\alpha_n^{T_1-1}Z_{M,T_1-1}^{[1]})\\
    \end{bmatrix}\\
    \vdots\\
    \begin{bmatrix}
        W_{1,\ell}+ (f_\ell-\alpha_n)(Z_{1,0}^{[\ell]}+\alpha_nZ_{1,1}^{[\ell]}+\dotsc+\alpha_n^{T_1-1}Z_{1,T_1-1}^{[\ell]})\\
        W_{2,\ell}+ (f_\ell-\alpha_n)(Z_{2,0}^{[\ell]}+\alpha_nZ_{2,1}^{[\ell]}+\dotsc+\alpha_n^{T_1-1}Z_{2,T_1-1}^{[\ell]})\\
        \vdots\\
        W_{M,\ell} + (f_\ell-\alpha_n)(Z_{M,0}^{[\ell]}+\alpha_nZ_{M,1}^{[\ell]}+\dotsc+\alpha_n^{T_1-1}Z_{M,T_1-1}^{[\ell]})\\
    \end{bmatrix}
    \end{bmatrix}, 
\end{align}
where $W_{i,j}$ is the $j$th bit of submodel $i$, $Z_{i,j}^{[k]}$ is the $(j+1)$st noise term for the $k$th bit of $W_i$, and $\{f_i\}_{i=1}^\ell$, $\{\alpha_n\}_{n=1}^N$ are globally known distinct constants chosen from $\bF_q$, such that each $\alpha_n$ and $f_i-\alpha_n$ for all $i\in\{1,\dotsc,\ell\}$ and $n\in\{1,\dotsc,N\}$ are coprime with $q$. Reading and writing to $\ell$ bits of the required submodel is explained in the rest of this section. The same procedure is followed $\frac{L}{\ell}$ times for the entire PRUW process, where $L$ is the total length of each submodel. 

\textbf{Reading phase:} Assume that the user requires to update $W_\theta$. Then, the user sends the following query to database $n$, in order to read the existing version of $W_\theta$,
\begin{align}\label{query}
    Q_n=\begin{bmatrix}
        \frac{1}{f_1-\alpha_n}e_M(\theta)+\Tilde{Z}_{1,0}+\alpha_n\Tilde{Z}_{1,1}+\alpha_n^{T_2-1}\Tilde{Z}_{1,T_2-1}\\
        \frac{1}{f_2-\alpha_n}e_M(\theta)+\Tilde{Z}_{2,0}+\alpha_n\Tilde{Z}_{2,1}+\alpha_n^{T_2-1}\Tilde{Z}_{2,T_2-1}\\
        \vdots\\
        \frac{1}{f_\ell-\alpha_n}e_M(\theta)+\Tilde{Z}_{\ell,0}+\alpha_n\Tilde{Z}_{\ell,1}+\alpha_n^{T_2-1}\Tilde{Z}_{\ell,T_2-1}
    \end{bmatrix}, \quad n\in\{1,\dotsc,N\},
\end{align}
where $e_M(\theta)$ is the all zeros vector of size $M\times 1$ with a 1 at the $\theta$th position and $\Tilde{Z}_{i,j}$s are random noise vectors of size $M\times 1$. Database $n$ then generates the answer given by,
\begin{align}
    A_n=&S_n^TQ_n\\
    =&\frac{1}{f_1-\alpha_n}W_{\theta,1}+\frac{1}{f_2-\alpha_n}W_{\theta,2}+\dotsc+\frac{1}{f_\ell-\alpha_n}W_{\theta,\ell}\nonumber\\
    &+\phi_0+\alpha_n\phi_1+\dotsc+\alpha_n^{T_1+T_2-1}\phi_{T_1+T_2-1},
\end{align}
where $\phi_i$s are combinations of noise terms that do not depend on $n$. The answers received from the $N$ databases in matrix form is given as follows,
\begin{align}\label{mat2}
    \begin{bmatrix}
        A_1\\A_2\\\vdots\\A_N
    \end{bmatrix}
    =
    \begin{bmatrix}
        \frac{1}{f_1-\alpha_1} & \frac{1}{f_2-\alpha_1} & \dotsc & \frac{1}{f_\ell-\alpha_1} & 1 & \alpha_1 & \dotsc & \alpha_1^{T_1+T_2-1}\\
        \frac{1}{f_1-\alpha_2} & \frac{1}{f_2-\alpha_2} & \dotsc & \frac{1}{f_\ell-\alpha_2} & 1 & \alpha_2 & \dotsc & \alpha_2^{T_1+T_2-1}\\
        \vdots & \vdots & \vdots & \vdots & \vdots & \vdots & \vdots & \vdots\\
        \frac{1}{f_1-\alpha_N} & \frac{1}{f_2-\alpha_N} & \dotsc & \frac{1}{f_\ell-\alpha_N} & 1 & \alpha_N & \dotsc & \alpha_N^{T_1+T_2-1}\\
    \end{bmatrix}
    \begin{bmatrix}
        W_{\theta,1}\\ \vdots\\ W_{\theta,\ell}\\ \phi_0\\\phi_1\\ \vdots\\ \phi_{T_1+T_2+1}
    \end{bmatrix}.
\end{align}
Since the matrix is invertible, the $\ell$ bits of $W_\theta$ can be retrieved using \eqref{mat2}. The reading cost is given by,
\begin{align} \label{readingcost}
    C_R=\frac{N}{\ell}=\frac{N}{N-T_1-T_2}.
\end{align}

\textbf{Writing phase:} In the writing phase, the user sends a single bit to each database (per subpacket), which is a combination of the updates of the $\ell$ bits of $W_\theta$ and $T_3$ random noise bits. The combined update bit is a polynomial of $\alpha_n$, which allows the databases to privately decompose it into the $\ell$ individual update bits, with the help of the queries received in the reading phase. Finally, these incremental updates are added to the existing storage to obtain the updated storage. As explained later in this section, the above stated decomposition performed at the databases introduces a few extra terms, which are added to the $T_1$ random noise terms in storage. From the crypto lemma, the updated $T_1$ noise terms are also independent and uniformly distributed (i.e., random noise). The reason behind over-designing the system to have extra noise terms in storage is to have a number of noise terms that matches the number of extra terms introduced by the \emph{decomposition} performed at the databases in the writing phase. The combined single update bit that the user sends to database $n$, $n\in\{1,\dotsc,N\}$, is given by,
\begin{align}
    U_n=\sum_{i=1}^\ell \Tilde{\Delta}_{\theta,i} \prod_{j=1,j\neq i}^\ell (f_j-\alpha_n)+\prod_{j=1}^\ell (f_j-\alpha_n)(Z_0+\alpha_nZ_1+\dotsc+\alpha_n^{T_3-1}Z_{T_3-1}),
\end{align}
where $Z_i$s are random noise bits, $\Tilde{\Delta}_{\theta,i}=\frac{\Delta_{\theta,i}}{\prod_{j=1,j\neq i}^\ell (f_j-f_i)}$ with $\Delta_{\theta,i}$ being the update for the $i$th bit of $W_\theta$. Once database $n$ receives $U_n$, it calculates the incremental update that needs to be added to the existing storage in order to obtain the new and updated storage. This calculation requires the following two definitions and two lemmas.

\begin{definition}(Scaling matrix)
\begin{align}
    D_n=\begin{bmatrix}
        (f_1-\alpha_n)I_M & 0 & \dotsc & 0\\
        0 & (f_2-\alpha_n)I_M & \dotsc & 0\\
        \vdots & \vdots & \vdots & \vdots\\
        0 & 0 & \dotsc & (f_\ell-\alpha_n)I_M\\
    \end{bmatrix}, \quad n\in\{1,\dotsc,N\}. 
\end{align}
\end{definition}

\begin{definition}\label{nullshaper}(Null shaper)
\begin{align}
\Omega_n&=\begin{bmatrix}
    \left(\frac{\prod_{r\in\mathcal{F}} (\alpha_r-\alpha_n)}{\prod_{r\in\mathcal{F}} (\alpha_r-f_1)}\right)I_M & 0 & \dotsc & 0\\
    0 & \left(\frac{\prod_{r\in\mathcal{F}} (\alpha_r-\alpha_n)}{\prod_{r\in\mathcal{F}} (\alpha_r-f_2)}\right)I_M & \dotsc & 0\\
    \vdots & \vdots & \vdots & \vdots\\
    0 & 0 & \dotsc & \left(\frac{\prod_{r\in\mathcal{F}} (\alpha_r-\alpha_n)}{\prod_{r\in\mathcal{F}} (\alpha_r-f_\ell)}\right)I_M
\end{bmatrix}, \quad n\in\{1,\dotsc,N\},
\end{align}
where $\mathcal{F}$ is any subset of databases satisfying $|\mathcal{F}|=2T_1-N-T_3+1$. 
\end{definition}

\begin{lemma}\label{lem1}
    \begin{align}
        \frac{U_n}{f_k-\alpha_n}=\frac{1}{f_k-\alpha_n}\Delta_{\theta,k}+P_{\alpha_n}(\ell+T_3-2), \quad k\in\{1,\dotsc,\ell\},
    \end{align}
    where $P_{\alpha_n}(\ell+T_3-2)$ is a plynomial in $\alpha_n$ of degree $\ell+T_3-2$. The coefficients of $\alpha_n^{i}$s in $P_{\alpha_n}(\ell+T_3-2)$ are fixed for all $n$.
\end{lemma}    

\begin{lemma}\label{lem2}
\begin{align}
    \left(\frac{\prod_{r\in\mathcal{F}} (\alpha_r-\alpha_n)}{\prod_{r\in\mathcal{F}} (\alpha_r-f_k)}\right)\frac{1}{f_k-\alpha_n}=\frac{1}{f_k-\alpha_n}+P_{\alpha_n}(|\mathcal{F}|-1), \quad k\in\{1,\dotsc,\ell\},
\end{align}
where $P_{\alpha_n}(|\mathcal{F}|-1)$ is a polynomial in $\alpha_n$ of degree $|\mathcal{F}|-1$.    
\end{lemma}

The proofs of Lemma~\ref{lem1} and Lemma~\ref{lem2} are given in Appendix~\ref{pf1} and Appendix~\ref{pf2}, respectively. With these definitions and lemmas, the incremental update is calculated by,
\begin{align}
    \bar{U}_n&=
    D_n\times \Omega_n \times U_n\times Q_n\label{update_beg}\\
    &=D_n \times \Omega_n \times 
    \begin{bmatrix}
        \frac{U_n}{f_1-\alpha_n}e_M(\theta)+U_n(\Tilde{Z}_{1,0}+\alpha_n\Tilde{Z}_{1,1}+\alpha_n^{T_2-1}\Tilde{Z}_{1,T_2-1})\\
        \frac{U_n}{f_2-\alpha_n}e_M(\theta)+U_n(\Tilde{Z}_{2,0}+\alpha_n\Tilde{Z}_{2,1}+\alpha_n^{T_2-1}\Tilde{Z}_{2,T_2-1})\\
        \vdots\\
        \frac{U_n}{f_\ell-\alpha_n}e_M(\theta)+U_n(\Tilde{Z}_{\ell,0}+\alpha_n\Tilde{Z}_{\ell,1}+\alpha_n^{T_2-1}\Tilde{Z}_{\ell,T_2-1})
    \end{bmatrix}.\label{update2}
\end{align}
Using Lemma~\ref{lem1}, 
\begin{align}\label{eqq1}
    \bar{U}_n=&D_n\times \Omega_n \times 
    \begin{bmatrix}
        \frac{1}{f_1-\alpha_n}\Delta_{\theta,1}e_M(\theta)+(\xi_0^{[1]}+\xi_1^{[1]}\alpha_n+\dotsc+\xi_{\ell+T_3-2}^{[1]}\alpha_n^{\ell+T_3-2}) e_M(\theta)\\ \vdots\\ \frac{1}{f_\ell-\alpha_n}\Delta_{\theta,\ell}e_M(\theta)+(\xi_0^{[\ell]}+\xi_1^{[\ell]}\alpha_n+\dotsc+\xi_{\ell+T_3-2}^{[\ell]}\alpha_n^{\ell+T_3-2})e_M(\theta)
    \end{bmatrix}\nonumber\\
    & +D_n\times \Omega_n \times
    \begin{bmatrix}
    \begin{bmatrix}
        (\Tilde{\xi}_{1,0}^{[1]}+\Tilde{\xi}_{1,1}^{[1]}\alpha_n+\dotsc+\Tilde{\xi}_{1,\ell+T_2+T_3-2}^{[1]}\alpha_n^{\ell+T_2+T_3-2})\\ \vdots\\(\Tilde{\xi}_{M,0}^{[1]}+\Tilde{\xi}_{M,1}^{[1]}\alpha_n+\dotsc+\Tilde{\xi}_{M,\ell+T_2+T_3-2}^{[1]}\alpha_n^{\ell+T_2+T_3-2})
    \end{bmatrix}
        \\ \vdots\\
        \begin{bmatrix}
            (\Tilde{\xi}_{1,0}^{[\ell]}+\Tilde{\xi}_{1,1}^{[\ell]}\alpha_n+\dotsc+\Tilde{\xi}_{1,\ell+T_2+T_3-2}^{[\ell]}\alpha_n^{\ell+T_2+T_3-2})\\ \vdots\\(\Tilde{\xi}_{M,0}^{[\ell]}+\Tilde{\xi}_{M,1}^{[\ell]}\alpha_n+\dotsc+\Tilde{\xi}_{M,\ell+T_2+T_3-2}^{[\ell]}\alpha_n^{\ell+T_2+T_3-2})
        \end{bmatrix}
    \end{bmatrix}\\
    =&D_n \times \begin{bmatrix}
        \left(\frac{\prod_{r\in\mathcal{F}} (\alpha_r-\alpha_n)}{\prod_{r\in\mathcal{F}} (\alpha_r-f_1)}\right)\frac{1}{f_1-\alpha_n}\Delta_{\theta,1}e_M(\theta) \\ \left(\frac{\prod_{r\in\mathcal{F}} (\alpha_r-\alpha_n)}{\prod_{r\in\mathcal{F}} (\alpha_r-f_2)}\right)\frac{1}{f_2-\alpha_n}\Delta_{\theta,2}e_M(\theta) \\ \vdots\\ \left(\frac{\prod_{r\in\mathcal{F}} (\alpha_r-\alpha_n)}{\prod_{r\in\mathcal{F}} (\alpha_r-f_\ell)}\right)\frac{1}{f_\ell-\alpha_n}\Delta_{\theta,\ell} e_M(\theta) 
    \end{bmatrix}\label{final}\nonumber\\
     &+D_n\times \begin{bmatrix}
    \begin{bmatrix}
        \Tilde{\eta}_{1,0}^{[1]}+\Tilde{\eta}_{1,1}^{[1]}\alpha_n+\dotsc+\Tilde{\eta}_{1,\ell+T_2+T_3-2+|\mathcal{F}|}^{[1]}\alpha_n^{\ell+T_2+T_3-2+|\mathcal{F}|}\\\vdots\\\Tilde{\eta}_{M,0}^{[1]}+\Tilde{\eta}_{M,1}^{[1]}\alpha_n+\dotsc+\Tilde{\eta}_{M,\ell+T_2+T_3-2+|\mathcal{F}|}^{[1]}\alpha_n^{\ell+T_2+T_3-2+|\mathcal{F}|}
    \end{bmatrix}
        \\ \vdots\\
    \begin{bmatrix}
        \Tilde{\eta}_{1,0}^{[\ell]}+\Tilde{\eta}_{1,1}^{[\ell]}\alpha_n+\dotsc+\Tilde{\eta}_{1,\ell+T_2+T_3-2+|\mathcal{F}|}^{[\ell]}\alpha_n^{\ell+T_2+T_3-2+|\mathcal{F}|}\\ \vdots\\\Tilde{\eta}_{M,0}^{[\ell]}+\Tilde{\eta}_{M,1}^{[\ell]}\alpha_n+\dotsc+\Tilde{\eta}_{M,\ell+T_2+T_3-2+|\mathcal{F}|}^{[\ell]}\alpha_n^{\ell+T_2+T_3-2+|\mathcal{F}|}
    \end{bmatrix}
    \end{bmatrix}. 
\end{align}
From Lemma~\ref{lem2},
\begin{align}
     \bar{U}_n=\begin{bmatrix}
        \Delta_{\theta,1}e_M(\theta) \\ \Delta_{\theta,2}e_M(\theta) \\ \vdots\\ \Delta_{\theta,\ell} e_M(\theta) 
    \end{bmatrix}+\begin{bmatrix}
    \begin{bmatrix}
        (f_1-\alpha_n)(\hat{\eta}_{1,0}^{[1]}+\hat{\eta}_{1,1}^{[1]}\alpha_n+\dotsc+\hat{\eta}_{1,T_1-1}^{[1]}\alpha_n^{T_1-1})\\\vdots\\(f_1-\alpha_n)(\hat{\eta}_{M,0}^{[1]}+\hat{\eta}_{M,1}^{[1]}\alpha_n+\dotsc+\hat{\eta}_{M,T_1-1}^{[1]}\alpha_n^{T_1-1})
    \end{bmatrix}
        \\ \vdots\\
    \begin{bmatrix}
        (f_\ell-\alpha_n)(\hat{\eta}_{1,0}^{[\ell]}+\hat{\eta}_{1,1}^{[\ell]}\alpha_n+\dotsc+\hat{\eta}_{1,T_1-1}^{[\ell]}\alpha_n^{T_1-1})\\ \vdots\\(f_\ell-\alpha_n)(\hat{\eta}_{M,0}^{[\ell]}+\hat{\eta}_{M,1}^{[\ell]}\alpha_n+\dotsc+\hat{\eta}_{M,T_1-1}^{[\ell]}\alpha_n^{T_1-1})
    \end{bmatrix}
    \end{bmatrix}\label{x}, 
\end{align}
where \eqref{eqq1} and \eqref{final} are due to the fact that $U_n$ and the diagonal elements of $\Omega_n$ are polynomials in $\alpha_n$ of degrees $\ell+T_3-1$ and $|\mathcal{F}|$, respectively. The polynomial coefficients $\xi_i^{[j]}$, $\Tilde{\xi}_i^{[j]}$, $\Tilde{\eta}_i^{[j]}$ and $\hat{\eta}_i^{[j]}$ are combined noise terms that do not depend on $n$. \eqref{x} is immediate from $|\mathcal{F}|=2T_1-N-T_3+1$ and $\ell=N-T_1-T_2$. Note that for databases $n\in\mathcal{F}$, $\Omega_n=0$, which makes the incremental update of those databases equal to zero. This means that the user could save the writing cost by not sending the update bit $U_n$ in the writing phase to those databases in $\mathcal{F}$. For each database $n\in\{1,\dotsc,N\}\backslash\mathcal{F}$, the incremental update in \eqref{x} is in the same format as the storage in \eqref{storage3}. Therefore, the updated storage is given by,
\begin{align}\label{finalupdate}
    S_n^{[t]}=S_n^{[t-1]}+\bar{U}_n, \quad n\in\{1,\dotsc,N\}\backslash\mathcal{F},
\end{align}
while $S_n^{[t]}=S_n^{[t-1]}$ for $n\in\mathcal{F}$, where $S_n^{[t-1]}$ and $S_n^{[t]}$ are the storages of database $n$ before and after the update, respectively.\footnote{Note that $W_\theta$ is still updated in databases $n\in\mathcal{F}$ even though the noise added storage has not changed. This is because the zeros of the incremental update polynomials occur at those $\alpha_n$s that correspond to $n\in\mathcal{F}$.} The writing cost of this scheme is given by,
\begin{align} \label{writingcost}
    C_W=\frac{N-|\mathcal{F}|}{\ell}=\frac{2N-2T_1+T_3-1}{N-T_1-T_2}.
\end{align}

\subsubsection{Total Communication Cost and Optimal Values of $T_1$, $T_2$, $T_3$}
From \eqref{readingcost} and \eqref{writingcost}, the total communication cost is,
\begin{align}\label{totcost}
    C_T=C_R+C_W=\frac{3N-2T_1+T_3-1}{N-T_1-T_2}.
\end{align}
The general scheme described in Section~\ref{generalscheme1} and the total cost in \eqref{totcost} are presented for arbitrary $T_1$, $T_2$, $T_3$ satisfying $T_i\geq1$ for $i=1,2,3$, and $\frac{N+T_3-1}{2}\leq T_1\leq N-T_2-1$, where the last condition is derived from $|\mathcal{F}|\geq0$ and $\ell\geq1$. In this subsection, we present the optimum values of $T_1$, $T_2$, $T_3$ that minimize the total cost for a given number of databases $N$. It is clear that the total cost in \eqref{totcost} increases with $T_2$ and $T_3$. Therefore, the optimum values of $T_2$ and $T_3$ such that the privacy constraints are satisfied are $T_2^*=T_3^*=1$. Then, the resulting total cost is,
\begin{align}
    C_T=\frac{3N-2T_1}{N-T_1-1},
\end{align}
which is increasing in $T_1$, since $\frac{dC_T}{dT_1}=\frac{N+2}{(N-T_1-1)^2}>0$. Thus, the optimum value of $T_1$ satisfying the constraint of $\frac{N+T_3-1}{2}\leq T_1\leq N-T_2-1$ with $T_2^*=T_3^*=1$ is $T_1^*=\left\lceil\frac{N}{2}\right\rceil$. The corresponding optimum subpacketization is $\ell^*=\lfloor\frac{N}{2}\rfloor-1$ and the optimum reading and writing costs are given in \eqref{read_p1} and \eqref{write_p1}, respectively.

\subsubsection{Proof of Privacy and Security}\label{privacy_proof}

The following facts are required for the proofs of privacy and security. In the proposed scheme, the submodel index $\theta$ is indicated by $e_M(\theta)$. However, the queries sent to each of the databases are independent from $e_M(\theta)$ due to the random noise terms added to it, from Shannon's one-time-pad theorem. Similarly, the submodel values $W_{i,j}$ are independent from the storage $S_n$ of each database and the values of updates $\Delta_{i,j}$ are independent from the uploads in the writing phase $U_n$, due to the random noise terms added.

\emph{Privacy of the submodel index:} Based on the chain rule, the mutual information term in the condition for the privacy of submodel index in \eqref{prvcy_id} can be written as,
\begin{align}
    I(\theta;Q_n^{[t]},U_n^{[t]}|S_n^{[0:t-1]},Q_n^{[1:t-1]},U_n^{[1:t-1]})&=I(\theta;Q_n^{[t]}, U_n^{[t]},S_n^{[0:t-1]},Q_n^{[1:t-1]},U_n^{[1:t-1]})\nonumber\\
    &\quad -I(\theta;S_n^{[0:t-1]},Q_n^{[1:t-1]},U_n^{[1:t-1]})\\
    &=I(\theta;S_n^{[0:t-1]},Q_n^{[1:t]},U_n^{[1:t]})\nonumber\\
    &\quad-I(\theta;S_n^{[0:t-1]},Q_n^{[1:t-1]},U_n^{[1:t-1]}).\label{line2}
\end{align}
Note that for any $m\in\{1,\dotsc,M\}$, $\bar{u}_n\in\mathbb{F}_q^t$ and $\bar{r}_n,\bar{s}_n\in\mathbb{F}_q^{M\ell t}$,
\begin{align}
    P(\theta=m|&Q_n^{[1:t]}=\bar{r}_n,U_n^{[1:t]}=\bar{u}_n,S_n^{[0:t-1]}=\bar{s}_n)\nonumber\\
    &=\frac{P(Q_n^{[1:t]}=\bar{r}_n,U_n^{[1:t]}=\bar{u}_n,S_n^{[0:t-1]}=\bar{s}_n|\theta=m)P(\theta=m)}{P(Q_n^{[1:t]}=\bar{r}_n,U_n^{[1:t]}=\bar{u}_n,S_n^{[0:t-1]}=\bar{s}_n)}.
\end{align}
Even though $e_M(\theta)$, $\Delta_{i,j}$ and $W_{i,j}$ are functions of $\theta$, all $Q_n$, $U_n$ and $S_n$ terms are independent of $\theta$ since they are simply random noise vectors/scalars and are independent of $e_M(\theta)$, $\Delta_{i,j}$ and $W_{i,j}$ from Shannon's one-time-pad theorem. Therefore,
\begin{align}
    P(\theta=m|&Q_n^{[1:t]}=\bar{r}_n,U_n^{[1:t]}=\bar{u}_n,S_n^{[0:t-1]}=\bar{s}_n)\nonumber\\
    &=\frac{P(Q_n^{[1:t]}=\bar{r}_n,U_n^{[1:t]}=\bar{u}_n,S_n^{[0:t-1]}=\bar{s}_n)P(\theta=m)}{P(Q_n^{[1:t]}=\bar{r}_n,U_n^{[1:t]}=\bar{u}_n,S_n^{[0:t-1]}=\bar{s}_n)}\\
    &=P(\theta=m),
\end{align}
which proves that the first term on the right side of \eqref{line2} is zero, which results in the privacy condition in \eqref{prvcy_id}.

\emph{Privacy of the values of updates:} The mutual information term in the condition given in \eqref{prvcy_u} can be written as,
\begin{align}
    I(\Delta_\theta^{[t]};Q_n^{[t]},U_n^{[t]}|Q_n^{[1:t-1]},S_n^{[0:t-1]},U_n^{[1:t-1]})&=I(\Delta_\theta^{[t]};Q_n^{[t]},U_n^{[t]},Q_n^{[1:t-1]},S_n^{[0:t-1]},U_n^{[1:t-1]})\nonumber\\
    &\quad-I(\Delta_\theta^{[t]};Q_n^{[1:t-1]},S_n^{[0:t-1]},U_n^{[1:t-1]})\\
    &=I(\Delta_\theta^{[t]};Q_n^{[1:t]},S_n^{[0:t-1]},U_n^{[1:t]})\nonumber\\
    &\quad-I(\Delta_\theta^{[t]};Q_n^{[1:t-1]},S_n^{[0:t-1]},U_n^{[1:t-1]})\label{line2_u}
\end{align}
Note that for any $\tilde{q}\in\mathbb{F}_q^\ell$, $\bar{u}_n\in\mathbb{F}_q^t$ and $\bar{r}_n,\bar{s}_n\in\mathbb{F}_q^{M\ell t}$,
\begin{align}
    P(\Delta_\theta^{[t]}=\tilde{q}|&Q_n^{[1:t]}=\bar{r}_n,S_n^{[0:t-1]}=\bar{s}_n,U_n^{[1:t]}=\bar{u}_n)\nonumber\\
    &=\frac{P(Q_n^{[1:t]}=\bar{r}_n,S_n^{[0:t-1]}=\bar{s}_n,U_n^{[1:t]}=\bar{u}_n|\Delta_\theta^{[t]}=\tilde{q})P(\Delta_\theta^{[t]}=\tilde{q})}{P(Q_n^{[1:t]}=\bar{r}_n,S_n^{[0:t-1]}=\bar{s}_n,U_n^{[1:t]}=\bar{u}_n)}.
\end{align}
As before, all $U_n$, $Q_n$ and $S_n$ values are random noise terms and are independent of $\Delta_{\theta}^{[t]}$ from Shannon's one-time-pad theorem. Therefore,
\begin{align}
    P(\Delta_\theta^{[t]}=\tilde{q}|&Q_n^{[1:t]}=\bar{r}_n,S_n^{[0:t-1]}=\bar{s}_n,U_n^{[1:t]}=\bar{u}_n)\nonumber\\
    &=\frac{P(Q_n^{[1:t]}=\bar{r}_n,S_n^{[0:t-1]}=\bar{s}_n,U_n^{[1:t]}=\bar{u}_n)P(\Delta_\theta^{[t]}=\tilde{q})}{P(Q_n^{[1:t]}=\bar{r}_n,S_n^{[0:t-1]}=\bar{s}_n,U_n^{[1:t]}=\bar{u}_n)}\\
    &=P(\Delta_\theta^{[t]}=\tilde{q}),
\end{align}
which proves that the first term in the right side of \eqref{line2_u} is zero, which proves the condition in \eqref{prvcy_u}.

\emph{Security of the stored submodels:} The condition on the security of submodels given in \eqref{security} is satisfied by the scheme as a direct consequence of Shannon's one-time-pad theorem, due to the random noises added to the submodels at $t=0$.

\section{PRUW with Top $r$ Sparsification}\label{topr}

In this section, we formally describe the problem of PRUW with top $r$ sparsification in FSL, and present a scheme that performs it along with an example.

\subsection{Problem Formulation}\label{problem}

We consider $N$ non-colluding databases storing $M$ independent submodels, each having $P$ subpackets. At a given time instance $t$, a given user reads, updates and writes one of the $M$ submodels, while not revealing any information about the updated submodel index or the values of updates to any of the databases. The submodels, queries and updates consist of symbols from a large enough finite field $\bF_q$.

In the PRUW process in FSL, users keep reading from and writing to required submodels in an iterative manner. With top $r$ sparsification, each user only writes to a selected $r$ fraction of subpackets of the updating submodel, that contains the most significant $r$ fraction of updates.\footnote{In the update stage (model training) users typically work in continuous fields (real numbers) and make $1-r$ of the updates equal to zero (i.e., not update) based on the concept of top $r$ sparsification in learning. These updates are then converted to symbols in $\bF_q$, to be sent to the databases. We assume that the zeros in the continuous field are converted to zeros in the finite field.} \footnote{We assume that all parameters in the most significant $r$ fraction of subpackets have non-zero updates.} This significantly reduces the writing cost. Therefore, a given user who reads the same submodel at time $t+1$ only has to download the union of each $r$ fraction of subpackets updated by all users at time $t$. Let the cardinality of this union be $Pr'$, where $0\leq r'\leq1$. This reflects sparsification in the downlink with a rate of $r'$. For cases where $r'$ increases exponentially with an increasing number of users at a given time, there are downlink sparsification protocols such as \cite{GGS} that limit the value of $r'$ in order to reduce the communication cost. Precisely, in this work, we assume that each user only updates $Pr$ subpackets that correspond to the most significant $r$ fraction of updates in the writing phase, and only downloads $Pr'$ subpackets sent by the databases in the reading phase, of the required submodel.

\begin{figure}[t]
    \centering
    \includegraphics[scale=0.8]{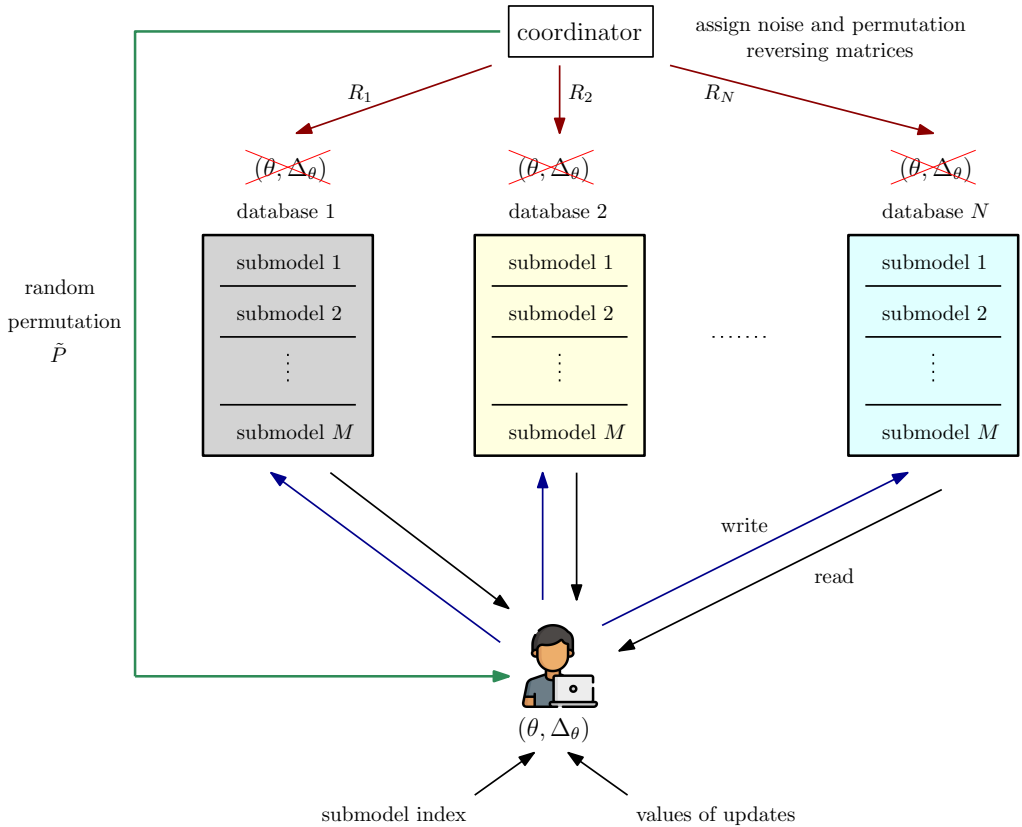}
    \caption{PRUW with top $r$ sparsification: system model.}
    \label{cord}
\end{figure}

The reduction in the communication cost of the PRUW process with sparsification results from communicating only a selected set of updates (parameters) and their positions to the databases (users) in the writing (reading) phase. However, this leaks information about $1-r$ of the updates in the writing phase, as their values (zero) are revealed to the databases. This requires the basic PRUW scheme to be modified in order to satisfy information-theoretic privacy of the updating submodel index and the values of updates while performing top $r$ sparsification to reduce the communication cost. The system model is shown in Figure~\ref{cord}, which is the same as the model of basic PRUW, with the explicit indication of a coordinator. The coordinator exists in the basic PRUW also, where it is used to initialize the storage with identical random noise terms in all databases in the basic PRUW. In PRUW with sparsification, it is also utilized in guaranteeing the privacy of zero-valued updates.

The three components in PRUW with sparsification that need to be kept private are: 1) index of the submodel updated by each user, 2) values of the updates, and 3) indices (positions) of the sparse updates. Note that 3 is a requirement that is implied by 2. The formal descriptions of the privacy constraints are given below. The constraints are presented in the perspective of a single user at time $t$, even though multiple independent users update the model simultaneously.

\emph{Privacy of the submodel index:} No information on the index of the submodel being updated, $\theta$, is allowed to leak to any of the databases at each time instance $t$. For each $n$, 
\begin{align}\label{prvcy_id2}
    I(\theta;Q_n^{[t]},Y_n^{[t]}|Q_n^{[1:t-1]},S_n^{[0:t-1]})=0, \quad  t\in \mathbb{N},
\end{align}
where $Q_n$ and $Y_n$ are the queries and updates/coordinates sent by a given user to database $n$ in the reading and writing phases, at the corresponding time instances indicated in square brackets and $S_n$ is the content of database $n$.

\emph{Privacy of the values of updates:} No information on the values of the updates is allowed to leak to any of the databases, based on the information obtained by the user. For each $n$, 
\begin{align}\label{value_prvcy_sparse}
    I(\Delta_{\theta}^{[t]};Q_n^{[t]},Y_n^{[t]}|Q_n^{[1:t-1]},S_n^{[0:t-1]})=0, \quad  t\in \mathbb{N},
\end{align}
where $\Delta_{\theta}^{[t]}$ is the update of submodel $\theta$ generated by a given user at time $t$.

\emph{Security of submodels:} This is defined by the same security condition in \eqref{security}.

\emph{Correctness in the reading phase:} The user should be able to correctly decode the sparse set of subpackets (denoted by $J$) of the required submodel, determined by the downlink sparsification protocol, from the answers received in the reading phase, i.e., 
\begin{align}
H(W_{\theta,J}^{[t-1]}|Q_{1:N}^{[t]},A_{1:N}^{[t]},\theta)=0, \quad t\in\mathbb{N},
\end{align}
where $W_{\theta,J}^{[t-1]}$ is the set of subpackets in  set $J$ of submodel $W_\theta$ at time $t-1$ and $A_n^{[t]}$ is the answer from database $n$ at time $t$.

\emph{Correctness in the writing phase:} Let $\theta$ be the updating submodel index and $J'$ be the set of most significant $Pr$ subpackets of $W_{\theta}$ updated by a given user. Then, the subpacket $s$ of submodel $m$ at time $t$ given by $W_{m}^{[t]}(s)$ is correctly updated as,
\begin{align}
    W_{m}^{[t]}(s)=
    \begin{cases}
    W_{m}^{[t-1]}(s)+\Delta_{m}^{[t]}(s), & \text{if $m=\theta$ and $s\in J'$}\\
    W_{m}^{[t-1]}(s), & \text{if $m\neq\theta$ or $s\notin J'$}
    \end{cases},
\end{align}
where $\Delta_{m}^{[t]}(s)$ is the corresponding update of $W_{m}^{[t-1]}(s)$. The reading and writing costs are defined the same as in Section~\ref{basicproblem}.

\subsection{Main Result}\label{main}

In this section, we provide the achievable reading and writing costs of the scheme proposed to perform top $r$ sparsification in FSL, while guaranteeing information-theoretic privacy of the updating submodel index and the values of the updates (which includes the indices of sparse updates). The key component of the proposed scheme is a novel permutation technique, which requires the databases to store certain \emph{noise-added permutation reversing matrices}. We propose two cases of the scheme based on the structure and the size of the \emph{noise-added permutation reversing matrices}. Theorem~\ref{thm1} summarizes the results of the two cases.  

\begin{theorem}\label{thm1}
In a private FSL setting with $N$ databases, $M$ submodels (each of size $L$), $P$ subpackets in each submodel, and $r$ and $r'$ sparsification rates in the uplink and downlink, respectively, the following reading and writing costs are achievable with the corresponding sizes of the noise-added permutation reversing matrices. The reading and writing costs are,
\begin{align}
    C_R&=\frac{4r'+\frac{4}{N}(1+r')\log_qP}{1-\frac{2}{N}}\label{re2}\\
    C_W&=\frac{4r(1+\log_q P)}{1-\frac{2}{N}},\label{wr2}
\end{align}
with noise-added permutation reversing matrices of size $O\left(\frac{L^2}{N^2}\right)$  and,
\begin{align}
    C_R&=\frac{2r'+\frac{2}{N}(1+r')\log_qP}{1-\frac{2}{N}}\label{re1}\\
    C_W&=\frac{2r(1+\log_q P)}{1-\frac{2}{N}},\label{wr1}
\end{align}
with noise-added permutation reversing matrices of size $O(L^2)$.
\end{theorem}

\begin{remark}
If sparsification is not considered in the PRUW process, the lowest achievable reading and writing costs are given by $C_R=C_W=\frac{2}{1-\frac{2}{N}}$; see Theorem~\ref{thm0}. Therefore, sparsification with smaller values of $r$ and $r'$ results in significantly reduced communication costs as shown in Theorem~\ref{thm1}.
\end{remark}

\begin{remark}
    The reading and writing costs double (approximately) as the size of the noise-added permutation reversing matrices reduces from $O(L^2)$ to $O\left(\frac{L^2}{N^2}\right)$.
\end{remark}

\subsection{Proposed Scheme}\label{proposedscheme}

The scheme is similar to what is presented in Section~\ref{basicscheme} with the additional component of sparse uploads and downloads. In the writing (reading) phase of the scheme in Section~\ref{basicscheme}, the updates (values) of all parameters in a given subpacket are combined into a single bit. Thus, a user sends (receives) $P$ bits per database, where $P$ is the number of subpackets in a submodel. In this section, using similar concepts as in Section~\ref{basicscheme}, the user only downloads and uploads $Pr'\ll P$ and $Pr\ll P$ bits corresponding to the respective sparse subpackets in the reading and writing phases, respectively, which significantly reduces the communication cost. However, revealing the indices of the subpackets with no update (all zeros) in the writing phase leaks privacy, as the values of those updates (zero) are directly known by the databases.\footnote{The sparse set of subpackets in the downlink (reading phase) is determined by the databases with no additional information from the users. Therefore, privacy leakage from the sparse subpacket indices can only occur in the writing phase.} Therefore, to send the indices of the sparse updates privately to the databases in the process of top $r$ sparsification, we use a permutation technique, which is the key component of the proposed scheme. The basic idea of this technique is to \emph{add noise} to the sparse subpacket indices, to hide the real indices from the databases. Note that basic PRUW adds noise to storage, queries and updates, while PRUW with sparsification \emph{adds noise} to the sparse subpacket indices, in addition to the storage, queries and updates. This is analogous to the case with \emph{normal} and \emph{timing} channels, where the \emph{normal} channels add noise to the values while the \emph{timing} channels add noise to the timings. Basic PRUW is analogous to a normal channel while PRUW with sparsification is analogous to a channel that combines characteristics of both normal and timing channels. 

The process of \emph{adding noise} to the sparse subpacket indices is as follows. In the writing phase, each user sends a random set of indices corresponding to the $Pr$ sparse subpackets (with non-zero updates) instead of sending the real indices (noisy indices). This random set of indices is generated by the users based on a specific random permutation of all subpacket indices, which is not known by the databases. However, in order to guarantee the correctness of the writing process, the permutation needs to be reversed, and the databases should be able to place the received updates at the correct positions. This is accomplished by the use of \emph{noise-added permutation reversing matrices}, stored at the databases. These permutation reversing matrices rearrange the permuted indices of the sparse subpackets received by the users in the correct order in such a way that the databases do not learn the underlying permutation (or the real indices of the sparse updates). Despite having access to the \emph{noise-added permutation reversing matrices}, the databases have zero information on the underlying permutation being reversed, from Shannon's one-time-pad theorem. The \emph{noise-added permutation reversing matrices} convert the \emph{noise} in the sparse subpacket indices (timing channel) into added noise in the incremental update calculation (normal channel). These extra noise terms in the incremental update calculation require extra noise terms to be added to storage, for the correctness of the writing phase, which adversely affects the efficiency of the process. However, the fact that the process is carried out only on $r$ fraction of the original number of subpackets makes the overall process significantly efficient in communication cost. 

The random selection and assignment of the permutation (to users) and the \emph{noise-added permutation reversing matrices} (to databases) are performed by the same coordinator that assigns similar noise terms to all databases at the initialization stage of basic PRUW. Based on the structure and the size of the \emph{noise-added permutation reversing matrices} stored at each database, we have two cases for the scheme, which result in two different total communication costs. Cases~1 and 2 correspond to \emph{noise-added permutation reversing matrices} of sizes $O\left(\frac{L^2}{N^2}\right)$ and $O(L^2)$, respectively. The general scheme for case~1 is described in detail next, along with the respective modifications for case~2.

\subsubsection{General Scheme}

\textbf{Storage and initialization:} The storage of a single subpacket in database $n$ is,
\begin{align}
    S_n=\begin{bmatrix}\label{storagech3}
    \begin{bmatrix}
        W_{1,1}+ (f_1-\alpha_n)\sum_{i=0}^{x}\alpha_n^i Z_{1,i}^{[1]}\\
        W_{2,1}+ (f_1-\alpha_n)\sum_{i=0}^{x}\alpha_n^i Z_{2,i}^{[1]}\\
        \vdots\\
        W_{M,1}+ (f_1-\alpha_n)\sum_{i=0}^{x}\alpha_n^i Z_{M,i}^{[1]}\\
    \end{bmatrix}\\
    \vdots\\
    \begin{bmatrix}
        W_{1,\ell}+ (f_\ell-\alpha_n)\sum_{i=0}^{x}\alpha_n^i Z_{1,i}^{[\ell]}\\
        W_{2,\ell}+ (f_\ell-\alpha_n)\sum_{i=0}^{x}\alpha_n^i Z_{2,i}^{[\ell]}\\
        \vdots\\
        W_{M,\ell} + (f_\ell-\alpha_n)\sum_{i=0}^{x}\alpha_n^i Z_{M,i}^{[\ell]}\\
    \end{bmatrix}
    \end{bmatrix}, 
\end{align}
where $\ell$ is the subpacketization, $W_{i,j}$ is the $j$th bit of the given subpacket of the $i$th submodel $W_i$, $Z_{i,j}^{[k]}$ is the $(j+1)$st noise term for the $k$th bit of $W_i$, and $\{f_i\}_{i=1}^\ell$, $\{\alpha_n\}_{n=1}^N$ are globally known distinct constants chosen from $\bF_q$, such that each $\alpha_n$ and $f_i-\alpha_n$ for all $i\in\{1,\dotsc,\ell\}$ and $n\in\{1,\dotsc,N\}$ are coprime with $q$. The degree of the noise polynomial in storage (value of $x$) for cases~1 and 2 are $x=2\ell$ and $x=\ell+1$, respectively. 

In PRUW, at time $t=0$, it should be ensured that all noise terms in storage are the same in all databases. This is handled by the coordinator in Figure~\ref{cord}. We make use of this coordinator again in PRUW with top $r$ sparsification as follows. In the reading and writing phases, the user only reads and writes parameters/updates corresponding to a subset of subpackets ($\ll P$) without revealing their true indices. The coordinator is used to privately shuffle the true non-zero subpacket indices as explained next.

At the beginning of the FSL system design, $t=0$, the coordinator picks a random permutation of indices $\{1,\dotsc,P\}$ out of all $P!$ options, denoted by $\tilde{P}$, where $P$ is the number of subpackets. The coordinator sends $\tilde{P}$ to all users involved in the PRUW process. Then, the coordinator sends the corresponding noise-added permutation reversing matrix to database $n$, $n\in\{1,\dotsc,N\}$, given by $R_n$, whose explicit forms are given below for the two cases. Each user sends the sparse updates to databases in the form (update, position), based on the order specified by $\tilde{P}$, and the databases can reverse the permutations using $R_n$, without knowing the permutation explicitly.  

\emph{Case~1:} The noise-added permutation reversing matrix is given by,
\begin{align}\label{rearrange1}
    R_n=R+\prod_{i=1}^\ell (f_i-\alpha_n)\bar{Z},
\end{align}
where $R$ is the permutation reversing matrix and $\bar{Z}$ is a random noise matrix, both of size $P\times P$.\footnote{Since $P=\frac{L}{\ell}$ and $\ell=O(N)$, $R_n$ is of $O(P^2)$ which is $O\left(\frac{L^2}{N^2}\right)$.} For example, for a case where $P=3$, the matrix $R$ for a random permutation given by $\tilde{P}=\{2,3,1\}$ is given by,
\begin{align}
    R=\begin{bmatrix}
     0 & 0 & 1\\
     1 & 0 & 0\\
     0 & 1 & 0
    \end{bmatrix}
\end{align}
For each database, $R_n$ is a random noise matrix from Shannon's one-time-pad theorem, from which nothing can be learned about the random permutation $\tilde{P}$. The matrix $R_n$ is fixed at database $n$ at all time instances.

\emph{Case~2:} The noise-added permutation reversing matrix is given by,
\begin{align}\label{rearrange2}
    R_n=\tilde{R}_n+\tilde{Z},
\end{align}
where $\tilde{R}_n$ is the permutation reversing matrix as in case~1 (i.e., $R$) with all its entries multiplied (element-wise) by the diagonal matrix,
\begin{align}
    \begin{bmatrix}
     \frac{1}{f_1-\alpha_n} & 0 & \dotsc & 0\\
     0 & \frac{1}{f_2-\alpha_n} & \dotsc & 0\\
     \vdots & \vdots & \vdots & \vdots &\\
     0 & 0 & \dotsc & \frac{1}{f_{\ell}-\alpha_n}
    \end{bmatrix}.
\end{align}
Therefore, $R_n$ is of size $P\ell\times P\ell=L\times L$. $\tilde{Z}$ is a random noise matrix of the same size. For the same example with $P=3$ and $\tilde{P}=\{2,3,1\}$, the matrix $\tilde{R}_n$ is given by,
\begin{align}\label{permrev2}
    \tilde{R}_n=\begin{bmatrix}
     0_{\ell\times\ell} & 0_{\ell\times\ell} & \begin{bmatrix}
     \frac{1}{f_1-\alpha_n} & \dotsc & 0\\
     \vdots & \vdots & \vdots\\
     0 & \dotsc & \frac{1}{f_{\ell}-\alpha_n}
    \end{bmatrix}\\
     \begin{bmatrix}
     \frac{1}{f_1-\alpha_n} & \dotsc & 0\\
     \vdots & \vdots & \vdots\\
     0 & \dotsc & \frac{1}{f_{\ell}-\alpha_n}
    \end{bmatrix} & 0_{\ell\times\ell} & 0_{\ell\times\ell}\\
     0_{\ell\times\ell} & \begin{bmatrix}
     \frac{1}{f_1-\alpha_n} & \dotsc & 0\\
     \vdots & \vdots & \vdots\\
     0 & \dotsc & \frac{1}{f_{\ell}-\alpha_n}
    \end{bmatrix} & 0_{\ell\times\ell}
    \end{bmatrix}.
\end{align}

\textbf{Reading phase:} The process of reading (downlink) a subset of parameters of a given submodel without revealing the submodel index or the parameter indices within the submodel to databases is explained in this section.\footnote{The privacy constraints of the problem only imply the privacy of the submodel index in the reading phase. However, the privacy constraints applicable to the writing phase in the previous iteration imply the privacy of the sparse subpacket indices of the current reading phase.} In the proposed scheme, all communications between the users and databases take place only in terms of the permuted subpacket indices. The users at time $t-1$ send the permuted indices of the sparse subpackets to databases in the writing phase, and the databases work only with these permuted indices of all users to identify the sparse set of subpackets for the next downlink, and send the permuted indices of the selected set of $Pr'$ sparse subpackets to all users at time $t$. Precisely, let $\tilde{V}$ be the set of permuted indices of the $Pr'$ subpackets chosen by the databases (e.g., union of permuted indices received by all users at time $t-1$) at time $t$, to be sent to the users in the reading phase. One designated database sends $\tilde{V}$ to each user at time $t$, from which the users find the real indices of the subpackets in $\tilde{V}$, using the known permutation $\tilde{P}$, received by the coordinator at the initialization stage. The next steps of the reading phase at time $t$ are as follows. Note that the following steps are identical in both cases. However, the equations given next correspond to case~1, followed by the respective calculations of case~2, separately after the calculations of case~1.
\begin{enumerate}
    \item The user sends a query to each database $n$, $n\in\{1,\dotsc,N\}$ to privately specify the required submodel $W_{\theta}$ given by,
    \begin{align}\label{query_p2}
    Q_n=\begin{bmatrix}
        \frac{1}{f_1-\alpha_n}e_M(\theta)+\Tilde{Z}_{1}\\
        \frac{1}{f_2-\alpha_n}e_M(\theta)+\Tilde{Z}_{2}\\
        \vdots\\
        \frac{1}{f_\ell-\alpha_n}e_M(\theta)+\Tilde{Z}_{\ell}
    \end{bmatrix},
    \end{align}
    where $e_M(\theta)$ is the all zeros vector of size $M\times1$ with a $1$ at the $\theta$th position and $\Tilde{Z}_{i}$ are random noise vectors of the same size.
    
    \item In order to send the non-permuted version of the $i$th, $i\in\{1,\dotsc,|\tilde{V}|\}$, sparse subpacket (i.e., $V(i)=\tilde{P}(\tilde{V}(i))$) from the set $\tilde{V}$, database $n$ picks the column $\tilde{V}(i)$ of the permutation reversing matrix $R_n$ given in \eqref{rearrange1} indicated by $R_n(:,\tilde{V}(i))$ and calculates the corresponding query given by,
    \begin{align}
        Q_n^{[V(i)]}&=\begin{bmatrix}
            R_n(1,\tilde{V}(i))Q_n\\
            \vdots\\
            R_n(P,\tilde{V}(i))Q_n
        \end{bmatrix}=\begin{bmatrix}
            (R(1,\tilde{V}(i))+\prod_{i=1}^\ell (f_i-\alpha_n)\bar{Z}(1,\tilde{V}(i)))Q_n\\
            \vdots\\
            (R(P,\tilde{V}(i))+\prod_{i=1}^\ell (f_i-\alpha_n)\bar{Z}(P,\tilde{V}(i)))Q_n
        \end{bmatrix}\\
        &=\begin{bmatrix}
            1_{\{V(i)=1\}}\begin{bmatrix}
                \frac{1}{f_1-\alpha_n}e_M(\theta)\\
                \vdots\\
                \frac{1}{f_{\ell}-\alpha_n}e_M(\theta)\\
            \end{bmatrix}+P_{\alpha_n}(\ell)\\
            \vdots\\
            1_{\{V(i)=P\}}\begin{bmatrix}
                \frac{1}{f_1-\alpha_n}e_M(\theta)\\
                \vdots\\
                \frac{1}{f_{\ell}-\alpha_n}e_M(\theta)\\
            \end{bmatrix}+P_{\alpha_n}(\ell)
        \end{bmatrix},
    \end{align}
    where $P_{\alpha_n}(\ell)$ are noise vectors consisting of polynomials in $\alpha_n$ of degree $\ell$.
    \item Then, the user downloads (non-permuted) subpacket $V(i)=\tilde{P}(\tilde{V}(i))$, $i\in\{1,\dotsc,|\tilde{V}|\}$ of the required submodel using the answers received by the $N$ databases given by,
    \begin{align}
        A_n^{[V(i)]}&=S_n^TQ_n^{[V(i)]}\\
        &=\frac{1}{f_1-\alpha_n}W_{\theta,1}^{[V(i)]}+\dotsc+\frac{1}{f_{\ell}-\alpha_n}W_{\theta,\ell}^{[V(i)]}+P_{\alpha_n}(\ell+x+1),
    \end{align}
    from which the $\ell$ bits of subpacket $V(i)$, $i\in\{1,\dotsc,|\tilde{V}|\}$ can be obtained from the $N$ answers, given that $N=\ell+\ell+x+2=4\ell+2$ is satisfied. Thus, the subpacketization is $\ell=\frac{N-2}{4}$, and the reading cost is,
    \begin{align}
        C_R&=\frac{P\log_qP+|\tilde{V}|(N+\log_qP)}{L}\\
        &=\frac{P\log_qP+Pr'(N+\log_qP)}{P\times\frac{N-2}{4}}\\
        &=\frac{4r'+\frac{4}{N}(1+r')\log_qP}{1-\frac{2}{N}},
    \end{align}
    where $r'$, $0\leq r'\leq 1$ is the sparsification rate in the downlink characterized by $|\tilde{V}|=P\times r'$.
\end{enumerate} 

\emph{Calculations of case~2:} The steps described above for case~1 are the same for case~2 as well, with the following modifications in the equations. The query sent by the user to database $n$, $n\in\{1,\dotsc,N\}$ in step~1 is given by,
\begin{align}\label{querycase2}
    Q_n=\begin{bmatrix}
        \hat{Q}_1=e_M(\theta)+(f_1-\alpha_n)\Tilde{Z}_{1}\\
        \hat{Q}_2=e_M(\theta)+(f_2-\alpha_n)\Tilde{Z}_{2}\\
        \vdots\\
        \hat{Q}_\ell=e_M(\theta)+(f_\ell-\alpha_n)\Tilde{Z}_{\ell}
    \end{bmatrix},
\end{align}
with the same notation. Then, in step~2, to download the (non-permuted) subpacket $V(i)=\tilde{P}(\tilde{V}(i))$ each database $n$ uses the following procedure. Denote the $P\ell\times\ell$ sized submatrix of $R_n$ (in \eqref{rearrange2}) that includes the first $\ell$ columns of $R_n$ by $R_n^{[1]}$, and the submatrix that includes the second $\ell$ columns of $R_n$ by $R_n^{[2]}$, and so on, i.e., $R_n^{[s]}=R_n(:,(s-1)\ell+1:s\ell)$. Now, to download subpacket $V(i)$, database $n$ picks $R_n^{[\tilde{V}(i)]}$, computes the sum of the columns in $R_n^{[\tilde{V}(i)]}$ as, 
\begin{align}
    \hat{R}_n^{[\tilde{V}(i)]}=\sum_{j=1}^\ell R_n^{[\tilde{V}(i)]}(:,j)=\sum_{j=1}^\ell R_n(:,(\tilde{V}(i)-1)\ell+j),
\end{align}
and calculates the corresponding query as,
\begin{align}
    Q_n^{[V(i)]}&=\begin{bmatrix}
    \begin{bmatrix}
     \hat{R}_n^{[\tilde{V}(i)]}(1)\hat{Q}_1\\
     \hat{R}_n^{[\tilde{V}(i)]}(2)\hat{Q}_2\\
     \vdots\\
     \hat{R}_n^{[\tilde{V}(i)]}(\ell)\hat{Q}_\ell
    \end{bmatrix}\\
    \begin{bmatrix}
     \hat{R}_n^{[\tilde{V}(i)]}(\ell+1)\hat{Q}_1\\
     \hat{R}_n^{[\tilde{V}(i)]}(\ell+2)\hat{Q}_2\\
     \vdots\\
     \hat{R}_n^{[\tilde{V}(i)]}(2\ell)\hat{Q}_\ell
    \end{bmatrix}\\
    \vdots\\
    \begin{bmatrix}
     \hat{R}_n^{[\tilde{V}(i)]}((P-1)\ell+1)\hat{Q}_1\\
     \hat{R}_n^{[\tilde{V}(i)]}((P-1)\ell+2)\hat{Q}_2\\
     \vdots\\
     \hat{R}_n^{[\tilde{V}(i)]}(P\ell)\hat{Q}_\ell
    \end{bmatrix}
    \end{bmatrix}=\begin{bmatrix}
            1_{\{V(i)=1\}}\begin{bmatrix}
                \frac{1}{f_1-\alpha_n}e_M(\theta)\\
                \vdots\\
                \frac{1}{f_{\ell}-\alpha_n}e_M(\theta)\\
            \end{bmatrix}+P_{\alpha_n}(1)\\
            1_{\{V(i)=2\}}\begin{bmatrix}
                \frac{1}{f_1-\alpha_n}e_M(\theta)\\
                \vdots\\
                \frac{1}{f_{\ell}-\alpha_n}e_M(\theta)\\
            \end{bmatrix}+P_{\alpha_n}(1)\\
            \vdots\\
            1_{\{V(i)=P\}}\begin{bmatrix}
                \frac{1}{f_1-\alpha_n}e_M(\theta)\\
                \vdots\\
                \frac{1}{f_{\ell}-\alpha_n}e_M(\theta)\\
            \end{bmatrix}+P_{\alpha_n}(1)
        \end{bmatrix},
\end{align}
where $P_{\alpha_n}(1)$ is vector polynomial in $\alpha_n$ of degree $1$ of size $M\ell\times1$. Then, in step~3, database $n$ sends the answers to the queries in the same way as,
\begin{align}
        A_n^{[V(i)]}&=S_n^TQ_n^{[V(i)]}\\
        &=\frac{1}{f_1-\alpha_n}W_{\theta,1}^{[V(i)]}+\dotsc+\frac{1}{f_{\ell}-\alpha_n}W_{\theta,\ell}^{[V(i)]}+P_{\alpha_n}(x+2),
\end{align}
where $W_{i,j}^{[k]}$ is the $j$th bit of submodel $i$ in subpacket $k$. The $\ell$ bits of $W_{\theta}$ in subpacket $V(i)$ are obtained when $N=\ell+x+3=2\ell+4$ is satisfied, which gives the subpacketization of case~2 as $\ell=\frac{N-4}{2}$, that results in the reading cost given by, 
\begin{align}
    C_R&=\frac{P\log_qP+|\tilde{V}|(N+\log_qP)}{L}\\
    &=\frac{P\log_qP+Pr'(N+\log_qP)}{P\times\frac{N-4}{2}}\\
    &=\frac{2r'+\frac{2}{N}(1+r')\log_qP}{1-\frac{4}{N}},
\end{align}
with the same notation used for case~1. 
 
\textbf{Writing phase:} Similar to the presentation of the reading phase, we describe the general scheme that is valid for both cases, along with the equations relevant to case~1, and provide the explicit equations corresponding to case~2 at the end. The writing phase of the PRUW scheme with top $r$ sparsification consists of the following steps.

\begin{enumerate}
\item The user generates combined updates (one bit per subpacket) of the non-zero subpackets and has zero as the combined update of the rest of the $P(1-r)$ subpackets. The update of subpacket $s$ for database $n$ is given by,\footnote{A permuted version of these updates is sent to the databases.}
\begin{align}\label{update1}
    U_n(s)=\begin{cases}
    0, & \text{$s\in B^c$},\\
    \sum_{i=1}^\ell \Tilde{\Delta}_{\theta,i}^{[s]} \prod_{j=1,j\neq i}^\ell (f_j-\alpha_n) +\prod_{j=1}^\ell (f_j-\alpha_n)Z_s, & \text{$s\in B$},
    \end{cases}
\end{align}
where $B$ is the set of subpacket indices with non-zero updates, $Z_s$ is a random noise bit and $\Tilde{\Delta}_{\theta,i}^{[s]}=\frac{\Delta_{\theta,i}^{[s]}}{\prod_{j=1,j\neq i}^\ell (f_j-f_i)}$ with $\Delta_{\theta,i}^{[s]}$ being the update for the $i$th bit of subpacket $s$ of $W_\theta$. 
\item The user permutes the updates of subpackets using $\tilde{P}$. The permuted combined updates are given by,
\begin{align}
    \hat{U}_n(i)=U_n(\tilde{P}(i)),\quad i=1,\dotsc,P.
\end{align}
\item Then, the user sends the following (update, position) pairs to each database $n$,
\begin{align}
    Y_n^{[j]}=(\hat{U}_n^{[j]}, k^{[j]}), \quad j=1,\dotsc,Pr,
\end{align}
where $k^{[j]}$ is the $j$th non-zero permuted subpacket index based on $\tilde{P}$ and $\hat{U}_n^{[j]}$ is the corresponding combined update.
\item Based on the received (update, position) pairs, each database constructs an update vector $\hat{V}_n$ of size $P\times 1$ with $\hat{U}_n^{[j]}$ placed as the $k^{[j]}$th entry and zeros elsewhere,
\begin{align}\label{permute}
    \hat{V}_n=\sum_{j=1}^{Pr} \hat{U}_n^{[j]}e_P(k^{[j]})=\hat{U}_n.
\end{align}
\item $\hat{V}_n$ in \eqref{permute} contains the combined updates of the form \eqref{update1} arranged in a random permutation given by $\tilde{P}$. The databases are unable to determine the true indices of all zero subpackets since $\tilde{P}$ is not known by the databases. However, for correctness in the writing phase, the updates in $\hat{V}_n$ must be rearranged in the correct order. This is done with the noise-added permutation reversing matrix given in \eqref{rearrange1} as,
\begin{align}
    T_n&=R_n\hat{V}_n=R\hat{V}_n+\prod_{i=1}^\ell (f_i-\alpha_n)P_{\alpha_n}(\ell),
\end{align}
where $P_{\alpha_n}(\ell)$ is a $P\times1$ vector containing noise polynomials in $\alpha_n$ of degree $\ell$, $R\hat{V}_n$ contains all updates of all subpackets (including zeros) in the correct order, while $\prod_{i=1}^\ell (f_i-\alpha_n)P_{\alpha_n}(\ell)$ contains random noise, that hides the indices of the zero update subpackets. 
\item The incremental update is calculated in the same way as described in Section~\ref{basicscheme} in each subpacket as,
\begin{align}
    \bar{U}_n(s)\!\!&=D_n\times T_n(s)\times Q_n\\
    &=D_n\times U_n(s)\times Q_n+D_n\times P_{\alpha_n}(2\ell)\label{first}\\
    &=\begin{cases}
    \!\!\begin{bmatrix}
        \Delta_{\theta,1}^{[s]}e_M(\theta) \\  \vdots\\ \Delta_{\theta,\ell}^{[s]} e_M(\theta) 
    \end{bmatrix}\!\!+\!\!\begin{bmatrix}
    \!(f_1\!-\alpha_n)P_{\alpha_n}(2\ell)\!
        \\ \vdots\\
    \!(f_\ell\!-\alpha_n)P_{\alpha_n}(2\ell)\!
    \end{bmatrix}, & s\in B,\\
    \!\!\begin{bmatrix}
    (f_1-\alpha_n)P_{\alpha_n}(2\ell)
        \\ \vdots\\
    (f_{\ell}-\alpha_n)P_{\alpha_n}(2\ell)
    \end{bmatrix}, & s\in B^c,
    \end{cases}\label{next}
\end{align}
where $P_{\alpha_n}(2\ell)$ here are noise vectors of size $M\ell\times1$ in \eqref{first} and $M\times1$ in \eqref{next} with polynomials in $\alpha_n$ of degree $2\ell$ and $D_n$ is the scaling matrix given by, 
\begin{align}
    D_n=\begin{bmatrix}
        (f_1-\alpha_n)I_M & \dotsc & 0\\
        \vdots & \vdots & \vdots\\
        0 & \dotsc & (f_\ell-\alpha_n)I_M\\
    \end{bmatrix}, 
\end{align}
for all $n$. $\bar{U}_n(s)$ is in the same format as the storage in \eqref{storagech3} with $x=2\ell$ (case~1) and hence can be added to the existing storage to obtain the updated storage, i.e.,
\begin{align}
    S_n^{[t]}(s)=S_n^{[t-1]}(s)+\bar{U}_n(s), \quad s=1,\dotsc,P.
\end{align}
Note that the degree $\ell$ noise polynomials in $\alpha_n$ (noise matrix) in the noise-added permutation reversing matrix in \eqref{rearrange1} introduces $\ell$ extra noise terms in the incremental update calculation, compared to the basic PRUW scheme in which the incremental update has a noise polynomial in $\alpha_n$ of degree $\ell$. In other words, the permutation technique requires $\ell$ dimensions from the $N$ dimensional space which reduces the number of dimensions left for data downloads and uploads to guarantee the privacy of the sparsification process. The writing cost of the scheme is given by,\footnote{Note that the \emph{upload cost} of the query vector from the reading phase, which is of size $M\ell\times1$ and is not considered in the writing cost calculation since $\frac{M\ell}{L}$ is negligible.}
\begin{align}
    C_W&=\frac{PrN(1+\log_q P)}{L}\\
    &= \frac{PrN(1+\log_q P)}{P\times\frac{N-2}{4}}\\
    &=\frac{4r(1+\log_q P)}{1-\frac{2}{N}}.
\end{align}
\end{enumerate}

\emph{Calculations of case~2:} Steps~1-4 in the general scheme are valid for case~2 with the same equations. In step~5, the updates of permuted subpackets $\hat{V}_n$ are privately arranged in the correct order as follows. Using the noise added permutation reversing matrix $R_n$ in \eqref{rearrange2}, database $n$, $n\in\{1,\dotsc,N\}$ calculates,
\begin{align}
    T_n&=R_n\times [\hat{V}_n(1)1_{\ell},\hat{V}_n(2)1_{\ell},\dotsc \hat{V}_n(P)1_{\ell}]^T\\
    &=(\tilde{R}_n+\tilde{Z})[\hat{V}_n(1)1_{\ell},\hat{V}_n(2)1_{\ell},\dotsc \hat{V}_n(P)1_{\ell}]^T\\
    &=\begin{bmatrix}
     U_n(1)\begin{bmatrix}
      \frac{1}{f_1-\alpha_n}\\
      \vdots\\
      \frac{1}{f_{\ell}-\alpha_n}
     \end{bmatrix}\\
      U_n(2)\begin{bmatrix}
      \frac{1}{f_1-\alpha_n}\\
      \vdots\\
      \frac{1}{f_{\ell}-\alpha_n}
     \end{bmatrix}\\
     \vdots\\
      U_n(P)\begin{bmatrix}
      \frac{1}{f_1-\alpha_n}\\
      \vdots\\
      \frac{1}{f_{\ell}-\alpha_n}
     \end{bmatrix}\\
    \end{bmatrix}+P_{\alpha_n}(\ell)=\begin{bmatrix}
     \begin{bmatrix}
      \frac{\Delta_{\theta,1}^{[1]}}{f_1-\alpha_n}\\
      \vdots\\
      \frac{\Delta_{\theta,\ell}^{[1]}}{f_{\ell}-\alpha_n}
     \end{bmatrix}\\
      \begin{bmatrix}
      \frac{\Delta_{\theta,1}^{[2]}}{f_1-\alpha_n}\\
      \vdots\\
      \frac{\Delta_{\theta,\ell}^{[2]}}{f_{\ell}-\alpha_n}
     \end{bmatrix}\\
     \vdots\\
      \begin{bmatrix}
      \frac{\Delta_{\theta,1}^{[P]}}{f_1-\alpha_n}\\
      \vdots\\
      \frac{\Delta_{\theta,\ell}^{[P]}}{f_{\ell}-\alpha_n}
     \end{bmatrix}\\
    \end{bmatrix}+P_{\alpha_n}(\ell),
\end{align}
where $1_{\ell}$ is an all ones vector of size $\ell\times1$ and $P_{\alpha_n}(\ell)$ here is a vector polynomial in $\alpha_n$ of degree $\ell$ of size $P\ell\times1$. Note that many $U_n(i)$s in the above calculation are zero due to sparsification. The last equality is derived from the application of Lemma~\ref{lem1} on expressions of the form $\frac{U_n(i)}{f_j-\alpha_n}$. Recall that $\Delta_{\theta,j}^{[i]}=0$, $j\in\{1,\dotsc,\ell\}$ for all subpackets $i$, that are not within the $Pr$ selected subpackets with non-zero updates.

Now that the updates are privately arranged in the correct order, it remains only to place the updates at the intended submodel in the storage. Note that the updates of the first subpacket are in the first $\ell$ rows of $T_n$, the updates of the second subpacket are in the next $\ell$ rows of $T_n$, and so on. Therefore, we divide $T_n$, based on its correspondence to subpackets as,
\begin{align}\label{tnsp}
    T_n^{[s]}=T_n((s-1)\ell+1:s\ell),
\end{align}
for $s\in\{1,\dotsc,P\}$. With this initialization, for step~6, each database calculates the incremental update of subpacket $s$, $s\in\{1,\dotsc,P\}$ using the the query in the reading phase \eqref{querycase2} as,
\begin{align}
    \bar{U}_n(s)&=D_n\times\begin{bmatrix}
     T_n^{[s]}(1)\hat{Q}_{1}\\
     \vdots\\
     T_n^{[s]}(\ell)\hat{Q}_{\ell}
    \end{bmatrix}=D_n\times\begin{bmatrix}
     \left(\frac{\Delta_{\theta,1}^{[s]}}{f_1-\alpha_n}+P_{\alpha_n}(\ell)\right)(e_M(\theta)+(f_1-\alpha_n)\tilde{Z}_1)\\
     \vdots\\
     \left(\frac{\Delta_{\theta,\ell}^{[s]}}{f_{\ell}-\alpha_n}+P_{\alpha_n}(\ell)\right)(e_M(\theta)+(f_{\ell}-\alpha_n)\tilde{Z}_1)
    \end{bmatrix}\\
    &=\begin{bmatrix}
     \Delta_{\theta,1}^{[s]}e_M(\theta)+(f_1-\alpha_n)P_{\alpha}(\ell+1)\\
     \vdots\\
     \Delta_{\theta,\ell}^{[s]}e_M(\theta)+(f_{\ell}-\alpha_n)P_{\alpha}(\ell+1)
    \end{bmatrix},
\end{align}
with the same notation used in case~1. Since the incremental update is in the same form as the storage in \eqref{storagech3} with $x=\ell+1$, $\bar{U}_n(s)$ for $s\in\{1,\dotsc,P\}$ is added to the existing storage to obtain the updated version similar to case~1. The resulting writing cost is given by,
\begin{align}
    C_W&=\frac{PrN(1+\log_q P)}{L}\\
    &= \frac{PrN(1+\log_q P)}{P\times\frac{N-4}{2}}\\
    &=\frac{2r(1+\log_q P)}{1-\frac{4}{N}}.
\end{align}

\begin{remark}
This problem can also be solved by considering a classical FSL setting without sparsification with $P$ submodels, i.e., $M=P$, and by using the private FSL scheme in Section~\ref{basicscheme} to update the sparse $Pr$ submodels. However, in this case the  normalized cost of sending the queries $Q_n$ given by $\frac{NM\ell}{L}=\frac{NP\ell}{L}=N$ is large, and cannot be neglected.  
\end{remark}

\subsubsection{Example}\label{eg}

Assume that there are $N=10$ databases containing $M$ submodels, each with $P=5$ subpackets. The coordinator first picks a random permutation of $\{1,\dotsc,5\}$ out of the $5!$ options available. Let the realization of the permutation be $\tilde{P}=\{2,5,1,3,4\}$.

\emph{Case~1:} The subpacketization is $\ell=\frac{N-2}{4}=2$ and the storage of database $n$ consists of the model given by,\footnote{Here we have only presented the storage of a single subpacket.}
\begin{align}
    S_n=\begin{bmatrix}\label{storageg}
    \begin{bmatrix}
        W_{1,1}+ (f_1-\alpha_n)\sum_{i=0}^4 \alpha_n^{i} Z_{1,i}^{[1]}\\
        \vdots\\
        W_{M,1}+ (f_1-\alpha_n)\sum_{i=0}^4 \alpha_n^{i} Z_{M,i}^{[1]}\\
    \end{bmatrix}\\
    \begin{bmatrix}
        W_{1,2}+ (f_2-\alpha_n)\sum_{i=0}^4 \alpha_n^{i} Z_{1,i}^{[2]}\\
        \vdots\\
        W_{M,2}+ (f_2-\alpha_n)\sum_{i=0}^4 \alpha_n^{i} Z_{M,i}^{[2]}\\
    \end{bmatrix}\\
    \end{bmatrix}, 
\end{align}
since the degree of the noise polynomial is $2\ell=4$. The permutation reversing matrix is given by,
\begin{align}\label{exR}
    R_n=\begin{bmatrix}
        0 & 0 & 1 & 0 & 0\\
        1 & 0 & 0 & 0 & 0\\
        0 & 0 & 0 & 1 & 0\\
        0 & 0 & 0 & 0 & 1\\
        0 & 1 & 0 & 0 & 0\\
    \end{bmatrix}+\prod_{i=1}^2(f_i-\alpha_n)\bar{Z},
\end{align}
where $\bar{Z}$ is a random noise matrix of size $5\times5$. The coordinator places matrix $R_n$ at database $n$ at the beginning of the process and sends $\tilde{P}$ to each user. Assume that a given user wants to update submodel $\theta$ at time $t$. In the reading phase, the user only downloads the sparse set of subpackets indicated by the permuted set of indices $\tilde{V}=\{2,3\}$, which is determined by the databases. One designated database sends these permuted indices to each of the users at time $t$. Then, the user obtains the real indices of the subpackets in $\tilde{V}$, using $V(i)=\tilde{P}(\tilde{V}(i))$ for $i=1,2$, i.e., $V=\{5,1\}$. The user sends the query specifying the requirement of submodel $\theta$ given by,
\begin{align}
    Q_n=\begin{bmatrix}
     \frac{1}{f_1-\alpha_n}e_M(\theta)+\tilde{Z}_1\\
     \frac{1}{f_2-\alpha_n}e_M(\theta)+\tilde{Z}_2
    \end{bmatrix}
\end{align}
to database $n$. Then, each database privately calculates the non-permuted query vector for each subpacket $V(i)$ using the noise added permutation reversing matrix and the query received. The query for subpacket $V(1)=5$ is,
\begin{align}\label{query_right}
    Q_n^{[5]}&=\begin{bmatrix}
        R_n(1,\tilde{V}(1))Q_n\\\vdots\\R_n(P,\tilde{V}(1))Q_n
    \end{bmatrix}=\begin{bmatrix}
        0_{2M}\\0_{2M}\\0_{2M}\\0_{2M}\\Q_n
    \end{bmatrix}+P_{\alpha_n}(2)
\end{align}
where $P_{\alpha_n}(2)$ is a vector of size $10M\times1$ consisting of polynomials in $\alpha_n$ of degree $2$ and $0_{2M}$ is the all zeros vector of size $2M\times1$. Then, the answer from database $n$ corresponding to subpacket $V(1)=5$ is given by,
\begin{align}
   A_n^{[5]}&=S_n^TQ_n^{[5]}=\frac{1}{f_1-\alpha_n}W_{\theta,1}^{[5]}+\frac{1}{f_{2}-\alpha_n}W_{\theta,2}^{[5]}+P_{\alpha_n}(3\times2+1),
\end{align}
from which the $2$ bits of subpacket $5$ of submodel $\theta$ can be correctly obtained by using the $N=10$ answers from the ten databases. Similarly, the user can obtain subpacket 1 of $W_{\theta}$ by picking column $\tilde{V}(2)=3$ of $R_n$ in \eqref{exR} in the calculation of \eqref{query_right} and following the same process.

Once the user downloads and trains $W_{\theta}$, the user generates the $r$ fraction of subpackets with non-zero updates. Let the subpacket indices with non-zero updates be 1 and 4. The noisy updates generated by the user to be sent to database $n$ according to \eqref{update1} is given by $U_n=[U_n(1),0,0,U_n(4),0]^T$ in the correct order. The user then permutes $U_n$ based on the given permutation $\tilde{P}$, i.e., $\hat{U}_n(i)=U_n(\tilde{P}(i))$ for $i=\{1,\dotsc,5\}$,
\begin{align}
    \hat{U}_n&=[0,0,U_n(1),0,U_n(4)]^T\label{perm}.
\end{align}
The user sends the values and the positions of the non-zero updates as $(U_n(1),3)$ and $(U_n(4),5)$ based on the permuted order. Each database receives these pairs and reconstructs \eqref{perm},
\begin{align}\label{perm22}
    \hat{V}_n&=U_n(1)e_5(3)+U_n(4)e_5(5)=\hat{U}_n.
\end{align}
To rearrange the updates back in the correct order privately, database $n$ multiplies $\hat{V}_n$ by the permutation reversing matrix,
\begin{align}
    T_n&=R_n\times \hat{V}_n\\
    &=\begin{bmatrix}
        0 & 0 & 1 & 0 & 0\\
        1 & 0 & 0 & 0 & 0\\
        0 & 0 & 0 & 1 & 0\\
        0 & 0 & 0 & 0 & 1\\
        0 & 1 & 0 & 0 & 0\\
    \end{bmatrix}\hat{V}_n+\prod_{i=1}^2(f_i-\alpha_n)\bar{Z}\times \hat{V}_n\\
    &=[U_n(1),0,0,U_n(4),0]^T+\prod_{i=1}^2(f_i-\alpha_n)P_{\alpha_n}(2)\label{reorder},
\end{align}
since $U_n(1)$ and $U_n(4)$ are of the form $\sum_{i=1}^2 \Tilde{\Delta}_{\theta,i} \prod_{j=1,j\neq i}^2 (f_j-\alpha_n)+\prod_{j=1}^2 (f_j-\alpha_n)Z=P_{\alpha_n}(2)$. The incremental update of subpacket $s$, is calculated by,
\begin{align}
    \bar{U}_n(s)&=D_n\times T_n(s)\times Q_n\\
    &=\begin{cases}
    \begin{bmatrix}
        \Delta_{1,1}^{[s]}e_M(\theta)\\ \Delta_{1,2}^{[s]} e_M(\theta) 
    \end{bmatrix}+\begin{bmatrix}
        (f_1-\alpha_n)P_{\alpha_n}(4)\\
        (f_2-\alpha_n)P_{\alpha_n}(4)
    \end{bmatrix}, & s=1,4\\
    \begin{bmatrix}
        (f_1-\alpha_n)P_{\alpha_n}(4)\\
        (f_2-\alpha_n)P_{\alpha_n}(4)
    \end{bmatrix}, & s=2,3,5
    \end{cases}
\end{align}
using Lemma~\ref{lem1}, where $P_{\alpha_n}(4)$ are vectors of size $M\times1$ consisting of noise polynomials in $\alpha_n$ of degree $4$. Since the incremental update is in the same format as the storage in \eqref{storageg}, the existing storage can be updated as $S_n^{[t]}(s)=S_n^{[t-1]}(s)+\bar{U}_n(s)$ for $s=1,\dotsc,5$, where $S_n^{[t]}(s)$ is the storage of subpacket $s$ in \eqref{storageg} at time $t$. 

\emph{Case~2:} For this case, the subpacketization is $\ell=\frac{N-4}{2}=3$ and the storage of the model is given by,
\begin{align}
    S_n=\begin{bmatrix}\label{storageg2}
    \begin{bmatrix}
        W_{1,1}+ (f_1-\alpha_n)\sum_{i=0}^4 \alpha_n^{i} Z_{1,i}^{[1]}\\
        \vdots\\
        W_{M,1}+ (f_1-\alpha_n)\sum_{i=0}^4 \alpha_n^{i} Z_{M,i}^{[1]}\\
    \end{bmatrix}\\
    \begin{bmatrix}
        W_{1,2}+ (f_2-\alpha_n)\sum_{i=0}^4 \alpha_n^{i} Z_{1,i}^{[2]}\\
        \vdots\\
        W_{M,2}+ (f_2-\alpha_n)\sum_{i=0}^4 \alpha_n^{i} Z_{M,i}^{[2]}\\
    \end{bmatrix}\\
    \begin{bmatrix}
        W_{1,3}+ (f_3-\alpha_n)\sum_{i=0}^4 \alpha_n^{i} Z_{1,i}^{[3]}\\
        \vdots\\
        W_{M,3}+ (f_3-\alpha_n)\sum_{i=0}^4 \alpha_n^{i} Z_{M,i}^{[3]}\\
    \end{bmatrix}\\
    \end{bmatrix}, 
\end{align}
since the degree of the noise polynomial $x=\ell+1=4$ and the permutation reversing matrix stored in database $n$, $n\in\{1,\dotsc,N\}$ is given by,
\begin{align}\label{exR2}
    R_n=\begin{bmatrix}
        0_{3\times3} & 0_{3\times3} & \Gamma & 0_{3\times3} & 0_{3\times3}\\
        \Gamma & 0_{3\times3} & 0_{3\times3} & 0_{3\times3} & 0_{3\times3}\\
        0_{3\times3} & 0_{3\times3} & 0_{3\times3} & \Gamma & 0_{3\times3}\\
        0_{3\times3} & 0_{3\times3} & 0_{3\times3} & 0_{3\times3} & \Gamma\\
        0_{3\times3} & \Gamma & 0_{3\times3} & 0_{3\times3} & 0_{3\times3}\\
    \end{bmatrix}+\tilde{Z},
\end{align}
where $\Gamma=\begin{bmatrix}
         \frac{1}{f_1-\alpha_n} & 0 & 0\\
         0 & \frac{1}{f_2-\alpha_n} & 0\\
         0 & 0 & \frac{1}{f_3-\alpha_n}\\
        \end{bmatrix}$, and $0_{3\times3}$ is the all zeros matrix of size $3\times3$. For the same example where users need to read the permuted subpackets $\tilde{V}=\{2,3\}$, a designated database sends $\tilde{V}$ to each user, from which the user obtains the non-permuted subpacket indices $V=\{5,1\}$ using $\tilde{P}$. The user sends the following query to specify the required submodel index $\theta$,
\begin{align}\label{querycase2eg}
    Q_n=\begin{bmatrix}
        \hat{Q}_1=e_M(\theta)+(f_1-\alpha_n)\Tilde{Z}_{1}\\
        \hat{Q}_2=e_M(\theta)+(f_2-\alpha_n)\Tilde{Z}_{2}\\
        \hat{Q}_3=e_M(\theta)+(f_3-\alpha_n)\Tilde{Z}_{3}
    \end{bmatrix}.
\end{align}
To read subpacket $V(1)=5$, database $n$ first computes the sum of the $\ell=3$ columns of the $\tilde{V}(1)=2$nd submatrix of $R_n$ given by,
\begin{align}
    \hat{R}_n^{[\tilde{V}(1)]}=\hat{R}_n^{[2]}=\sum_{j=1}^3R_n(:,3+i)=\begin{bmatrix}
     0_{3}\\
     0_{3}\\
     0_{3}\\
     0_{3}\\
     \frac{1}{f_1-\alpha_n}\\
     \frac{1}{f_2-\alpha_n}\\
     \frac{1}{f_3-\alpha_n}
    \end{bmatrix}+\hat{Z},
\end{align}
where $0_3$ is the all zeros vector of size $3\times1$, $\hat{Z}$ is a random vector of size $15\times1$. Then, each database computes the specific query for $V(1)=5$ given by,
\begin{align}\label{qexample}
    Q_n^{[5]}=\begin{bmatrix}
    \begin{bmatrix}
     \hat{R}_n^{[2]}(1)\hat{Q}_1\\
     \hat{R}_n^{[2]}(2)\hat{Q}_2\\
     \hat{R}_n^{[2]}(3)\hat{Q}_3
    \end{bmatrix}\\
    \vdots\\
    \begin{bmatrix}
     \hat{R}_n^{[2]}(13)\hat{Q}_1\\
     \hat{R}_n^{[2]}(14)\hat{Q}_2\\
     \hat{R}_n^{[2]}(15)\hat{Q}_3
    \end{bmatrix}\\
    \end{bmatrix}=\begin{bmatrix}
     0\times\hat{Q}_1 \\
     0\times\hat{Q}_1 \\
     0\times\hat{Q}_3 \\
     \vdots\\
     0\times\hat{Q}_1 \\
     0\times\hat{Q}_1 \\
     0\times\hat{Q}_3 \\
     \frac{1}{f_1-\alpha_n}\hat{Q}_1\\
     \frac{1}{f_2-\alpha_n}\hat{Q}_2\\
     \frac{1}{f_3-\alpha_n}\hat{Q}_3
    \end{bmatrix}+P_{\alpha_n}(1)=\begin{bmatrix}
     0_{3M}\\
     0_{3M}\\
     0_{3M}\\
     0_{3M}\\
     \frac{1}{f_1-\alpha_n}e_M(\theta)\\
     \frac{1}{f_2-\alpha_n}e_M(\theta)\\
     \frac{1}{f_3-\alpha_n}e_M(\theta)
    \end{bmatrix}+P_{\alpha_n}(1),
\end{align}
where the polynomial vectors $P_{\alpha_n}(1)$ are resulted by the multiplications of the form $\hat{Z}_{i}\hat{Q}_j$ and by the residual terms of the calculations of the form $\frac{1}{f_i-\alpha_n}\hat{Q}_j$. Note that the two $P_{\alpha_n}(1)$ vectors in \eqref{qexample} are not the same, and they are both some random vector polynomials in $\alpha_n$ of degree 1 of size $15M\times1$. Each database $n$,  $n\in\{1,\dotsc,N\}$ then sends the answers to this query given by,
\begin{align}
    A_n^{[5]}=S_n^{T}Q_n^{[5]}=\frac{1}{f_1-\alpha_n}W_{\theta,1}^{[5]}+\frac{1}{f_2-\alpha_n}W_{\theta,2}^{[5]}+\frac{1}{f_3-\alpha_n}W_{\theta,3}^{[5]}+P_{\alpha_n}(6),
\end{align}
from which the three bits of subpacket 5 can be obtained since $N=3+6+1=10$.

For the same example considered in case~1, the user sends the two updates corresponding to subpackets~1 and 4, along with the permuted positions, from which the databases compute $\hat{V}_n=[0,0,U_n(1),0,U_n(4)]^T$ given in \eqref{perm22}, where $U_n(1)$ and $U_n(4)$ are of the form $\sum_{i=1}^3 \Tilde{\Delta}_{\theta,i} \prod_{j=1,j\neq i}^3 (f_j-\alpha_n)+\prod_{j=1}^3 (f_j-\alpha_n)Z=P_{\alpha_n}(3)$. Then, database $n$, $n\in\{1,\dotsc,N\}$ rearranges the updates in the correct order as,
\begin{align}
    T_n&=R_n\times [\hat{V}_n(1)1_{3},\hat{V}_n(2)1_{3},\hat{V}_n(3)1_{3},\hat{V}_n(4)1_{3},\hat{V}_n(5)1_{3}]^T\\
    &=\left(\begin{bmatrix}
        0_{3\times3} & 0_{3\times3} & \Gamma & 0_{3\times3} & 0_{3\times3}\\
        \Gamma & 0_{3\times3} & 0_{3\times3} & 0_{3\times3} & 0_{3\times3}\\
        0_{3\times3} & 0_{3\times3} & 0_{3\times3} & \Gamma & 0_{3\times3}\\
        0_{3\times3} & 0_{3\times3} & 0_{3\times3} & 0_{3\times3} & \Gamma\\
        0_{3\times3} & \Gamma & 0_{3\times3} & 0_{3\times3} & 0_{3\times3}\\
    \end{bmatrix}+\bar{Z}\right)\times[0_{3\times1},0_{3\times1},U_n(1)1_{3},0_{3\times1},U_n(4)1_{3}]^T\\
    &=\begin{bmatrix}
     U_n(1)\begin{bmatrix}
     \frac{1}{f_1-\alpha_n}\\
     \frac{1}{f_2-\alpha_n}\\
     \frac{1}{f_{3}-\alpha_n}
     \end{bmatrix}\\
     0_{3\times1}\\
     0_{3\times1}\\
     U_n(4)\begin{bmatrix}
     \frac{1}{f_1-\alpha_n}\\
     \frac{1}{f_2-\alpha_n}\\
     \frac{1}{f_{3}-\alpha_n}
     \end{bmatrix}\\
     0_{3\times1}
    \end{bmatrix}+P_{\alpha_n}(3)=\begin{bmatrix}
    \begin{bmatrix}
      \frac{\Delta_{\theta,1}^{[1]}}{f_1-\alpha_n}\\
      \frac{\Delta_{\theta,2}^{[1]}}{f_2-\alpha_n}\\
      \frac{\Delta_{\theta,3}^{[1]}}{f_{3}-\alpha_n}
     \end{bmatrix}\\
    0_{3\times1}\\
    0_{3\times1}\\
    \begin{bmatrix}
      \frac{\Delta_{\theta,1}^{[4]}}{f_1-\alpha_n}\\
      \frac{\Delta_{\theta,2}^{[4]}}{f_2-\alpha_n}\\
      \frac{\Delta_{\theta,3}^{[4]}}{f_{3}-\alpha_n}
     \end{bmatrix}\\
     0_{3\times1}
    \end{bmatrix}+P_{\alpha_n}(3),
\end{align}
where the last equality is obtained by using Lemma~\ref{lem1}. Since the subpacketization is $\ell=3$, we divide $T_n$ into blocks of 3 elements each (subpackets) as shown in \eqref{tnsp}. For example,  $T_n^{[1]}=\begin{bmatrix}
      \frac{\Delta_{\theta,1}^{[4]}}{f_1-\alpha_n}\\
      \frac{\Delta_{\theta,2}^{[4]}}{f_2-\alpha_n}\\
      \frac{\Delta_{\theta,3}^{[4]}}{f_{3}-\alpha_n}
     \end{bmatrix}+P_{\alpha_n}(3)$ and $T_n^{[2]}=P_{\alpha_n}(3)$, where $P_{\alpha_n}(3)$ is a vector polynomial of size $3\times1$. Then, as an example, the incremental update of the first subpacket is calculated as,
\begin{align}
    \bar{U}_n(1)&=D_n\times\begin{bmatrix}
     T_n^{[1]}(1)\hat{Q}_{1}\\
     T_n^{[1]}(2)\hat{Q}_{2}\\
     T_n^{[1]}(3)\hat{Q}_{3}
    \end{bmatrix}=D_n\times\begin{bmatrix}
     \left(\frac{\Delta_{\theta,1}^{[1]}}{f_1-\alpha_n}+P_{\alpha_n}(3)\right)(e_M(\theta)+(f_1-\alpha_n)\tilde{Z}_1)\\
     \left(\frac{\Delta_{\theta,2}^{[1]}}{f_2-\alpha_n}+P_{\alpha_n}(3)\right)(e_M(\theta)+(f_2-\alpha_n)\tilde{Z}_2)\\
     \left(\frac{\Delta_{\theta,3}^{[1]}}{f_{3}-\alpha_n}+P_{\alpha_n}(3)\right)(e_M(\theta)+(f_{3}-\alpha_n)\tilde{Z}_3)
    \end{bmatrix}\\
    &=\begin{bmatrix}
     \Delta_{\theta,1}^{[1]}e_M(\theta)+(f_1-\alpha_n)P_{\alpha}(4)\\
     \Delta_{\theta,2}^{[1]}e_M(\theta)+(f_2-\alpha_n)P_{\alpha}(4)\\
     \Delta_{\theta,3}^{[1]}e_M(\theta)+(f_{3}-\alpha_n)P_{\alpha}(4)
    \end{bmatrix},
\end{align}
where $P_{\alpha_n}(4)$ is a vector polynomial in $\alpha_n$ of degree $4$, of size $M\times1$. The above incremental update is directly added to the first subpacket of the existing storage in \eqref{storageg2} since both are of the same format.

\subsubsection{Proof of Privacy}

\emph{Privacy of submodel index:} The privacy constraint in \eqref{prvcy_id2} can be written as,
\begin{align}\label{id2_pf}
    I(\theta;Q_n^{[t]},Y_n^{[t]}|S_n^{[0:t-1]},Q_n^{[1:t-1]})&=I(\theta;S_n^{[0:t-1]},Q_n^{[1:t]},Y_n^{[t]})-I(\theta;S_n^{[0:t-1]},Q_n^{[1:t-1]})\\
    &=I(\theta;S_n^{[0:t-1]},Q_n^{[1:t]},\hat{U}_n^{[t]},k^{[t]}),
\end{align}
since $Y_n=(\hat{U}_n,k)$, based on the proposed scheme and since the second term in \eqref{id2_pf} is zero as all $S_n$ and $Q_n$ terms are random noise and are independent of $\theta$ from Shannon's one-time-pad theorem. For each $m\in\{1,\dotsc,M\}$ and arbitrary realizations of storage, queries, updates and permuted positions ($\bar{s}_n,\bar{r}_n,\bar{u}_n,\bar{k}$), consider the aposteriori probability,
\begin{align}
    P(\theta=m&|S_n^{[0:t-1]}=\bar{s}_n,Q_n^{[1:t]}=\bar{r}_n,\hat{U}_n^{[t]}=\bar{u}_n,k^{[t]}=\bar{k})\nonumber\\
    &=\frac{P(\theta=m,S_n^{[0:t-1]}=\bar{s}_n,Q_n^{[1:t]}=\bar{r}_n,\hat{U}_n^{[t]}=\bar{u}_n,k^{[t]}=\bar{k})}{P(S_n^{[0:t-1]}=\bar{s}_n,Q_n^{[1:t]}=\bar{r}_n,\hat{U}_n^{[t]}=\bar{u}_n,k^{[t]}=\bar{k})}\\
    &=\frac{P(\theta=m,k^{[t]}=\bar{k})P(S_n^{[0:t-1]}=\bar{s}_n,Q_n^{[1:t]}=\bar{r}_n,\hat{U}_n^{[t]}=\bar{u}_n)}{P(S_n^{[0:t-1]}=\bar{s}_n,Q_n^{[1:t]}=\bar{r}_n,\hat{U}_n^{[t]}=\bar{u}_n)P(k^{[t]}=\bar{k})}\label{113}\\
    &=\frac{P(k^{[t]}=\bar{k}|\theta=m)P(\theta=m)}{P(k^{[t]}=\bar{k})}\label{main_pf2},
\end{align}
where \eqref{113} is due to the fact that $S_n$, $Q_n$ and $\hat{U}_n$ are random noise terms that are independent of $\theta$ and $k^{[t]}$. Note that for all realizations of updates $\delta$ and permutations $\tilde{p}$,
\begin{align}
    P(k^{[t]}=\bar{k}|\theta=m)&=\sum_{\delta}\sum_{\tilde{p}} P(k^{[t]}=\bar{k}, \tilde{P}=\tilde{p},\Delta_{\theta}^{[t]}=\delta|\theta=m)\\
    &=\sum_{\delta}\sum_{\tilde{p}} P(k^{[t]}=\bar{k}|\tilde{P}=\tilde{p},\Delta_{\theta}^{[t]}=\delta,\theta=m)P(\tilde{P}=\tilde{p},\Delta_{\theta}^{[t]}=\delta|\theta=m)\\
    &=\sum_{\delta}P(\Delta_\theta^{[t]}=\delta|\theta=m)\sum_{\tilde{p}}1_{\{\tilde{P}=\tilde{p},\Delta_\theta^{[t]}=\delta,k^{[t]}=\bar{k}\}}P(\tilde{P}=\tilde{p})\label{117}\\
    &=\frac{(Pr)!(P-Pr)!}{P!}\\
    &=\frac{1}{\binom{P}{Pr}},
\end{align}
where \eqref{117} is from the fact that the randomly selected permutation $\tilde{P}$ is independent of the updating submodel index and the values of updates. Moreover,
\begin{align}
    P(k^{[t]}=\bar{k})&=\sum_{\delta}\sum_{\tilde{p}} P(k^{[t]}=\bar{k}|\Delta_\theta=\delta, \tilde{P}=\tilde{p})P(\Delta_\theta=\delta,\tilde{P}=\tilde{p})\\
    &=\sum_{\delta}\sum_{\tilde{p}} 1_{\{k^{[t]}=\bar{k},\Delta_\theta=\delta, \tilde{P}=\tilde{p}\}}P(\Delta_\theta=\delta)P(\tilde{P}=\tilde{p})\label{116}\\
    &=\sum_{\delta}P(\Delta_\theta=\delta)\frac{(Pr)!(P-Pr)!}{P!}\\
    &=\frac{1}{\binom{P}{Pr}},
\end{align}
where \eqref{116} is due to the fact that the randomly chosen permutation is independent of the values of updates generated, and that the permuted positions $k^{[t]}$ only depend on the values of updates and the permutation. Therefore, from \eqref{main_pf2},
\begin{align}
    P(\theta=m|S_n^{[0:t-1]}=\bar{s}_n,Q_n^{[1:t]}=\bar{r}_n,\hat{U}_n^{[t]}=\bar{u}_n,k^{[t]}=\bar{k})
    &=\frac{P(k^{[t]}=\bar{k}|\theta=m)P(\theta=m)}{P(k^{[t]}=\bar{k})}\\
    &=\frac{\frac{1}{\binom{P}{Pr}}P(\theta=m)}{\frac{1}{\binom{P}{Pr}}}\\
    &=P(\theta=m),
\end{align}
which proves that $I(\theta;Q_n^{[t]},Y_n^{[t]}|S_n^{[0:t-1]},Q_n^{[1:t-1]})=I(\theta;S_n^{[0:t-1]},Q_n^{[1:t]},\hat{U}_n^{[t]},k^{[t]})=0$.

\emph{Privacy of the values of updates:} The privacy constraint in \eqref{value_prvcy_sparse} can be written as,
\begin{align}\label{p2_prvcy}
    I(\Delta_{\theta}^{[t]};Q_n^{[t]},Y_n^{[t]}|S_n^{[0:t-1]},Q_n^{[1:t-1]})&=I(\Delta_{\theta}^{[t]};S_n^{[0:t-1]},Q_n^{[1:t]},Y_n^{[t]})-I(\Delta_{\theta}^{[t]};S_n^{[0:t-1]},Q_n^{[1:t-1]}).
\end{align}
The second term in \eqref{p2_prvcy} is zero since the storage $S_n$ and queries $Q_n$ are random noise terms that are independent of the submodel index and the values of updates based on Shannon's one-time-pad theorem. Therefore,
\begin{align}
    I(\Delta_{\theta}^{[t]};Q_n^{[t]},Y_n^{[t]}|S_n^{[0:t-1]},Q_n^{[1:t-1]})
    &=I(\Delta_{\theta}^{[t]};S_n^{[0:t-1]},Q_n^{[1:t]},Y_n^{[t]})\\
    &=I(\Delta_{\theta}^{[t]};S_n^{[0:t-1]},Q_n^{[1:t]},\hat{U}_n^{[t]},k^{[t]}).
\end{align}
For any set of sparse updates of submodel $\theta$ given by, $\tilde{q}\in\mathbb{F}_q^L$ and arbitrary realizations of storage, queries, updates and positions ($\bar{s}_n,\bar{r}_n,\bar{u}_n,\bar{k}$), consider the aposteriori probability,
\begin{align}
    P(\Delta_{\theta}^{[t]}=\tilde{q}&|S_n^{[0:t-1]}=\bar{s}_n,Q_n^{[1:t]}=\bar{r}_n,\hat{U}_n^{[t]}=\bar{u}_n,k^{[t]}=\bar{k})\nonumber\\
    &=\frac{P(\Delta_{\theta}^{[t]}=\tilde{q},S_n^{[0:t-1]}=\bar{s}_n,Q_n^{[1:t]}=\bar{r}_n,\hat{U}_n^{[t]}=\bar{u}_n,k^{[t]}=\bar{k})}{P(S_n^{[0:t-1]}=\bar{s}_n,Q_n^{[1:t]}=\bar{r}_n,\hat{U}_n^{[t]}=\bar{u}_n,k^{[t]}=\bar{k})}\\
    &=\frac{P(\Delta_{\theta}^{[t]}=\tilde{q},k^{[t]}=\bar{k})P(S_n^{[0:t-1]}=\bar{s}_n,Q_n^{[1:t]}=\bar{r}_n,\hat{U}_n^{[t]}=\bar{u}_n)}{P(S_n^{[0:t-1]}=\bar{s}_n,Q_n^{[1:t]}=\bar{r}_n,\hat{U}_n^{[t]}=\bar{u}_n)P(k^{[t]}=\bar{k})},
\end{align}
since $S_n$, $Q_n$ and $\hat{U}_n$ are random noise terms that are independent of the actual values of updates and the submodel index. Therefore,
\begin{align}\label{main_p2_prvcy}
    P(\Delta_{\theta}^{[t]}=\tilde{q}&|S_n^{[0:t-1]}=\bar{s}_n,Q_n^{[1:t]}=\bar{r}_n,\hat{U}_n^{[t]}=\bar{u}_n,k^{[t]}=\bar{k})
    =\frac{P(k^{[t]}=\bar{k}|\Delta_{\theta}^{[t]}=\tilde{q})P(\Delta_{\theta}^{[t]}=\tilde{q})}{P(k^{[t]}=\bar{k})},
\end{align}

Note that for all possible realizations of permutations $\tilde{p}$,
\begin{align}
    P(k^{[t]}=\bar{k}|\Delta_{\theta}^{[t]}=\tilde{q})&=\sum_{\tilde{p}} P(k^{[t]}=\bar{k}, \tilde{P}=\tilde{p}|\Delta_{\theta}^{[t]}=\tilde{q})\\
    &=\sum_{\tilde{p}} P(k^{[t]}=\bar{k}|\tilde{P}=\tilde{p},\Delta_{\theta}^{[t]}=\tilde{q})P(\tilde{P}=\tilde{p}|\Delta_{\theta}^{[t]}=\tilde{q})\\
    &=\sum_{\tilde{p}} 1_{\{k^{[t]}=\bar{k},\Delta_{\theta}^{[t]}=\tilde{q},\tilde{P}=\tilde{p}\}}P(\tilde{P}=\tilde{p})\label{indep}\\
    &=\frac{(Pr)!(P-Pr)!}{P!}\\
    &=\frac{1}{\binom{P}{Pr}},
\end{align}
where \eqref{indep} is due to the fact that the permutation is independently and randomly selected, irrespective of the values of updates. Therefore, from \eqref{main_p2_prvcy},
\begin{align}\
    P(\Delta_{\theta}^{[t]}=\tilde{q}|S_n^{[0:t-1]}=\bar{s}_n,Q_n^{[1:t]}=\bar{r}_n,\hat{U}_n^{[t]}=\bar{u}_n,k^{[t]}=\bar{k})
    &=\frac{\frac{1}{\binom{P}{Pr}}P(\Delta_{\theta}^{[t]}=\tilde{q})}{\frac{1}{\binom{P}{Pr}}}=P(\Delta_{\theta}^{[t]}=\tilde{q}),
\end{align}
which proves that $I(\Delta_{\theta}^{[t]};Q_n^{[t]},Y_n^{[t]}|S_n^{[0:t-1]},Q_n^{[1:t-1]})=I(\Delta_{\theta}^{[t]};S_n^{[0:t-1]},Q_n^{[1:t]},Y_n^{[t]})=0$ since all realizations of the random variables considered in the calculation are arbitrary. 

\emph{Security of the stored submodels:} The same arguments provided in Section~\ref{privacy_proof} for the security of submodels is valid in this section as well. 

\section{PRUW with Random Sparsification}\label{rndm}

In this section, we investigate how the communication cost of a PRUW process can be reduced by performing random sparsification, where pre-determined amounts of randomly chosen parameters and updates are not downloaded and uploaded in the reading and writing phases, respectively. This process introduces some amount of distortion in the two phases since a pre-determined amount of downloads and uploads are made zero (not communicated) irrespective of their real values. We study the behavior of the communication cost with the level of distortion (random sparsification rate) allowed. Our results characterize the rate-distortion trade-off in PRUW.  

\subsection{Problem Formulation}\label{rndmproblem}

We consider the basic PRUW setting described in Section~\ref{basicproblem} with $N$ non-colluding databases storing $M$ independent submodels $\{W_1,\ldots,W_M\}$ of size $L$, each containing random symbols from $\bF_q$. At each time instance $t$, a user updates an arbitrary submodel without revealing its index or the values of updates. Pre-determined amounts of distortion (random sparsification rates in the uplink and downlink) are allowed in the reading and writing phases ($\tilde{D}_r$ and $\tilde{D}_w$, respectively), in order to reduce the communication cost. 

\emph{Distortion in the reading phase:} A distortion of no more than $\tilde{D}_r$ is allowed in the reading phase, i.e., $D_r\leq \tilde{D}_r$, with
\begin{align}\label{rerror}
    D_{r}=\frac{1}{L}\sum_{i=1}^{L}1_{W_{\theta,i}\neq \hat{W}_{\theta,i}}
\end{align}
where $W_{\theta,i}$, $\hat{W}_{\theta,i}$ are the actual and downloaded versions of the $i$th bit of the required submodel $W_\theta$.

\emph{Distortion in the writing phase:} A distortion of no more than $\tilde{D}_w$ is allowed in the writing phase, i.e., $D_w\leq \tilde{D}_w$, with
\begin{align}\label{werror}
    D_{w}=\frac{1}{L}\sum_{i=1}^{L}1_{\Delta_{\theta,i}\neq \hat{\Delta}_{\theta,i}}
\end{align}
where $\Delta_{\theta,i}$ and $\hat{\Delta}_{\theta,i}$ are the actual and uploaded versions of the $i$th bit of the update to the required submodel.

The goal of this work is to find schemes that result in the lowest total communication cost under given distortion budgets in the reading and writing phases in the PRUW setting, i.e., a rate-distortion trade-off in PRUW. The privacy constraints on the updating submodel index and the values of updates as well as the security constraint on the submodels are the same as \eqref{prvcy_id}, \eqref{prvcy_u} and \eqref{security}, respectively. The correctness conditions are defined as follows.  

\emph{Correctness in the reading phase:} The user should be able to correctly decode the sparse set of parameters (denoted by $G$) of the required submodel $W_\theta$ from the answers received in the reading phase, i.e., 
\begin{align}
H(W_{\theta,G}^{[t-1]}|Q_{1:N}^{[t]},A_{1:N}^{[t]},\theta)=0, \quad t\in\mathbb{N},
\end{align}
where $W_{\theta,G}^{[t-1]}$ is the set of parameters in  set $G$ of submodel $W_\theta$ at time $t-1$, $Q_n^{[t]}$ is the query sent to database $n$ at time $t$ and $A_n^{[t]}$ is the corresponding answer.

\emph{Correctness in the writing phase:} Let $\theta$ be the updating submodel index and $G'$ be the sparse set of parameters with non-zero updates of $W_{\theta}$ in the writing phase. Then, the $i$th parameter of submodel $m$ at time $t$, $t\in\mathbb{N}$ given by $W_{m,i}^{[t]}$ is correctly updated as,
\begin{align}
    W_{m,i}^{[t]}=
    \begin{cases}
    W_{m,i}^{[t-1]}+\Delta_{m,i}^{[t]}, & \text{if $m=\theta$ and $i\in G'$}\\
    W_{m,i}^{[t-1]}, & \text{if $m\neq\theta$ or $i\notin G'$}
    \end{cases},
\end{align}
where $\Delta_{m,i}^{[t]}$ is the corresponding update of $W_{m,i}^{[t-1]}$.

In the reading phase, users privately send queries to download a randomly selected set of parameters of the required submodel, and in the writing phase, users privately send updates to be added to a randomly selected set of parameters of the existing submodels while ensuring the distortions resulted by sparse downloads and uploads in the two phases are within the allowed budgets ($\tilde{D}_r$, $\tilde{D}_w$). The reading, writing and total costs are defined the same as in Section~\ref{basicproblem}.

\subsection{Main Result}\label{main_p3}

\begin{theorem}
    For a PRUW setting with $N$ non-colluding databases containing $M$ independent submodels, where $\tilde{D}_r$ and $\tilde{D}_w$ amounts of distortion are allowed in the reading and writing phases, respectively, the following reading and writing costs are achievable,
\begin{align}\label{main_res_p3}
    C_R=\begin{cases}
    \frac{2}{1-\frac{2}{N}}(1-\tilde{D}_r), & \text{even $N$}\\
    \frac{2-\frac{2}{N}}{1-\frac{3}{N}}(1-\tilde{D}_r), & \text{odd $N$, \ $\tilde{D}_r<\tilde{D}_w$}\\
    \frac{2}{1-\frac{3}{N}}(1-\tilde{D}_r), & \text{odd $N$, \ $\tilde{D}_r\geq\tilde{D}_w$}
    \end{cases},\\
    C_W=\begin{cases}
    \frac{2}{1-\frac{2}{N}}(1-\tilde{D}_w), & \text{even $N$}\\
    \frac{2}{1-\frac{3}{N}}(1-\tilde{D}_w), & \text{odd $N$, \ $\tilde{D}_r<\tilde{D}_w$}\\
    \frac{2-\frac{2}{N}}{1-\frac{3}{N}}(1-\tilde{D}_w), & \text{odd $N$, \ $\tilde{D}_r\geq\tilde{D}_w$}
    \end{cases}.
\end{align}
\end{theorem}

\begin{remark}
The total communication cost decreases linearly with the increasing amounts of distortion allowed in the reading and writing phases.
\end{remark}

\subsection{Proposed Scheme}\label{proposedscheme_p3}

The proposed scheme is an extension of the scheme presented in Section~\ref{basicscheme}. The scheme in Section~\ref{basicscheme} considers $\lfloor\frac{N}{2}\rfloor-1$ bits of the required submodel at a time (called subpacketization) and reads from and writes to $\lfloor\frac{N}{2}\rfloor-1$ bits using a single bit in each of the reading and writing phases with no error. In this section, we consider larger subpackets with more bits, i.e., $\ell\geq\lfloor\frac{N}{2}\rfloor-1$, and correctly read from/write to only $\lfloor\frac{N}{2}\rfloor-1$ selected bits in each subpacket using single bits in the two phases. The rest of the  $\ell-\lfloor\frac{N}{2}\rfloor+1$ bits in each subpacket account for the distortion in each phase, which is maintained under the allowed distortion budgets. The privacy of the updating submodel index as well as the values of updates is preserved in this scheme, while also not revealing the indices of the distorted uploads/downloads. 

The proposed scheme consists of the following three tasks: 1) calculating the optimum reading and writing subpacketizations $\ell_r^*$ and $\ell_w^*$ based on the given distortion budgets $\tilde{D}_r$ and $\tilde{D}_w$, 2) specifying the scheme, i.e., storage, reading/writing queries and single bit updates, for given values of $\ell_r^*$ and $\ell_w^*$, and 3) in cases where the subpacketizations calculated in task 1 are non-integers, the model is divided into two sections and two different integer-valued subpacketizations are assigned to the two sections in such a way that the resulting distortion is within the given budgets. Then, task 2 is performed in each of the two sections.

For task 2, note that the scheme in Section~\ref{basicscheme} allocates distinct constants $f_i$, $i\in\{1,\dotsc,\ell\}$ to the $i$th bit of each subpacket in all submodels (see \eqref{storage3}) in the storage, which makes it possible to combine all parameters/updates in a given subpacket to a single bit in a way that the parameters/updates can be correctly and privately decomposed. However, in this scheme, since there may be two subpacketizations in the two phases (reading and writing), we need to ensure that each subpacket in both phases consists of bits with distinct associated $f_i$s. In order to do this, we associate distinct $f_i$s with each consecutive $\max\{\ell_r^*,\ell_w^*\}$ bits in a cyclic manner so that each subpacket in both phases have distinct $f_i$s. The proposed scheme is explained in detail next, along with an example.

The scheme is defined on a single subpacket in each of the two phases, and is applied repeatedly on all subpackets. Since the number of bits correctly downloaded/updated remains constant at $\lfloor\frac{N}{2}\rfloor-1$ for a given $N$, the distortion in a subpacket of size $\ell$ is $\frac{\ell-\lfloor\frac{N}{2}\rfloor+1}{\ell}$. Note that this agrees with the definitions in \eqref{rerror} and \eqref{werror} since the same distortion is resulted by all subpackets.\footnote{Here, we assume that the integer-valued subpacketization $\ell$ is uniform throughout the storage, i.e., task 3 is not applicable. The extension to non-uniform subpacketizations (two different subpacketizations as in task 3) is derived from the same concept and is described in detail in Section~\ref{opt}.} Therefore, the optimum subpacketizations in the two phases, $\ell_r^*$ and $\ell_w^*$, are functions of $\tilde{D}_r$, $\tilde{D}_w$ and $N$, and will be calculated in Section~\ref{opt}. First, we describe the general scheme for any given $\ell_r^*$ and $\ell_w^*$. The scheme is studied under two cases, 1) $y=\ell_w^*>\ell_r^*$, and 2) $y=\ell_r^*\geq\ell_w^*$.

\subsubsection{Case 1: $y=\ell_w^*>\ell_r^*$}

\textbf{Storage and initialization:} The storage of $y=\max\{\ell_r^*, \ell_w^*\}=\ell_w^*$ bits of all submodels in database $n$, $n\in\{1,\dotsc,N\}$ is given by,
\begin{align}S_n=
    \begin{bmatrix}
    \begin{bmatrix}
    \frac{1}{f_{1}-\alpha_n}W_{1,1}+\sum_{j=0}^{\lfloor\frac{N}{2}\rfloor-1}\alpha_n^j Z_{1,j}^{[1]}\\\vdots\\\frac{1}{f_{1}-\alpha_n}W_{M,1}+\sum_{j=0}^{\lfloor\frac{N}{2}\rfloor-1}\alpha_n^j Z_{M,j}^{[1]}
    \end{bmatrix}\\\vdots\\
    \begin{bmatrix}
    \frac{1}{f_y-\alpha_n}W_{1,y}+\sum_{j=0}^{\lfloor\frac{N}{2}\rfloor-1}\alpha_n^j Z_{1,j}^{[y]}\\\vdots\\\frac{1}{f_y-\alpha_n}W_{M,y}+\sum_{j=0}^{\lfloor\frac{N}{2}\rfloor-1}\alpha_n^j Z_{M,j}^{[y]}
    \end{bmatrix}
    \end{bmatrix},\label{storage}
\end{align}
where $W_{i,j}$ is the $j$th bit of submodel $i$, $Z_{i,j}$s are random noise vectors of size $M\times1$ and $\{f_i\}_{i=1}^y$, $\{\alpha_n\}_{n=1}^N$ are globally known distinct constants from $\mathbb{F}_q$, such that each $\alpha_n$ and $f_i-\alpha_n$ for all $i\in\{1,\dotsc,\ell\}$ and $n\in\{1,\dotsc,N\}$ are coprime with $q$.

\begin{figure*}[t]
    \centering
    \includegraphics[scale=0.85]{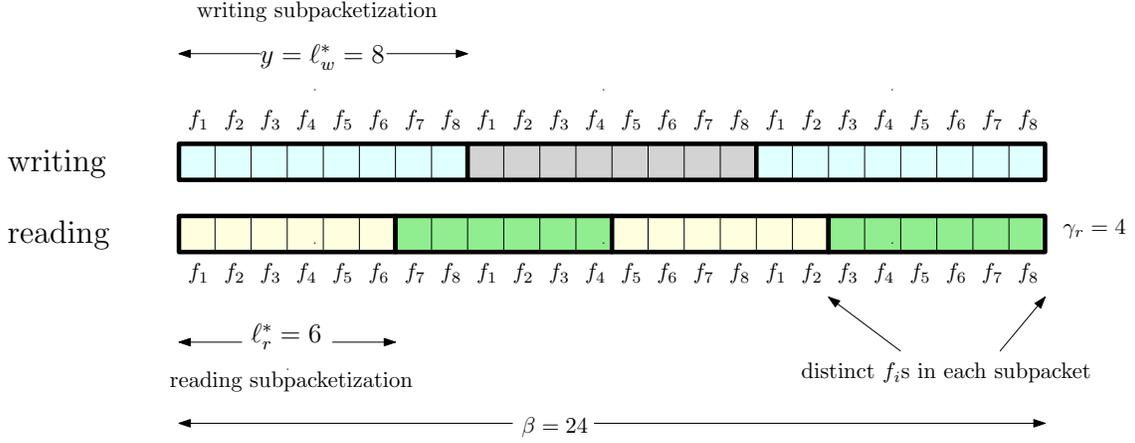}
    \caption{An example setting for case 1.}
    \label{fig1}
\end{figure*}

\textbf{Reading phase:} In this case, the user considers subpackets of size $\ell_r^*$ and only downloads $\lfloor\frac{N}{2}\rfloor-1$ bits of each subpacket. Note that each consecutive $y=\ell_w^*$ bits in storage are associated with distinct $f_i$s, which makes each consecutive set of $\ell_r^*$ (reading subpacket size) $f_i$s distinct as well (since $\ell^*_r\leq\ell^*_w$). However, not all reading subpackets have the same $f_i$ allocated to their $i$th bit due to the definition of the storage structure (cyclic allocation of $\ell_w^*$ distinct values of $f_i$). Therefore, we cannot define the reading query on a single subpacket and use it repeatedly, since the reading queries depend on $f_i$s. Thus, we define $\gamma_r=\frac{\text{lcm}\{\ell_r^*,\ell_w^*\}}{\ell_r^*}$ queries to read any $\gamma_r$ consecutive subpackets. Note that the \emph{super subpacket} which consists of any $\gamma_r$ consecutive reading subpackets have the same set of $f_i$s that occur in a cyclic manner in the storage. Therefore, the $\gamma_r$ queries can be defined once on a \emph{super subpacket}, and can be used repeatedly throughout the process. An example setting is given in Figure~\ref{fig1}, where the reading and writing subpacketizations are given by $\ell_r^*=6$, $\ell_w^*=8$ and the storage structure repeats at every $y=8$ bits. Each square in Figure~\ref{fig1} corresponds to a single bit of all submodels associated with the corresponding value of $f_i$. It shows three consecutive storage/writing subpackets on the top row. The same set of bits are viewed as $\gamma_r=\frac{\text{lcm}\{6,8\}}{6}=4$ reading subpackets, each of size $\ell_r^*=6$ in the bottom row. Note that each reading subpacket contains distinct $f_i$s, which are not the same across the four subpackets. However, it is clear that the structure of the \emph{super subpacket} which contains the four regular subpackets keeps repeating with the same set of $f_i$s in order. The reading phase has the following steps.

The user sends the following queries to database $n$ to obtain each of the arbitrary sets of $\lfloor\frac{N}{2}\rfloor-1$ bits of each subpacket in each set of $\gamma_r=\frac{\text{lcm}\{\ell_r^*,\ell_w^*\}}{\ell_r^*}$ consecutive, non-overlapping subpackets. Let $J_r^{[s]}$ be the set of $\lfloor\frac{N}{2}\rfloor-1$ parameter indices that are read correctly from subpacket $s$ for $s\in\{1,\dotsc,\gamma_r\}$. The query to download subpacket $s$ is,
\begin{align}\label{query_read2}
        Q_n(s)= \begin{bmatrix}
        e_M(\theta)1_{\{1\in J_r^{[s]}\}}+(f_{g((s-1)\ell_r^*+1)}-\alpha_n)\tilde{Z}_{s,1}\\
        \vdots\\
        e_M(\theta)1_{\{\ell_r^*\in J_r^{[s]}\}}+(f_{g(s\ell_r^*)}-\alpha_n)\tilde{Z}_{s,\ell_r^*}
    \end{bmatrix}, 
\end{align}
and the corresponding subpacket $s$ is,
\begin{align}
S_n(s)=
   \begin{bmatrix}
    \begin{bmatrix}
    \frac{1}{f_{g((s-1)\ell_r^*+1)}-\alpha_n}W_{1,1}^{[s]}+\sum_{j=0}^{\lfloor\frac{N}{2}\rfloor-1}\alpha_n^j Z_{1,j}^{[1]}(s)\\\vdots\\\frac{1}{f_{g((s-1)\ell_r^*+1)}-\alpha_n}W_{M,1}^{[s]}+\sum_{j=0}^{\lfloor\frac{N}{2}\rfloor-1}\alpha_n^j Z_{M,j}^{[1]}(s)
    \end{bmatrix}\\
    \vdots\\
    \begin{bmatrix}
    \frac{1}{f_{g(s\ell_r^*)}-\alpha_n}W_{1,\ell_r^*}^{[s]}+\sum_{j=0}^{\lfloor\frac{N}{2}\rfloor-1}\alpha_n^j Z_{1,j}^{[y]}(s)\\\vdots\\\frac{1}{f_{g(s\ell_r^*)}-\alpha_n}W_{M,\ell_r^*}^{[s]}+\sum_{j=0}^{\lfloor\frac{N}{2}\rfloor-1}\alpha_n^j Z_{M,j}^{[y]}(s)
    \end{bmatrix}
    \end{bmatrix},\label{extend}
\end{align}
where $e_M(\theta)$ is the all zeros vector of size $M\times1$ with a $1$ at the $\theta$th position, $\tilde{Z}_{i,j}$s are random noise vectors of size $M\times1$ and the function $g(\cdot)$ is defined as,
\begin{align}\label{gfunc}
    g(x)=\begin{cases}
    x\!\!\!\!\mod y, & \text{if $x\!\!\!\!\mod y\neq 0$}\\
    y, & \text{if $x\!\!\!\!\mod y=0$}
    \end{cases}
\end{align}

Note that the \emph{super subpacket} $S_n=[S_n^{[1]},\dotsc, S_n^{[\gamma_r]}]^T$ is the concatenation of $\frac{\text{lcm}\{\ell_r^*,\ell_w^*\}}{y}$ blocks of the form \eqref{storage}. The $\gamma_r$ answers received by database $n$, $n\in\{1,\dotsc,N\}$, are given by,
\begin{align}
    A_n(s)=&S_n(s)^TQ_n(s)\\
    =&\sum_{i=1}^{\ell_r^*} \left(\frac{1}{f_{g((s-1)\ell_r^*+i)}-\alpha_n}W_{\theta,i}^{[s]}\right)1_{\{i\in J_r^{[s]}\}}+P_{\alpha_n}(\lfloor\frac{N}{2}\rfloor)\label{ans1},
\end{align}
for each $s\in\{1,\dotsc,\gamma_r\}$, where $P_{\alpha_n}(\lfloor\frac{N}{2}\rfloor)$ is a polynomial in $\alpha_n$ of degree $\lfloor\frac{N}{2}\rfloor$. Since $|J_r^{[s]}|=\lfloor\frac{N}{2}\rfloor-1$ for each $s\in\{1,\dotsc,\gamma_r\}$, the required $\lfloor\frac{N}{2}\rfloor-1$ bits of each of the $\gamma_r$ subpackets can be correctly retrieved from $2\lfloor\frac{N}{2}\rfloor$ answers of the form \eqref{ans1} (corresponding to $2\lfloor\frac{N}{2}\rfloor$ databases). Note that when $N$ is odd, the user has to download answers from only $N-1$ databases, since $N-1$ equations of the form \eqref{ans1} with distinct $\alpha_n$s suffice to solve for the $\lfloor\frac{N}{2}\rfloor-1$ parameters of the required submodel when $N$ is odd. The resulting reading cost of the first case is given by,
\begin{align}
    C_R^{[1]}=\begin{cases}
    \frac{\gamma_r\times N}{\gamma_r\times \ell_r^*}=\frac{N}{\ell^*_r}, & \text{even $N$},\\
    \frac{\gamma_r\times (N-1)}{\gamma_r\times \ell_r^*}=\frac{N-1}{\ell^*_r}, & \text{odd $N$}.
    \end{cases}
\end{align}

For a better understanding of the reading phase, we present the queries and answers corresponding to the example in Figure~\ref{fig1} next. Assume that $N=6$ for this example and the set of $\lfloor\frac{N}{2}\rfloor-1=2$ parameter indices that are read correctly from the second subpacket (out of $\gamma_r=4$ subpackets) is given by $J_r^{[2]}=\{2,5\}$. Then, the query corresponding to the second subpacket is given by, 
\begin{align}\label{query_read_eg}
        Q_n(2)= \begin{bmatrix}
        &(f_7-\alpha_n)\tilde{Z}_{2,1}\\
        e_M(\theta)+&(f_8-\alpha_n)\tilde{Z}_{2,2}\\
        &(f_1-\alpha_n)\tilde{Z}_{2,3}\\
        &(f_2-\alpha_n)\tilde{Z}_{2,4}\\
        e_M(\theta)+&(f_3-\alpha_n)\tilde{Z}_{2,5}\\
        &(f_4-\alpha_n)\tilde{Z}_{2,6}
    \end{bmatrix}, 
\end{align}
which is used to obtain the $2$nd and $5$th elements of the second reading subpacket given by,
\begin{align}
S_n(2)=
   \begin{bmatrix}
    \frac{1}{f_7-\alpha_n}W^{[2]}_{\cdot,1}+\sum_{j=0}^2\alpha_n^j Z_{\cdot,j}^{[1]}(2)\\
    \frac{1}{f_8-\alpha_n}W^{[2]}_{\cdot,2}+\sum_{j=0}^2\alpha_n^j Z_{\cdot,j}^{[2]}(2)\\
    \frac{1}{f_1-\alpha_n}W^{[2]}_{\cdot,3}+\sum_{j=0}^2\alpha_n^j Z_{\cdot,j}^{[3]}(2)\\
    \frac{1}{f_2-\alpha_n}W^{[2]}_{\cdot,4}+\sum_{j=0}^2\alpha_n^j Z_{\cdot,j}^{[4]}(2)\\
    \frac{1}{f_3-\alpha_n}W^{[2]}_{\cdot,5}+\sum_{j=0}^2\alpha_n^j Z_{\cdot,j}^{[5]}(2)\\
    \frac{1}{f_4-\alpha_n}W^{[2]}_{\cdot,6}+\sum_{j=0}^2\alpha_n^j Z_{\cdot,j}^{[6]}(2)
    \end{bmatrix},
\end{align}
where $W^{[2]}_{\cdot,i}=[W^{[2]}_{1,i},\dotsc,W^{[2]}_{M,i}]^T$ and $Z_{\cdot,j}^{[i]}(2)=[Z_{1,j}^{[i]}(2),\dotsc,Z_{M,j}^{[i]}(2)]^T$. Then, the answer from database $n$, $n\in\{1,\dotsc,6\}$ for this specific subpacket ($s=2$) is given by, 
\begin{align}
    A_n(2)=&S_n(2)^TQ_n(2)=\frac{1}{f_8-\alpha_n}W^{[2]}_{\theta,2}+\frac{1}{f_3-\alpha_n}W^{[2]}_{\theta,5}+P_{\alpha_n}(3), 
\end{align}
where $P_{\alpha_n}(3)$ is a polynomial in $\alpha_n$ of degree $3$. The user can then find $W^{[2]}_{\theta,2}$ and $W^{[2]}_{\theta,5}$ by solving,
\begin{align}
    \begin{bmatrix}
        A_1(2)\\\vdots\\A_6(2)
    \end{bmatrix}=\begin{bmatrix}
        \frac{1}{f_8-\alpha_1} & \frac{1}{f_3-\alpha_1} & 1 & \alpha_1 & \alpha_1^2 & \alpha_1^3\\
        \vdots & \vdots & \vdots & \vdots & \vdots & \vdots\\
        \frac{1}{f_8-\alpha_6} & \frac{1}{f_3-\alpha_6} & 1 & \alpha_6 & \alpha_6^2 & \alpha_6^3\\
    \end{bmatrix}
    \begin{bmatrix}
        W^{[2]}_{\theta,2}\\W^{[2]}_{\theta,5}\\R_0\\R_1\\R_2\\R_3
    \end{bmatrix}.
\end{align}

\textbf{Writing phase:} Since the subpacketization in the writing phase is $y$, which is the same as the period of the cyclic structure of the storage in \eqref{storage}, a single writing query, specifying the submodel index and the correctly updated bit indices, defined on a single subpacket suffices to repeatedly update all subpackets, as the $f_i$s in all subpackets are identical. The writing query sent to database $n$, $n\in\{1,\dotsc,N\}$, is,
\begin{align}
    \tilde{Q}_n=
    \begin{bmatrix}
        \frac{1}{f_{1}-\alpha_n}e_M(\theta)1_{\{1\in J_w\}}+\hat{Z}_{1}\\
        \vdots\\
        \frac{1}{f_{y}-\alpha_n}e_M(\theta)1_{\{y\in J_w\}}+\hat{Z}_{y}
    \end{bmatrix},
\end{align}
where $J_w$ is the set of indices of the $\lfloor\frac{N}{2}\rfloor-1$ parameters of each subpacket that are updated correctly and $\hat{Z}$s are random noise vectors of size $M\times1$. Since $\tilde{Q}_n$ is sent only once, the same set of $J_w$ indices will be correctly updated in all subpackets. The user then sends a single bit combined update for each subpacket of the form \eqref{storage} given by,
\begin{align}
    U_n=\sum_{i\in J_w} \tilde{\Delta}_{\theta,i}\prod_{j\in J_w, j\neq i}(f_j-\alpha_n)+\prod_{j\in J_w}(f_j-\alpha_n)Z, \quad n\in\{1,\dotsc,N\} 
\end{align}
where $\tilde{\Delta}_{\theta,i}=\frac{\Delta_{\theta,i}}{\prod_{j\in J_w, j\neq i}(f_j-f_i)}$ and $Z$ is a random noise bit. Each database then calculates the incremental update as,
\begin{align}
    \tilde{U}_n&=U_n\times \tilde{Q}_n\label{step1}\\
    &=\begin{bmatrix}
        \frac{\Delta_{\theta,1}}{f_{1}-\alpha_n}e_M(\theta)1_{\{1\in J_w\}}+P_{\alpha_n}(\lfloor\frac{N}{2}\rfloor-1)\\
        \vdots\\
        \frac{\Delta_{\theta,y}}{f_{y}-\alpha_n}e_M(\theta)1_{\{y\in J_w\}}+P_{\alpha_n}(\lfloor\frac{N}{2}\rfloor-1)
    \end{bmatrix},\label{step2}
\end{align}
where $P_{\alpha_n}(\cdot)$ is a polynomial in $\alpha_n$ of degree in parenthesis,\footnote{Note that all $P_{\alpha_n}(\cdot)$ are not the same and each polynomial is resulted by the combination of all unwanted terms (noise subspace) resulting from the decomposition of combined updates.} and \eqref{step2} is obtained from \eqref{step1} by applying Lemma~\ref{lem1}. Since the incremental update in \eqref{step2} is in the same form as the storage in \eqref{storage}, \eqref{step2} is directly added to the existing storage to obtain the updated submodel as,
\begin{align}
    S_n^{[t]}=S_n^{[t-1]}+\bar{U}_n^{[t]},
\end{align}
for each $n\in\{1,\dotsc,N\}$ for both even and odd $N$. 

The writing cost of case~1 is given by,
\begin{align}
    C_W^{[1]}=\frac{N}{\ell_w^*}\label{wr1_p3}.
\end{align}

\subsubsection{Case~2: $y=\ell_r^*\geq \ell_w^*$}

\textbf{Storage and initialization:} The storage of $y=\max\{\ell_r^*, \ell_w^*\}=\ell_r^*$ bits of all submodels in database $n$, $n\in\{1,\dotsc,N\}$ is given by,
\begin{align}S_n=
    \begin{bmatrix}
    \begin{bmatrix}
    \frac{1}{f_{1}-\alpha_n}W_{1,1}+\sum_{j=0}^{\lceil\frac{N}{2}\rceil-1}\alpha_n^j Z_{1,j}^{[1]}\\\vdots\\\frac{1}{f_{1}-\alpha_n}W_{M,1}+\sum_{j=0}^{\lceil\frac{N}{2}\rceil-1}\alpha_n^j Z_{M,j}^{[1]}
    \end{bmatrix}\\
    \vdots\\
    \begin{bmatrix}
    \frac{1}{f_{y}-\alpha_n}W_{1,y}+\sum_{j=0}^{\lceil\frac{N}{2}\rceil-1}\alpha_n^j Z_{1,j}^{[y]}\\\vdots\\\frac{1}{f_{y}-\alpha_n}W_{M,y}+\sum_{j=0}^{\lceil\frac{N}{2}\rceil-1}\alpha_n^j Z_{M,j}^{[y]}
    \end{bmatrix}
    \end{bmatrix},\label{storage2}
\end{align}
where $W_{i,j}$ is the $j$th bit of submodel $i$ and the $Z$s are random noise vectors of size $M\times1$.

\textbf{Reading phase:} In the reading phase, each user correctly downloads $\lfloor\frac{N}{2}\rfloor-1$ bits from each subpacket while not downloading the rest of the $\ell_r^*-\lfloor\frac{N}{2}\rfloor+1$ bits. The user randomly picks the $\lfloor\frac{N}{2}\rfloor-1$ bits within the subpacket that are downloaded correctly and prepares the query to be sent to database $n$ as follows. Let $J_r$ be the set of indices of the $\lfloor\frac{N}{2}\rfloor-1$ bits that need to be downloaded correctly. Then,
\begin{align}\label{query_p3}
    Q_n=\begin{bmatrix}
        e_M(\theta)1_{\{1\in J_r\}}+(f_1-\alpha_n)\Tilde{Z}_{1}\\
        \vdots\\
        e_M(\theta)1_{\{y\in J_r\}}+(f_y-\alpha_n)\Tilde{Z}_{y}
    \end{bmatrix}, 
\end{align}
where $\tilde{Z}$ are random noise vectors of size $M\times1$. The answer of database $n$ is,
\begin{align}
    A_n&=S_n^TQ_n=\sum_{i=1}^y \left(\frac{1}{f_i-\alpha_n}W_{\theta,i}\right)1_{\{i\in J_r\}}+P_{\alpha_n}(\lceil\frac{N}{2}\rceil).
\end{align}
Since $|J_r|=\lfloor\frac{N}{2}\rfloor-1$, the user required $\lfloor\frac{N}{2}\rfloor-1$ bits of $W_{\theta}$ can be correctly downloaded using the answers received by the $N$ databases. The resulting reading cost of case~2 is given by,
\begin{align}
    C_R^{[2]}=\frac{N}{\ell_r^*}.
\end{align}

\textbf{Writing phase:} In the writing phase, the user considers subpackets of size $\ell_w^*$ and only updates $\lfloor\frac{N}{2}\rfloor-1$ out of the $\ell_w^*$ bits correctly, while making the updates of the rest of the $\ell_w^*-\lfloor\frac{N}{2}\rfloor+1$ bits zero. The $\lfloor\frac{N}{2}\rfloor-1$ bits that are correctly updated are chosen randomly. The following steps describe the writing process when $\ell_w^*\leq \ell_r^*=y$.

\begin{enumerate}
    \item A general writing query that specifies the submodel to which the update should be added, along with the positions of the $\lfloor\frac{N}{2}\rfloor-1$ non-zero updates in each subpacket is sent first. The same query from the reading phase \eqref{query_p3} can be used if the subpacketization and the indices of the correct $\lfloor\frac{N}{2}\rfloor-1$ bits within the subpacket are the same in both phases. However, for the strict case $\ell_r^*>\ell_w^*$, we need a new general query $\tilde{Q}_n$ for the writing phase. $\tilde{Q}_n$ consists of $\gamma_w=\frac{\text{lcm}\{\ell_r^*,\ell_w^*\}}{\ell_w^*}$ sub-queries, where each sub-query corresponds to a single subpacket of size $\ell_w^*$. These sub-queries are required since the storage structure of these $\gamma_w$ subpackets is not identical, which calls for $\gamma_w$ different queries, customized for each subpacket. An example setting for case~2 is given in Figure~\ref{fig2}, where $y=\ell_r^*=6$ in the storage given in \eqref{storage2}. However, $\ell_w^*=4$, which results in distinct sets of associated $f_i$s in every $\gamma_w=\frac{\text{lcm}\{\ell_r^*,\ell_w^*\}}{\ell_w^*}=3$ consecutive writing subpackets of size $\ell_w^*$. However, the \emph{super subpacket} containing these $\gamma_w=\frac{\text{lcm}\{\ell_r^*,\ell_w^*\}}{\ell_w^*}=3$ regular subpackets keep repeating with the same set of associated $f_i$s. Therefore, we can only send the $\gamma_w=\frac{\text{lcm}\{\ell_r^*,\ell_w^*\}}{\ell_w^*}=3$ sub-queries of $\tilde{Q}_n$ once to each database, which will be repeatedly used throughout the writing process. The general writing scheme that writes to each of the $\gamma_w$ consecutive subpackets is described in the next steps.
    \item Let $J_w^{[s]}$ be the set of indices of the correctly updated $\lfloor\frac{N}{2}\rfloor-1$ parameters in subpacket $s$ for $s\in\{1,\dotsc,\gamma_w\}$. Then, the sub-query $s$, $s\in\{1,\dotsc,\gamma_w\}$ of the writing query for database $n$ is given by,
    \begin{align}\label{query_write}
        \tilde{Q}_n(s)=\begin{bmatrix}
        \frac{1}{f_{g((s-1)\ell_w^*+1)}-\alpha_n}e_M(\theta)1_{\{1\in J_w^{[s]}\}}+\hat{Z}_{s,1}\\
        \vdots\\
        \frac{1}{f_{g(s\ell_w^*)}-\alpha_n}e_M(\theta)1_{\{\ell_w^*\in J_w^{[s]}\}}+\hat{Z}_{s,\ell_w^*}
    \end{bmatrix},  
    \end{align}
where $\hat{Z}$ are random noise vectors of size $M\times1$ and the function $g(\cdot)$ is defined as \eqref{gfunc}. For the example considered in Figure~\ref{fig2}, the sub-query corresponding to subpacket~2 if $J_w^{[2]}=\{1,3\}$ is given by,
\begin{align}
        \tilde{Q}_n(2)=\begin{bmatrix}
        \frac{1}{f_5-\alpha_n}e_M(\theta)+&\hat{Z}_{2,1}\\
        &\hat{Z}_{2,2}\\
        \frac{1}{f_1-\alpha_n}e_M(\theta)+&\hat{Z}_{2,3}\\
        &\hat{Z}_{2,4}\\
    \end{bmatrix}, 
\end{align}

Note that the values of $f_i$ in each individual section of $\tilde{Q}_n$ are distinct due to $\ell_w^*\leq y$ (in the example, the first section has $f_i=\{1,2,3,4\}$ and the second has $f_i=\{5,6,1,2\}$ and so on). This makes it possible for the user to send a single combined update bit (combining the updates of the $\lfloor\frac{N}{2}\rfloor-1$ non-zero updates in each subpacket) to each individual subpacket as described in Section~\ref{basicscheme}. The query $\tilde{Q}_n$ (consisting of $\gamma_w$ sub-queries) will only be sent once to each database. Therefore, the indices of the non-zero updates $J_w^{[s]}$, $s\in\{1,\dotsc,\gamma_w\}$ will be fixed at each consecutive non-overlapping group of $\gamma_w$ subpackets. 
\begin{figure*}[t]
    \centering
    \includegraphics[scale=1]{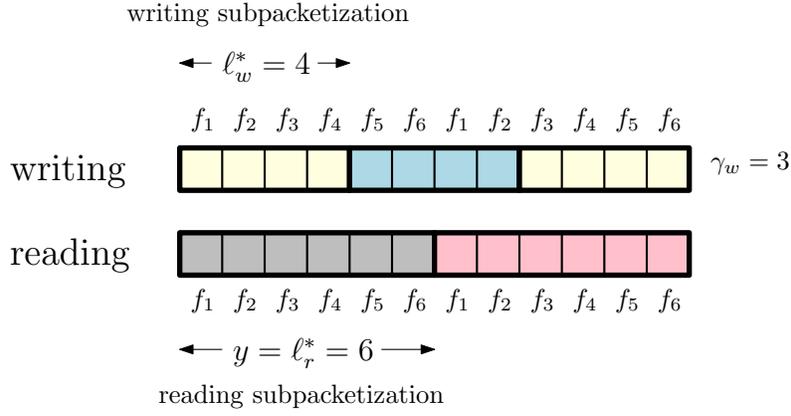}
    \caption{An example setting for case 2.}
    \label{fig2}
\end{figure*}

\item Next, the user sends a single combined update bit corresponding to each subpacket. The $\gamma_w$ combined updates sent to database $n$, $n\in\{1,\dotsc,N\}$ corresponding to a given set of $\gamma_w$ consecutive subpackets is given by,
\begin{align}\label{update}
    U_n(s)&=\sum_{i\in J_w^{[s]}} \tilde{\Delta}_{\theta,i}^{[s]}\prod_{j\in J_w^{[s]}, j\neq i}(f_{g((s-1)\ell_w^*+j)}-\alpha_n)+\prod_{j\in J_w^{[s]}}(f_{g((s-1)\ell_w^*+j)}-\alpha_n)Z_s, 
\end{align}
for each subpacket $s\in\{1,\dotsc,\gamma_w\}$, where $\tilde{\Delta}_{\theta,i}^{[s]}=\frac{\Delta_{\theta,i}^{[s]}}{\prod_{j\in J_w^{[s]}, j\neq i}(f_{g((s-1)\ell_w^*+j)}-f_{g((s-1)\ell_w^*+i)})}$ and $Z_s$ are random noise bits. Note that each $U_n(s)$ is a polynomial in $\alpha_n$ of degree $\lfloor\frac{N}{2}\rfloor-1$. For the example in Figure~\ref{fig2}, the combined update corresponding to subpacket~2 with $J_w^{[2]}=\{1,3\}$ is given by,
\begin{align}
    U_n(2)&=\tilde{\Delta}_{\theta,1}^{[2]}(f_1-\alpha_n)+\tilde{\Delta}_{\theta,3}^{[2]}(f_5-\alpha_n)+(f_1-\alpha_n)(f_5-\alpha_n)Z, 
\end{align}
where $\tilde{\Delta}_{\theta,1}^{[2]}=\frac{\Delta_{\theta,1}^{[2]}}{f_1-f_5}$ and $\tilde{\Delta}_{\theta,3}^{[2]}=\frac{\Delta_{\theta,3}^{[2]}}{f_5-f_1}$. 

\item Each database then calculates the incremental update of each subpacket as follows. The incremental update of subpacket $s$, $s\in\{1,\dotsc,\gamma_w\}$ is given by,
\begin{align}
    \tilde{U}_n(s)&=\begin{cases}
    U_n(s)\times \tilde{Q}_n(s), & \text{even $N$}\\
    \begin{bmatrix}
        \frac{\alpha_r-\alpha_n}{\alpha_r-f_{g((s-1)\ell_w^*+1)}} & 0 & \dotsc & 0\\
        0 & \frac{\alpha_r-\alpha_n}{\alpha_r-f_{g((s-1)\ell_w^*+2)}} & \dotsc & 0\\
        \vdots & \vdots & \vdots & \vdots\\
        0 & 0 & \dotsc & \frac{\alpha_r-\alpha_n}{\alpha_r-f_{g(s\ell_w^*)}}
    \end{bmatrix}\times U_n(s) \times \tilde{Q}_n(s), & \text{odd $N$}
    \end{cases}\label{step21}\\
    &=\begin{cases}
    \begin{bmatrix}
        \frac{\Delta_{\theta,1}^{[s]}}{f_{g((s-1)\ell_w^*+1)}-\alpha_n}e_M(\theta)1_{\{1\in J_w^{[s]}\}}+P_{\alpha_n}(\lfloor\frac{N}{2}\rfloor-1)\\
        \vdots\\
        \frac{\Delta_{\theta,\ell_w^*}^{[s]}}{f_{g(s\ell_w^*)}-\alpha_n}e_M(\theta)1_{\{y\in J_w^{[s]}\}}+P_{\alpha_n}(\lfloor\frac{N}{2}\rfloor-1)
    \end{bmatrix}, & \text{even $N$}\\
       \begin{bmatrix}
        \frac{\Delta_{\theta,1}^{[s]}}{f_{g((s-1)\ell_w^*+1)}-\alpha_n}e_M(\theta)1_{\{1\in J_w^{[s]}\}}+P_{\alpha_n}(\lfloor\frac{N}{2}\rfloor)\\
        \vdots\\
        \frac{\Delta_{\theta,\ell_w^*}^{[s]}}{f_{g(s\ell_w^*)}-\alpha_n}e_M(\theta)1_{\{y\in J_w^{[s]}\}}+P_{\alpha_n}(\lfloor\frac{N}{2}\rfloor)
    \end{bmatrix}, & \text{odd $N$}
    \end{cases}\label{step22}
\end{align}
where $r$ is a randomly chosen database out of the $N$ databases for odd $N$. Note that when $N$ is odd, the user can reduce the writing cost by not sending the combined updates to database $r$, since $\tilde{U}_r(s)=0$ for all $s$. The convention for the updates of each $i\notin J_w^{[s]}$ is $\Delta_{\theta,i}^{[s]}=0$. Lemmas~\ref{lem1} and \ref{lem2} are used to obtain \eqref{step22} from \eqref{step21}. Note that the concatenation of all $\gamma_w$ incremental updates of the form \eqref{step22} is in the same format as the concatenation of $\eta=\frac{\text{lcm}\{\ell_r^*,\ell_w^*\}}{y}$ reading subpackets (storage in \eqref{storage2}) since $g(\gamma_w\ell_w^*)=g(\text{lcm}\{\ell_r^*,\ell_w^*\})=y$, and therefore, can be added to the corresponding subpackets to obtain their updated versions, i.e.,
\begin{align}
    [S_n^{[t]}(1),\dotsc,S_n^{[t]}(\eta)]^T=[S_n^{[t-1]}(1),\dotsc,S_n^{[t-1]}(\eta)]^T+[\tilde{U}_n(1),\dotsc,\tilde{U}_n(\gamma_w)]^T,
\end{align}
where $[S_n^{[t]}(1),\dotsc,S_n^{[t]}(\eta)]^T$ contains $\eta$ consecutive $S_n$s of the form given in \eqref{storage2}.
\end{enumerate}

The writing cost of case 2 is given by,
\begin{align}
    C_W^{[2]}=\begin{cases}
    \frac{\gamma_w\times N}{\gamma_w\times\ell_w^*}=\frac{N}{\ell_w^*}, & \text{even $N$},\\
    \frac{\gamma_w\times (N-1)}{\gamma_w\times\ell_w^*}=\frac{N-1}{\ell_w^*}, & \text{odd $N$}.
    \end{cases}\label{wr1_p3_2}
\end{align}

\begin{remark}
For even $N$, both cases achieve reading and writing costs given by $\frac{N}{\ell_r^*}$ and $\frac{N}{\ell_w^*}$, respectively. However, when $N$ is odd, it is possible to achieve either a lower reading cost ($\frac{N-1}{\ell^*_r}$) with fewer noise terms in storage ($\lfloor\frac{N}{2}\rfloor-1$), or a lower writing cost ($\frac{N-1}{\ell^*_w}$) with an extra noise term in storage ($\lceil\frac{N}{2}\rceil-1$), with case~1 and case~2, respectively. In particular, when $N$ is odd, the total costs for the two options are given by $\frac{N-1}{\ell_r^*}+\frac{N}{\ell_w^*}=\frac{N(\ell_r^*+\ell_w^*)}{\ell_r^*\ell_w^*}-\frac{1}{\ell_r^*}$ and $\frac{N}{\ell_r^*}+\frac{N-1}{\ell_w^*}=\frac{N(\ell_r^*+\ell_w^*)}{\ell_r^*\ell_w^*}-\frac{1}{\ell_w^*}$, respectively. This justifies the extra noise term in storage for case~2 when $N$ is odd.
\end{remark}

\begin{remark}
Note that the cost of sending $Q_n$ and $\tilde{Q}_n$ is not considered in the above writing cost since they are sent only once to each database in the entire PRUW process (not per subpacket) and the maximum combined cost of $Q_n$ and $\tilde{Q}_n$ given by $\frac{M}{L}(\text{lcm}\{\ell_r^*,\ell_w^*\}+\max\{\ell_r^*,\ell_w^*\})$ is negligible since $L$ is very large. 
\end{remark}

\subsubsection{Calculation of Optimum $\ell_r^*$ and $\ell_w^*$ for Given ($\tilde{D}_r$, $\tilde{D}_w$)} \label{opt}

In order to minimize the total communication cost, the user correctly reads from and writes to only $\lfloor\frac{N}{2}\rfloor-1$ out of each of the $\ell_r^*$ and $\ell_w^*$ bits in reading and writing phases, respectively. This results in an error that needs to be kept within the given distortion budgets of $\tilde{D}_r$ and $\tilde{D}_w$. Note that $\min C_r+\min C_w\leq\min C_r+C_w$. In this section, we find the subpacketizations in the reading and writing phases ($\ell_r^*$, $\ell_r^*$) that achieve $\min C_r+\min C_w$ while being compatible with the proposed scheme. Note that each reading/writing cost in both cases is of the form $\frac{N}{\ell}$ or $\frac{N-1}{\ell}$, where $\ell$ is the respective subpacketization. Since only $\lfloor\frac{N}{2}\rfloor-1$ bits in a subpacket are read/written correctly, the subpacketization in general can be written as,
\begin{align}\label{subpckt_p3}
    \ell=\lfloor\frac{N}{2}\rfloor-1+i
\end{align}
for some $i\in\mathbb{Z}^+_0$. Therefore, the reading/writing costs of both cases are of the form $\frac{N}{\lfloor\frac{N}{2}\rfloor-1+i}$ or $\frac{N-1}{\lfloor\frac{N}{2}\rfloor-1+i}$ for some $i\in\mathbb{Z}^+_0$, both decreasing in $i$. For a subpacketization of the form $\ell=\lfloor\frac{N}{2}\rfloor-1+i$ (irrespective of reading or writing), the resulting distortion is given by,
\begin{align}
    D=\frac{i}{\lfloor\frac{N}{2}\rfloor-1+i},
\end{align}
if the same subpacketization is considered throughout the storage. Since the resulting distortion must satisfy $D\leq\tilde{D}$,\footnote{Here, $\tilde{D}$ refers to $\tilde{D}_r$ or $\tilde{D}_w$, based on the phase the subpacketization is defined for.} an upper bound on $i$ is derived as,
\begin{align}
    i\leq\frac{\tilde{D}}{1-\tilde{D}}\left(\lfloor\frac{N}{2}\rfloor-1\right).
\end{align}
Therefore, for given distortion budgets in the reading and writing phases ($\tilde{D}_r$,$\tilde{D}_w$), the optimum values of $i$ are given by,
\begin{align}
    i^*_r&=\frac{\tilde{D}_r}{1-\tilde{D}_r}\left(\lfloor\frac{N}{2}\rfloor-1\right)\label{ir*}\\ i^*_w&=\frac{\tilde{D}_w}{1-\tilde{D}_w}\left(\lfloor\frac{N}{2}\rfloor-1\right)\label{iw*},
\end{align}
which determine the optimum subpacketizations from \eqref{subpckt_p3}. For cases where $i^*_r\notin\mathbb{Z}^+_0$ or $i^*_w\notin\mathbb{Z}^+_0$, we divide all submodels into two sections, assign two separate integer-subpacketizations that guarantee the distortion budget, and apply the scheme on the two sections independently, which achieves the minimum costs in \eqref{main_res_p3}, after using an optimum ratio for the subsection lengths. To find the optimum ratio, we solve the following optimization problem. Let $\lambda_i$ be the fraction of each submodel with subpacketization $\ell_i=\lfloor\frac{N}{2}\rfloor-1+i$ for some $i=\eta_1,\eta_2\in \mathbb{Z}^+_0$. In this calculation, we drop the $r$ and $w$ subscripts which indicate the phase (reading/writing), since the calculation is the same for both phases.\footnote{Note that we focus on minimizing each individual cost (reading/writing cost) at a time since $\min C_r+\min C_w\leq\min C_r+C_w$.} The given $\tilde{D}_r$ and $\tilde{D}_w$ must be substituted for $\tilde{D}$ in the following calculation to obtain the specific results for the reading and writing phases, respectively. The optimum subpacketizations are obtained by solving,\footnote{Even though there are two types of reading and writing costs costs ($\frac{N}{\lfloor\frac{N}{2}\rfloor-1+i}$ and $\frac{N-1}{\lfloor\frac{N}{2}\rfloor-1+i}$), the optimization problem remains the same since the two costs are scaled versions of one another.}
\begin{align}
    \min & \quad \sum_{i=\eta_1,\eta_2}\lambda_i\frac{N}{\lfloor\frac{N}{2}\rfloor-1+i}\nonumber\\
    \text{s.t.} &\quad \sum_{i=\eta_1,\eta_2}\lambda_i\frac{i}{\lfloor\frac{N}{2}\rfloor-1+i}\leq \tilde{D} \nonumber\\
    &\quad \lambda_{\eta_1}+\lambda_{\eta_2}=1\nonumber\\
    &\quad \lambda_{\eta_1},\lambda_{\eta_2} \geq 0.
\end{align}

This problem has multiple solutions that give the same minimum total communication costs. As one of the solutions, consider $\eta_1=0$ and $\eta_2=\eta$, where $\eta=\lceil\frac{\tilde{D}}{1-\tilde{D}}(\lfloor\frac{N}{2}\rfloor-1)\rceil$, 
\begin{align}
    \lambda_0&=1-\frac{\tilde{D}}{\eta}\left(\lfloor\frac{N}{2}\rfloor-1+\eta\right),\label{p0}\\
    \lambda_\eta&=\frac{\tilde{D}}{\eta}\left(\lfloor\frac{N}{2}\rfloor-1+\eta\right)\label{p1}. 
\end{align}
This gives a minimum cost of $C_{\text{min}}=\frac{N}{\lfloor\frac{N}{2}\rfloor-1}(1-\tilde{D})$ which match the terms in \eqref{main_res_p3}, with $\tilde{D}=\tilde{D}_r$ and $\tilde{D}=\tilde{D}_w$. The optimality of the solution to the optimization problem is obvious since the resulting total cost is the same as what is achieved by the optimum subpacketizations characterized by \eqref{ir*} and \eqref{iw*}, with no segmentation of submodels.

\begin{figure*}[t]
    \centering
    \includegraphics[scale=1]{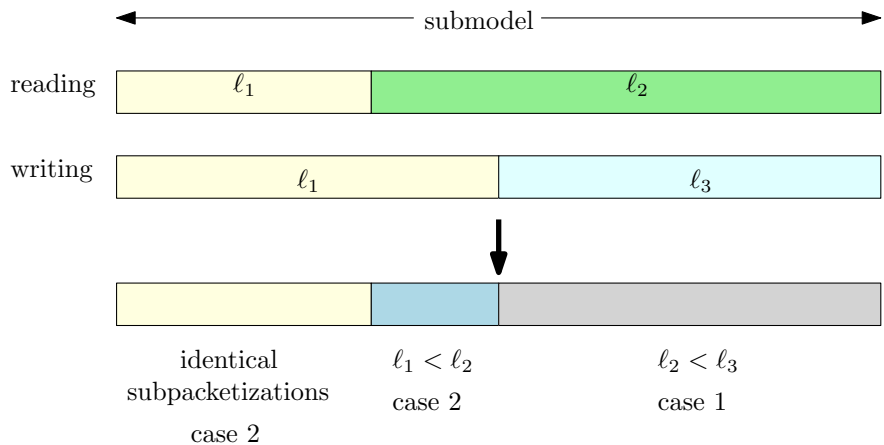}
    \caption{Storage of submodels}
    \label{fig3}
\end{figure*}

Next, we present the explicit expressions of optimum subpacketizations, with the optimum values of $i$ obtained above. For a setting with given $N$, $\tilde{D}_r$ and $\tilde{D}_w$, the reading and writing costs given in \eqref{main_res_p3} are achievable with corresponding subpacketizations given by,
\begin{align}
    \ell_r^*&=\begin{cases}
    \lfloor\frac{N}{2}\rfloor-1, &  \text{for $\lambda_0^{[r]}$ of submodel},\\
    \lfloor\frac{N}{2}\rfloor-1+\lceil\frac{\tilde{D}_r\left(\lfloor\frac{N}{2}\rfloor-1\right)}{1-\tilde{D}_r}\rceil, & \text{for $1-\lambda_0^{[r]}$ of submodel},
    \end{cases}
\end{align}
and
\begin{align}
    \ell_w^*&=\begin{cases}
    \lfloor\frac{N}{2}\rfloor-1, &  \text{for $\lambda_0^{[w]}$ of submodel},\\
    \lfloor\frac{N}{2}\rfloor-1+\lceil\frac{\tilde{D}_w\left(\lfloor\frac{N}{2}\rfloor-1\right)}{1-\tilde{D}_w}\rceil, & \text{for $1-\lambda_0^{[w]}$ of submodel},
    \end{cases}
\end{align}
where $\lambda_0^{[r]}$ and $\lambda_0^{[w]}$ are $\lambda_0$ in \eqref{p0} with $\tilde{D}$ replaced by $\tilde{D}_r$ and $\tilde{D}_w$, respectively. Once the subpacketizations of both reading and writing phases are determined based on the given distortion budgets, each section of all submodels is assigned a \emph{case}, based on the corresponding values of $\ell_r^*$ and $\ell_w^*$, which determines the specific form of storage from either \eqref{storage} or \eqref{storage2}. An example setting is shown in Figure~\ref{fig3}. Assume that the subpacketizations satisfy $\ell_1<\ell_2<\ell_3$, and therefore, for example the middle section which has a reading subpacketization of $\ell_2$ and a writing subpacketization of $\ell_1$ satisfying $\ell_1<\ell_2$, belongs to case~2 by definition. 

\subsubsection{Proof of Privacy}

The structures and sizes of the queries, updates and storage are determined at the initialization stage (when the subpacketizations are calculated and the storage is initialized), based on the given distortion budgets in the proposed scheme, and do not depend on each user's updating submodel index or the values of sparse updates. Moreover, the queries $Q_n$, updates $U_n$ and storage $S_n$ in this scheme are random noise terms that are independent of the values and positions of the sparse updates as well as the updating submodel index. Therefore, the proofs presented in Section~\ref{privacy_proof} for the privacy of submodel index, privacy of values of updates and security of submodels are valid in this section as well.

\section{Discussion and Conclusions}

In this paper, we first provided a basic PRUW scheme that results in the lowest known communication cost for PRUW and extended it to two cases, where two forms of sparsification are considered, in relation to private FSL, which is an application of PRUW. The first form is top $r$ sparsification, where only a selected number of parameters and updates are read and written in the reading and writing phases, respectively. These parameters/updates are chosen based on their significance. In order to satisfy the privacy constraint on the values of updates, we used a parameter/update permutation technique, which ensures the privacy of the indices of the sparse updating parameters, which in turn satisfies the privacy of the values of updates of all parameters. This permutation technique however requires additional \emph{noise-added permutation reversing matrices} to be stored in databases. Based on the structure and size of these matrices, the scheme is able to achieve asymptotic (large $N$) normalized reading and writing costs of $2r$ or $4r$, where $r$ is the sparsification rate.

The second scheme considers random sparsification in private FSL, and randomly chooses a pre-determined set of parameters (updates) in each reading (writing) subpacket to download (update). The problem setting is formulated in terms of a rate-distortion characterization, and the optimum reading and writing subpacketizations are calculated for given amounts of distortion allowed in the reading and writing phases, respectively. The resulting asymptotic normalized reading and writing costs are both equal to $2r$, where $r=1-\tilde{D}$, where $\tilde{D}$ is the distortion allowed. Since a fraction of $\tilde{D}$ parameters of the entire submodel are not read/updated, the sparsification rate for this case is $r=1-\tilde{D}$. It is clear that random sparsification outperforms (or performs equally) top $r$ sparsification in terms of the communication cost when similar sparsification rates are considered. However, random sparsification may not be as effective as top $r$ sparsification since it does not capture the most significant variations of the gradients in the SGD process, in relation to the underlying learning task. This may have an adverse effect on the model convergence time as well as on the accuracy of the trained model.   

\appendix
\section{Proof of Lemma~\ref{lem1}}\label{pf1}

\begin{Proof}
\begin{align}
    \frac{U_n}{f_k-\alpha_n}&=\frac{\sum_{i=1}^\ell \Tilde{\Delta}_{\theta,i} \prod_{j=1,j\neq i}^\ell (f_j-\alpha_n)+\prod_{j=1}^\ell (f_j-\alpha_n)(Z_0+\alpha_nZ_1+\dotsc+\alpha_n^{T_3-1}Z_{T_3-1})}{f_k-\alpha_n}\\
    &=\frac{\Tilde{\Delta}_{\theta,k} \prod_{j=1,j\neq k}^\ell (f_j-\alpha_n)}{f_k-\alpha_n}+\frac{\sum_{i=1,i\neq k}^\ell \Tilde{\Delta}_{\theta,i} \prod_{j=1,j\neq i}^\ell (f_j-\alpha_n)}{f_k-\alpha_n}\label{un2}\nonumber\\
    & \quad \quad+\frac{\prod_{j=1}^\ell(f_j-\alpha_n)(Z_0+\alpha_nZ_1+\dotsc+\alpha_n^{T_3-1}Z_{T_3-1})}{f_k-\alpha_n}.
\end{align}
Now consider,
\begin{align}
    \frac{\prod_{j=1,j\neq k}^\ell (f_j-\alpha_n)}{f_k-\alpha_n}&=\frac{(f_1-f_k+f_k-\alpha_n)}{f_k-\alpha_n}\prod_{j=2,j\neq k}^\ell (f_j-\alpha_n)\\
    &=\prod_{j=2,j\neq k}^\ell (f_j-\alpha_n)+(f_1-f_k)\frac{\prod_{j=2,j\neq k}^\ell (f_j-\alpha_n)}{f_k-\alpha_n}\\
    &=\prod_{j=2,j\neq k}^\ell (f_j-\alpha_n)+(f_1-f_k)\frac{(f_2-f_k+f_k-\alpha_n)}{f_k-\alpha_n}\prod_{j=3,j\neq k}^\ell (f_j-\alpha_n)\\
    &=\prod_{j=2,j\neq k}^\ell (f_j-\alpha_n)+(f_1-f_k)\prod_{j=3,j\neq k}^\ell (f_j-\alpha_n)\nonumber\\
    &\quad \quad \quad +(f_1-f_k)(f_2-f_k)\frac{\prod_{j=3,j\neq k}^\ell (f_j-\alpha_n)}{f_k-\alpha_n}\\
    & \quad \quad \quad \quad \quad \quad \vdots\nonumber\\
    &=\prod_{j=2,j\neq k}^\ell (f_j-\alpha_n)+(f_1-f_k)\prod_{j=3,j\neq k}^\ell (f_j-\alpha_n)+\dotsc\nonumber\\
    &\qquad+\prod_{i=1}^{k-2}(f_i-f_k)\prod_{j=k,j\neq k}^\ell (f_j-\alpha_n)+\prod_{i=1}^{k-1}(f_i-f_k)\frac{\prod_{j=k,j\neq k}^\ell(f_j-\alpha_n)}{f_k-\alpha_n}\\
    &=\prod_{j=2,j\neq k}^\ell (f_j-\alpha_n)+(f_1-f_k)\prod_{j=3,j\neq k}^\ell (f_j-\alpha_n)+\dotsc\nonumber\\
    &\qquad+\prod_{i=1}^{k-2}(f_i-f_k)\prod_{j=k+1}^\ell (f_j-\alpha_n)+\prod_{i=1}^{k-1}(f_i-f_k)\prod_{j=k+2}^\ell (f_j-\alpha_n)\nonumber\\
    &\qquad \qquad+\prod_{i=1,i\neq k}^{k+1}(f_i-f_k)\frac{\prod_{j=k+2}^\ell(f_j-\alpha_n)}{f_k-\alpha_n}\\
    & \quad \quad \quad \quad \quad \quad \vdots\nonumber\\
    &=\prod_{j=2,j\neq k}^\ell (f_j-\alpha_n)+(f_1-f_k)\prod_{j=3,j\neq k}^\ell (f_j-\alpha_n)+\dotsc\nonumber\\
    &\qquad+\prod_{i=1}^{k-1}(f_i-f_k)\prod_{j=k+2}^\ell (f_j-\alpha_n)+\prod_{i=1,i\neq k}^{k+1}(f_i-f_k)\prod_{j=k+3}^\ell(f_j-\alpha_n)+\dotsc\nonumber\\
    &\qquad\qquad+\prod_{i=1,i\neq k}^{\ell-2}(f_i-f_k)\frac{(f_{\ell-1}-f_k+f_k-\alpha_n)(f_\ell-\alpha_n)}{f_k-\alpha_n}\\
    &=\prod_{j=2,j\neq k}^\ell (f_j-\alpha_n)+(f_1-f_k)\prod_{j=3,j\neq k}^\ell (f_j-\alpha_n)+\dotsc\nonumber\\
    &\qquad+\prod_{i=1}^{k-1}(f_i-f_k)\prod_{j=k+2}^\ell (f_j-\alpha_n)+\prod_{i=1,i\neq k}^{k+1}(f_i-f_k)\prod_{j=k+3}^\ell(f_j-\alpha_n)+\dotsc\nonumber\\
    &\qquad\qquad+(f_\ell-\alpha_n)\prod_{i=1,i\neq k}^{\ell-2}(f_i-f_k)+\prod_{i=1,i\neq k}^{\ell-1}(f_i-f_k)\frac{(f_\ell-f_k+f_k-\alpha_n)}{f_k-\alpha_n}\\
    &=\prod_{j=2,j\neq k}^\ell (f_j-\alpha_n)+(f_1-f_k)\prod_{j=3,j\neq k}^\ell (f_j-\alpha_n)+\dotsc\nonumber\\
    &\qquad+\prod_{i=1}^{k-1}(f_i-f_k)\prod_{j=k+2}^\ell (f_j-\alpha_n)+\prod_{i=1,i\neq k}^{k+1}(f_i-f_k)\prod_{j=k+3}^\ell(f_j-\alpha_n)+\dotsc\nonumber\\
    &\qquad\qquad+(f_\ell-\alpha_n)\prod_{i=1,i\neq k}^{\ell-2}(f_i-f_k)+\prod_{i=1,i\neq k}^{\ell-1}(f_i-f_k)+\frac{\prod_{i=1,i\neq k}^{\ell}(f_i-f_k)}{f_k-\alpha_n}\\
    &=P_{\alpha_n}(\ell-2)+\frac{\prod_{i=1,i\neq k}^\ell (f_i-f_k)}{f_k-\alpha_n},
\end{align}
where $P_{\alpha_n}(\ell-2)$ is a polynomial in $\alpha_n$ of degree $\ell-2$. Therefore, from \eqref{un2},
\begin{align}
    \frac{U_n}{f_k-\alpha_n}&=\Tilde{\Delta}_{\theta,k}\left(P_{\alpha_n}(\ell-2)+\frac{\prod_{i=1,i\neq k}^\ell (f_i-f_k)}{f_k-\alpha_n}\right)+ P_{\alpha_n}(\ell+T_3-2),
\end{align}
since the second and third terms of \eqref{un2} result in a polynomial in $\alpha_n$ of degree $\ell+T_3-2$. Therefore,
\begin{align}
   \frac{U_n}{f_k-\alpha_n}=\frac{1}{f_k-\alpha_n}\Delta_{\theta,k}+P_{\alpha_n}(\ell+T_3-2).
\end{align}
\end{Proof}

\section{Proof of Lemma~\ref{lem2}}\label{pf2}

\begin{Proof}
\begin{align}
    \left(\frac{\prod_{r\in\mathcal{F}} (\alpha_r-\alpha_n)}{\prod_{r\in\mathcal{F}} (\alpha_r-f_k)}\right)\frac{1}{f_k-\alpha_n}&=\frac{1}{f_k-\alpha_n}\left(\frac{\prod_{r\in\mathcal{F}} (\alpha_r-f_k+f_k-\alpha_n)}{\prod_{r\in\mathcal{F}} (\alpha_r-f_k)}\right)\\
    &=\frac{1}{f_k-\alpha_n}\prod_{r\in\mathcal{F}}\left(1+\frac{f_k-\alpha_n}{\alpha_r-f_k}\right)\\
    &=\frac{1}{f_k-\alpha_n}+P_{\alpha_n}(|\mathcal{F}|-1).
\end{align}
\end{Proof}

\bibliographystyle{unsrt}
\bibliography{references-journal}

\begin{thebibliography}{10}

\bibitem{FL1}
H.~B. McMahan, E.~Moore, et~al.
\newblock Communication efficient learning of deep networks from decentralized
  data.
\newblock {\em AISTATS}, April 2017.

\bibitem{FL2}
Q.~Yang, Y.~Liu, T.~Chen, and Y.~Tong.
\newblock Federated machine learning: Concept and applications.
\newblock {\em ACM Trans. on Intelligent Systems and Technology}, 10(2):1--19,
  January 2019.

\bibitem{Advances}
P.~Kairouz, H.~B. McMahan, B.~Avent, A.~Bellet, M.~Bennis, et~al.
\newblock Advances and open problems in federated learning.
\newblock {\em Foundations and Trends in Machine Learning}, 14(1-2):1--210,
  June 2021.

\bibitem{magazine}
T.~Li, A.~K. Sahu, A.~S. Talwalkar, and V.~Smith.
\newblock Federated learning: Challenges, methods, and future directions.
\newblock {\em IEEE Signal Processing Magazine}, 37:50--60, May 2020.

\bibitem{comprehensive}
M.~Nasr, R.~Shokri, and A.~Houmansadr.
\newblock Comprehensive privacy analysis of deep learning: Passive and active
  white-box inference attacks against centralized and federated learning.
\newblock In {\em {IEEE} {SSP}}, May 2019.

\bibitem{MembershipInterference}
R.~Shokri, M.~Stronati, C.~Song, and V.~Shmatikov.
\newblock Membership inference attacks against machine learning models.
\newblock In {\em {IEEE} {SSP}}, May 2017.

\bibitem{featureLeakage}
L.~Melis, C.~Song, E.~De Cristofaro, and V.~Shmatikov.
\newblock Exploiting unintended feature leakage in collaborative learning.
\newblock In {\em {IEEE} {SSP}}, May 2019.

\bibitem{SecretSharer}
N.~Carlini, C.~Liu, U.~Erlingsson, J.~Kos, and D.~Song.
\newblock The secret sharer: Evaluating and testing unintended memorization in
  neural networks.
\newblock In {\em {USENIX}}, April 2019.

\bibitem{InvertingGradients}
J.~Geiping, H.~Bauermeister, H.~Droge, and M.~Moeller.
\newblock Inverting gradients--how easy is it to break privacy in federated
  learning?
\newblock In {\em {NeurIPS}}, December 2020.

\bibitem{DeepLeakage}
L.~Zhu, Z.~Liu, and S.~Han.
\newblock Deep leakage from gradients.
\newblock In {\em NeurIPS}, December 2019.

\bibitem{BeyondClassRepresentatives}
Z.~Wang, M.~Song, Z.~Zhang, Y.~Song, Q.~Wang, and H.~Qi.
\newblock Beyond inferring class representatives: User-level privacy leakage
  from federated learning.
\newblock In {\em IEEE Infocom}, April-May 2019.

\bibitem{PracticalSecureAgg}
K.~Bonawitz, V.~Ivanov, et~al.
\newblock Practical secure aggregation for privacy-preserving machine learning.
\newblock In {\em CCS}, October 2017.

\bibitem{DP}
C.~Dwork and A.~Roth.
\newblock The algorithmic foundations of differential privacy.
\newblock {\em Foundations and Trends in Theoretical Computer Science},
  9(3-4):211--407, August 2014.

\bibitem{reinforcement}
H.~Ono and T.~Takahashi.
\newblock Locally private distributed reinforcement learning.
\newblock Available online at arXiv:2001.11718.

\bibitem{avgDP}
Y.~Li, T.~Chang, and C.~Chi.
\newblock Secure federated averaging algorithm with differential privacy.
\newblock {\em IEEE MLSE}, September 2020.

\bibitem{cpSGD}
N.~Agarwal, A.~Suresh, F.~Yu, S.~Kumar, and H.~B. McMahan.
\newblock cp{SGD}: Communication-efficient and differentially-private
  distributed {SGD}.
\newblock In {\em NeurIPS}, December 2018.

\bibitem{PrivacyAmp}
B.~Balle, G.~Barthe, and M.~Gaboardi.
\newblock Privacy amplification by subsampling: Tight analyses via couplings
  and divergences.
\newblock In {\em NeurIPS}, December 2018.

\bibitem{cross}
M.~Heikkila, A.~Koskela, K.~Shimizu, S.~Kaski, and A.~Honkela.
\newblock Differentially private cross-silo federated learning.
\newblock Available online at arXiv:2007.05553.

\bibitem{language}
H.~B. McMahan, D.~Ramage, K.~Talwar, and L.~Zhang.
\newblock Learning differentially private recurrent language models.
\newblock In {\em ICLR}, May 2018.

\bibitem{DPFL}
S.~Asoodeh and F.~Calmon.
\newblock Differentially private federated learning: An information-theoretic
  perspective.
\newblock In {\em ICML-FL}, July 2020.

\bibitem{shuffle}
U.~Erlingsson, V.~Feldman, I.~Mironov, A.~Raghunathan, K.~Talwar, and
  A~Thakurta.
\newblock Amplification by shuffling: From local to central differential
  privacy via anonymity.
\newblock In {\em ACM-SIAM symposium on discrete algorithms}, January 2019.

\bibitem{PrivacyBlanket}
B.~Balle, J.~Bell, A.~Gascon, , and K.~Nissim.
\newblock The privacy blanket of the shuffle model.
\newblock In {\em CRYPTO}, August 2019.

\bibitem{shuffledDPFL}
A.~Girgis, D.~Data, et~al.
\newblock Shuffled model of differential privacy in federated learning.
\newblock In {\em AISTAT}, April 2021.

\bibitem{client}
R.~C. Geyer, T.~Klein, and M.~Nabi.
\newblock Differentially private federated learning: A client level
  perspective.
\newblock In {\em NeurIPS}, December 2017.

\bibitem{recent}
S.~Ulukus, S.~Avestimehr, M.~Gastpar, S.~A. Jafar, R.~Tandon, and C.~Tian.
\newblock Private retrieval, computing and learning: Recent progress and future
  challenges.
\newblock {\em IEEE JSAC}, 40(3):729--748, March 2022.

\bibitem{aggregation}
C.~Naim, R.~D'Oliveira, and S.~El Rouayheb.
\newblock Private multi-group aggregation.
\newblock {\em IEEE ISIT}, July 2021.

\bibitem{sparse1}
J.~Wangni, J.~Wang, et~al.
\newblock Gradient sparsification for communication-efficient distributed
  optimization.
\newblock In {\em NeurIPS}, December 2018.

\bibitem{GGS}
S.~Li, Q.~Qi, et~al.
\newblock {GGS}: General gradient sparsification for federated learning in edge
  computing.
\newblock In {\em IEEE ICC}, June 2020.

\bibitem{adaptive}
P~Han, S.~Wang, and K.~Leung.
\newblock Adaptive gradient sparsification for efficient federated learning: An
  online learning approach.
\newblock In {\em IEEE ICDCS}, November 2020.

\bibitem{conv}
S.~Shi, K.~Zhao, Q.~Wang, Z.~Tang, and X.~Chu.
\newblock A convergence analysis of distributed {SGD} with
  communication-efficient gradient sparsification.
\newblock In {\em IJCAI}, August 2019.

\bibitem{conv2}
D.~Alistarh, T.~Hoefler, M.~Johansson, S.~Khirirat, N.~Konstantinov, and
  C.~Renggli.
\newblock The convergence of sparsified gradient methods.
\newblock In {\em NeurIPS}, December 2018.

\bibitem{overtheair}
Y.~Sun, S.~Zhou, Z.~Niu, and D.~Gunduz.
\newblock Time-correlated sparsification for efficient over-the-air model
  aggregation in wireless federated learning.
\newblock Available online at arXiv:2202.08420.

\bibitem{rtopk}
L.~Barnes, H.~Inan, B.~Isik, and A.~Ozgur.
\newblock {rTop}-$k$: A statistical estimation approach to distributed {SGD}.
\newblock {\em IEEE JSAIT}, 1(3):897--907, November 2020.

\bibitem{timecorr}
E.~Ozfatura, K.~Ozfatura, and D.~Gunduz.
\newblock Time-correlated sparsification for communication-efficient federated
  learning.
\newblock In {\em IEEE ISIT}, July 2021.

\bibitem{qsl}
D.~Basu, D.~Data, C.~Karakus, and S.~Diggavi.
\newblock Qsparse-local-{SGD}: Distributed {SGD} with quantization,
  sparsification, and local computations.
\newblock {\em IEEE JSAIT}, 1(1):217--226, May 2020.

\bibitem{fedpaq}
A.~Reisizadeh, A.~Mokhtari, H.~Hassani, A.~Jadbabaie, and R.~Pedarsani.
\newblock Fedpaq: A communication-efficient federated learning method with
  periodic averaging and quantization.
\newblock In {\em AISTATS}, August 2020.

\bibitem{qsgd}
D.~Alistarh, D.~Grubic, J.~Li, R.~Tomioka, and M.~Vojnovic.
\newblock {QSGD}: Communication-efficient {SGD} via gradient quantization and
  encoding.
\newblock In {\em NeurIPS}, December 2017.

\bibitem{constraints}
N.~Shlezinger, M.~Chen, Y.~Eldar, H.~Poor, and S.~Cui.
\newblock Federated learning with quantization constraints.
\newblock In {\em IEEE ICASSP}, May 2020.

\bibitem{billion}
C.~Niu, F.~Wu, S.~Tang, L.~Hua, R.~Jia, C.~Lv, Z.~Wu, and G.~Chen.
\newblock Billion-scale federated learning on mobile clients: A submodel design
  with tunable privacy.
\newblock In {\em MobiCom}, April 2020.

\bibitem{secureFSL}
C.~Niu, F.~Wu, S.~Tang, L.~Hua, R.~Jia, C.~Lv, Z.~Wu, and G.~Chen.
\newblock Secure federated submodel learning.
\newblock Available online at arXiv:1911.02254.

\bibitem{paper1}
M.~Kim and J.~Lee.
\newblock Information-theoretic privacy in federated submodel learning.
\newblock {\em ICT express}, February 2022.

\bibitem{rw_jafar}
Z.~Jia and S.~A. Jafar.
\newblock {$X$}-secure {$T$}-private federated submodel learning.
\newblock In {\em IEEE ICC}, June 2021.

\bibitem{ourICC}
S.~Vithana and S.~Ulukus.
\newblock Efficient private federated submodel learning.
\newblock In {\em IEEE ICC}, May 2022.

\bibitem{dropout}
Z.~Jia and S.~A. Jafar.
\newblock ${X}$-secure ${T}$-private federated submodel learning with elastic
  dropout resilience.
\newblock {\em IEEE Trans. on Info. Theory}, 68(8):5418--5439, August 2022.

\bibitem{pruw}
S.~Vithana and S.~Ulukus.
\newblock Private read update write {(PRUW)} with storage constrained
  databases.
\newblock In {\em IEEE ISIT}, June 2022.

\bibitem{sparse}
S.~Vithana and S.~Ulukus.
\newblock Private federated submodel learning with sparsification.
\newblock In {\em IEEE ITW}, November 2022.
\newblock Also available online at arXiv:2205.15992.

\bibitem{rd}
S.~Vithana and S.~Ulukus.
\newblock Rate distortion tradeoff in private read update write in federated
  submodel learning.
\newblock In {\em Asilomar Conference}, October 2022.
\newblock Also available online at arXiv:2206.03468.

\bibitem{original}
B.~Chor, E.~Kushilevitz, O.~Goldreich, and M.~Sudan.
\newblock Private information retrieval.
\newblock {\em Journal of the ACM}, 45(6):965--981, November 1998.

\bibitem{PIR}
H.~Sun and S.~A. Jafar.
\newblock The capacity of private information retrieval.
\newblock {\em IEEE Trans. on Info. Theory}, 63(7):4075--4088, July 2017.

\bibitem{leaky}
I.~Samy, M.~Attia, R.~Tandon, and L.~Lazos.
\newblock Asymmetric leaky private information retrieval.
\newblock {\em IEEE Trans. on Info. Theory}, 67(8):5352--5369, August 2021.

\bibitem{ChaoTian}
C.~Tian, H.~Sun, and J.~Chen.
\newblock Capacity-achieving private information retrieval codes with optimal
  message size and upload cost.
\newblock {\em IEEE Trans. on Info. Theory}, 65(11):7613--7627, November 2019.

\bibitem{semanticPIR}
S.~Vithana, K.~Banawan, and S.~Ulukus.
\newblock Semantic private information retrieval.
\newblock {\em IEEE Trans. on Info. Theory}, 68(4):2635--2652, April 2022.

\bibitem{colluding}
H.~Sun and S.~A. Jafar.
\newblock The capacity of robust private information retrieval with colluding
  databases.
\newblock {\em IEEE Trans. on Info. Theory}, 64(4):2361--2370, April 2018.

\bibitem{coded}
K.~Banawan and S.~Ulukus.
\newblock The capacity of private information retrieval from coded databases.
\newblock {\em IEEE Trans. on Info. Theory}, 64(3):1945--1956, March 2018.

\bibitem{incorrectconjecture}
H.~Sun and S.~A. Jafar.
\newblock Private information retrieval from {MDS} coded data with colluding
  servers: Settling a conjecture by {F}reij-{H}ollanti et al.
\newblock {\em IEEE Trans. on Info. Theory}, 64(2):1000--1022, February 2018.

\bibitem{codedcolluded}
L.~Holzbaur, R.~Freij-Hollanti, J.~Li, and C.~Hollanti.
\newblock Towards the capacity of private information retrieval from coded and
  colluding servers.
\newblock {\em IEEE Trans. on Info. Theory}, 2021.

\bibitem{sideinfo}
S.~Kadhe, B.~Garcia, A.~Heidarzadeh, S.~El Rouayheb, and A.~Sprintson.
\newblock Private information retrieval with side information.
\newblock {\em IEEE Trans. on Info. Theory}, 66(4):2032--2043, April 2020.

\bibitem{singleDB}
S.~Li and M.~Gastpar.
\newblock Single-server multi-message private information retrieval with side
  information: the general cases.
\newblock In {\em IEEE ISIT}, June 2020.

\bibitem{byzantine}
K.~Banawan and S.~Ulukus.
\newblock The capacity of private information retrieval from {B}yzantine and
  colluding databases.
\newblock {\em IEEE Trans. on Info. Theory}, 65(2):1206--1219, February 2019.

\bibitem{SecureStorage}
H.~Yang, W.~Shin, and J.~Lee.
\newblock Private information retrieval for secure distributed storage systems.
\newblock {\em IEEE Trans. on Info. Forensics and Security}, 13(12):2953--2964,
  December 2018.

\bibitem{XSTPIR}
Z.~Jia and S.~A. Jafar.
\newblock {$X$}-secure {$T$}-private information retrieval from {MDS} coded
  storage with {B}yzantine and unresponsive servers.
\newblock {\em IEEE Trans. on Info. Theory}, 66(12):7427--7438, December 2020.

\bibitem{CodeColludeByzantinePIR}
R.~Tajeddine, O.~W. Gnilke, D.~Karpuk, R.~Freij-Hollanti, and C.~Hollanti.
\newblock Private information retrieval from coded storage systems with
  colluding, {B}yzantine, and unresponsive servers.
\newblock {\em IEEE Trans. on Info. Theory}, 65(6):3898--3906, June 2019.

\bibitem{smallfields}
J.~Xu and Z.~Zhang.
\newblock Building capacity-achieving {PIR} schemes with optimal
  sub-packetization over small fields.
\newblock In {\em IEEE ISIT}, June 2018.

\bibitem{Kumar_PIRarbCoded}
S.~Kumar, H.-Y. Lin, E.~Rosnes, and A.~G. i~Amat.
\newblock Achieving maximum distance separable private information retrieval
  capacity with linear codes.
\newblock {\em IEEE Trans. on Info. Theory}, 65(7):4243--4273, July 2019.

\bibitem{YamamotoPIR}
T.~Chan, S.~Ho, and H.~Yamamoto.
\newblock Private information retrieval for coded storage.
\newblock In {\em IEEE ISIT}, June 2015.

\bibitem{VardyConf2015}
A.~Fazeli, A.~Vardy, and E.~Yaakobi.
\newblock Codes for distributed {PIR} with low storage overhead.
\newblock In {\em IEEE ISIT}, June 2015.

\bibitem{MultiroundPIR}
H.~Sun and S.~A. Jafar.
\newblock Multiround private information retrieval: Capacity and storage
  overhead.
\newblock {\em IEEE Trans. on Info. Theory}, 64(8):5743--5754, August 2018.

\bibitem{SPIR}
H.~Sun and S.~A. Jafar.
\newblock The capacity of symmetric private information retrieval.
\newblock {\em IEEE Trans. on Info. Theory}, 65(1):329--322, January 2019.

\bibitem{MMPIR}
K.~Banawan and S.~Ulukus.
\newblock Multi-message private information retrieval: Capacity results and
  near-optimal schemes.
\newblock {\em IEEE Trans. on Info. Theory}, 64(10):6842--6862, October 2018.

\bibitem{evesdroppers}
Q.~Wang, H.~Sun, and M.~Skoglund.
\newblock The capacity of private information retrieval with eavesdroppers.
\newblock {\em IEEE Trans. on Info. Theory}, 65(5):3198--3214, May 2019.

\bibitem{cache}
Q.~Wang, H.~Sun, and M.~Skoglund.
\newblock Cache-aided private information retrieval.
\newblock {\em IEEE Trans. on Info. Theory}, 65(5):3198--3214, May 2019.

\bibitem{PSI}
Z.~Wang, K.~Banawan, and S.~Ulukus.
\newblock Private set intersection: A multi-message symmetric private
  information retrieval perspective.
\newblock {\em IEEE Trans. on Info. Theory}, 68(3):2001--2019, March 2022.

\bibitem{CSA}
Z.~Jia, H.~Sun, and S.~A. Jafar.
\newblock Cross subspace alignment and the asymptotic capacity of {$X$}-secure
  {$T$}-private information retrieval.
\newblock {\em IEEE Trans. on Info. Theory}, 65(9):5783--5798, September 2019.

\bibitem{OTP}
C.~E. Shannon.
\newblock Communication theory of secrecy systems.
\newblock {\em Bell System Technical Journal}, 28(4):656--715, October 1949.

\bibitem{wireless}
A.~Yener and S.~Ulukus.
\newblock Wireless physical layer security: Lessons learned from information
  theory.
\newblock {\em Proceedings of the IEEE}, 103(10):1814--1825, October 2015.

\bibitem{DOF}
J.~Xie and S.~Ulukus.
\newblock Secure degrees of freedom of multi-user networks: One-time-pads in
  the air via alignment.
\newblock {\em Proceedings of the IEEE}, 103(10):1857--1873, October 2015.

\end{thebibliography}
\end{document}